%% file: article.tex
\documentclass[journal=jpccck,manuscript=article,layout=traditional]{achemso}
\pdfoutput=1
\usepackage[T1]{fontenc} 
\usepackage{subcaption}
\usepackage{graphicx}
\usepackage{mathtools}
\usepackage{rotating}
\usepackage{listings}
\usepackage[version=4]{mhchem}
\usepackage{siunitx}
\usepackage[section]{placeins}
\usepackage{adjustbox}

\DeclareSIUnit[number-unit-product = {\,}]\cal{cal}
\DeclareSIUnit{\angstrom}{\textup{\AA}}

\DeclareMathOperator{\BO}{BO}

\SectionNumbersOn

\title{Managing Expectations and Imbalanced Training Data in Reactive Force Field Development: an Application to Water Adsorption on Alumina}

\author{Lo\"ic Dumortier}
\affiliation[IFPEN-Paris]{IFP Energies nouvelles, 1 et 4 Avenue de Bois-Pr\'eau, 92852 Rueil-Malmaison, France}
\alsoaffiliation[CMM]{Center for Molecular Modeling (CMM), Ghent University, Technologiepark-Zwijnaarde 46, B-9052, Zwijnaarde, Belgium}

\author{C\'eline Chizallet}
\affiliation[IFPEN-Solaize]{IFP Energies nouvelles, Rond-point de l'échangeur de Solaize, BP3, 69360 Solaize, France}

\author{Benoit Creton}
\affiliation[IFPEN-Paris]{IFP Energies nouvelles, 1 et 4 Avenue de Bois-Pr\'eau, 92852 Rueil-Malmaison, France}

\author{Theodorus de Bruin}
\affiliation[IFPEN-Paris]{IFP Energies nouvelles, 1 et 4 Avenue de Bois-Pr\'eau, 92852 Rueil-Malmaison, France}

\author{Toon Verstraelen}
\affiliation[CMM]{Center for Molecular Modeling (CMM), Ghent University, Technologiepark-Zwijnaarde 46, B-9052, Zwijnaarde, Belgium}
\email{toon.verstraelen@ugent.be}

\begin{document}
\begin{abstract}
ReaxFF is a computationally efficient model for reactive molecular dynamics simulations, which has been applied to a wide variety of chemical systems.
When ReaxFF parameters are not yet available for a chemistry of interest, they must be (re)optimized, for which one defines a set of training data that the new ReaxFF parameters should reproduce.
ReaxFF training sets typically contain diverse properties with different units, some of which are more abundant (by orders of magnitude) than others.
To find the best parameters, one conventionally minimizes a weighted sum of squared errors over all data in the training set.
One of the challenges in such numerical optimizations is to assign weights so that the optimized parameters represent a good compromise between all the requirements defined in the training set.
This work introduces a new loss function, called Balanced Loss, and a workflow that replaces weight assignment with a more manageable procedure.
The training data is divided into categories with corresponding ``tolerances'', i.e. acceptable root-mean-square errors for the categories, which define the expectations for the optimized ReaxFF parameters.
Through the Log-Sum-Exp form of Balanced Loss, the parameter optimization is also a validation of one's expectations, providing meaningful feedback that can be used to reconfigure the tolerances if needed.
The new methodology is demonstrated with a non-trivial parameterization of ReaxFF for water adsorption on alumina.
This results in a new force field that reproduces both rare and frequent properties of a validation set not used for training.
We also demonstrate the robustness of the new force field with a molecular dynamics simulation of water desorption from a \ce{\gamma-Al2O3} slab model.
\end{abstract}

\begin{tocentry}
    \centering
    \includegraphics{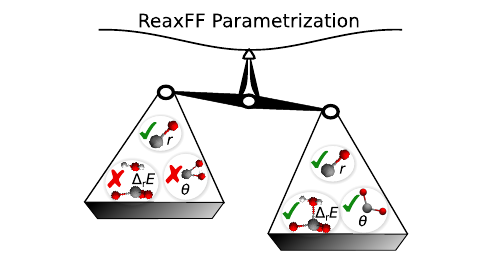}
\end{tocentry}
\section{Introduction}

Reactive force fields are widely used in Molecular Dynamics (MD) simulations because they combine low computational cost, close to that of Molecular Mechanics (MM) models, with the ability to describe chemical events, similar to more expensive Quantum Mechanics (QM) methods, such as Density Functional Theory (DFT).
Unlike hybrid QM/MM schemes,\cite{Brunk2015} reactive force fields handle many simultaneous chemical reactions throughout the simulation cell, not just in one predefined active site.
This is advantageous for the direct simulation of reaction networks\cite{Dntgen2015} of complex chemical processes such as combustion, \cite{Chenoweth2009} pyrolysis, \cite{Ding2013, Li2021} chemisorption, \cite{Fogarty_2010, Mueller2010} catalysis, \cite{Mueller2010} mechanochemistry, \cite{muller_2016} crack propagation, \cite{Rimsza2017} nucleation, \cite{Mao2017, Lei2019} and so on.
ReaxFF is one of the most established reactive force fields and is efficient enough to perform multi-nanosecond MD of systems with thousands of atoms using only a single high-performance compute node. \cite{van_duin_reaxff_2001, vanDuin2003, senftle_reaxff_2016}
Compared to other popular reactive force fields such as Tersoff, \cite{Tersoff1988} AIREBO-M, \cite{OConnor2015} or COMB3, \cite{Zhang2019} ReaxFF has been parameterized for more diverse chemical spaces. \cite{senftle_reaxff_2016}
More recently, machine learning potentials have also been proposed for reactive MD simulations.\cite{Rowe2020, Xue2021, Schreiner2022}
All of these models share the ambition to simulate complex chemical systems at a computational cost that scales like MM models.

The computational efficiency of reactive force fields comes at a price.
They are generally empirical models, sometimes inspired by physical principles, whose parameters must be fitted to reproduce a chemistry of interest.
Such parameterization is fraught with challenges: the collection of reference data sets for training and validation, the choice of numerical optimization algorithm, the selection of parameters to optimize, the computational burden of the parameter optimization, and so on.
Specifically for ReaxFF, many optimization algorithms have been proposed and tested, \cite{Pahari2011, Deetz2014, JaramilloBotero2014, Dittner2015, Trnka2017, Furman2018, shchygol_reaxff_2019, Sengul2021, kaymak_jaxreaxff_2022} while the design of reference data sets has received much less attention.
For example, ReaxFF parameters are rarely published with their training sets in a reusable form, save for a few exceptions. \cite{muller_2016, Trnka2017, shchygol_reaxff_2019, Sengul2021, Komissarov2023, FreitasGustavo2023}
However, these data are vital, as they specify the requirements for the optimized parameters and no models will ever outperform the data it was trained on.

A conventional ReaxFF training set consists of various target properties of relevant molecular or periodic structures, including internal coordinates, energy differences and atomic forces.
Reaction energies and barriers are obviously important for reactive force fields, but for a training set of $N$ systems, one has at most $N-1$ independent energy differences and many more internal coordinates and atomic forces.
In the context of machine learning potentials, this imbalance is addressed by weighting data categories (typically energies and forces) inversely proportional to their prevalence, \cite{Wang2018, Unke2019, CoolsCeuppens2022} but this practice is less established in the context of ReaxFF.
Moreover, in ReaxFF, such weights are often adjusted empirically.
For example, one gives more weight to an important subset of the training data in order to prioritize the performance of the trained model for that subset.\cite{Furman2018, fedkin_development_2019, brown_reaxff_2022}
Conversely, the model of interest may also be inherently limited for some subsets of reference data, making it pointless to give high weight to such subsets.
These subjective motivations mean that training set design requires expert judgment.
To make this task more accessible to a broader audience, this paper introduces a new loss function and an intuitive workflow for reweighting training data, called Balanced Loss.
It naturally takes into account data imbalance and inherent strengths and weaknesses of the model being trained.
A ReaxFF parameterization is used as a case study in this paper, because we believe ReaxFF can greatly benefit from Balanced Loss, but the methodology is general enough to be applied to other (even non-chemical) parameterizations with similar challenges.\cite{verstraelen_computation_2012, bureekaew_mofff_2013, Grimme2017, Komissarov2021, kaymak_jaxreaxff_2022, Jicum2022, wlodarczyk_mixing_2022}
As software tools and algorithms for (re)parameterizating (reactive) force fields improve,  \cite{Dittner2015, Furman2018, shchygol_reaxff_2019, komissarov_params_2021, Sengul2021, kaymak_jaxreaxff_2022, FreitasGustavo2022} we expect that more practitioners to face the challenge of data imbalance, also for machine learning potentials that are trained on increasingly large and diverse data sets. \cite{Smith2017, Chen2022, Takamoto2022}

Alumina provides a great test case for ReaxFF parameterization because it is a versatile and widely used material in the chemical industry with a complex chemistry, \cite{hart_alumina_1990, lefevre_hydration_2002, arrouvel_effects_2004, lagauche_thermodynamic_2017, pigeon_revisiting_2022} also at scales out of reach for DFT.
Alumina has many known polymorphs, including \ce{\gamma-Al2O3}, \ce{\alpha-Al2O3}, \ce{\delta-Al2O3} and \ce{\theta-Al2O3}, \cite{levin_metastable_1998} of which \ce{\gamma-Al2O3} is the most relevant for catalytic applications. \cite{trueba_gamma-alumina_2005, euzen_alumina_2002}
For example, alumina selectively adsorbs unwanted elements such as sulfur and can be used as a catalyst for the dehydration of alcohols to ethers and olefins. \cite{knozinger_dehydration_1968, knozinger_dehydration_1972, phung_study_2014, kohl_chapter_1997, larmier_mechanistic_2015, jain_catalytic_1967}
However, the main application of \ce{\gamma-Al2O3} is in the automotive and petrochemical industries, where it serves as a support for other heterogeneous catalysts such as metals, metal sulfides or metal oxides. \cite{trueba_gamma-alumina_2005, pascal_catalysis_2013, coperet_surface_2016, boudart_catalysis_1969}
Despite their massive use in industry, the design of alumina-supported catalysts is an empirical process, partly due to the limited fundamental understanding of the materials involved.
For example, the exact structure of \ce{\gamma-Al2O3} is still under discussion due to its poor crystallinity. \cite{trueba_gamma-alumina_2005, prins_structure_2020}
Also, the microscopic mechanisms at the water-alumina interface during support preparation, metal phase impregnation, shaping and use as a catalyst remain unclear. \cite{VALERO2020539}

The formation, stability and structure of \ce{\gamma-Al2O3} are controlled by hydration and dehydration processes. \cite{krokidis_theoretical_2001, wischert_gamma-alumina_2012, lagauche_thermodynamic_2017, pigeon_revisiting_2022, digne_hydroxyl_2002, digne_use_2004}
The $\gamma$ phase is formed upon dehydration of boehmite at temperatures between \SI{700}{\kelvin} and \SI{800}{\kelvin}.
Once formed, the $\gamma$ polymporph remains stable up to \SI{1100}{\kelvin} under dry conditions. \cite{paglia_boehmite_2004, trueba_gamma-alumina_2005, euzen_alumina_2002}
\ce{\gamma-Al2O3} transitions to other polymorphs upon further increase in temperature and/or water partial pressure.
Having both Lewis acid and basic sites on the surface, alumina can react with water in several ways depending on temperature, water partial pressure (for gas/solid interfaces) and pH (for liquid/solid interfaces).
\cite{lefevre_hydration_2002, arrouvel_effects_2004, wischert_gamma-alumina_2012, lagauche_thermodynamic_2017, reocreux_reactivity_2019, pigeon_revisiting_2022}
Water can adsorb without dissociation by forming an Al-O bond.
One of the O-H bonds of water may then dissociate and react with a surface Al-O pair, resulting in two hydroxyl groups, called aluminols.
Dissociative adsorption is reported to be more prevalent at crystal surface defects, leading to an ``etching''-like degradation at these positions. \cite{MARDILOVICH1995131, Brown1999}
It is clear that the chemistry at the \ce{H2O}/\ce{\gamma-Al2O3}interface is highly complex and depends on an interplay of multiple microscopic mechanisms and external conditions.

Molecular simulation of the \ce{H2O}/\ce{\gamma-Al2O3} interface is a promising but ambitious method to improve our understanding of widely used supported catalysts and to pave the way towards their rational design.
DFT has often been used to model the \ce{H2O}/\ce{\gamma-Al2O3} interface. \cite{arrouvel_effects_2004, batista_beyond_2019, krokidis_theoretical_2001, pigeon_revisiting_2022,digne_hydroxyl_2002, digne_use_2004, batista_structure_2023}
Ideally, sufficiently large atomistic models are considered to avoid artificial spatial correlations, to introduce defects at low concentrations, to include both support and catalyst, and to mimic realistic water concentrations. \cite{Pitman2012, Rimsza2016, Porter2021}
Because larger models also have a larger configurational space, with many local minima on the potential energy surface, their properties can no longer be simulated with static calculations and one should resort to MD to sample all relevant configurations. \cite{vanGunsteren1990}
Linear-scaling DFT implementations\cite{Nakata2020, Khne2020} have enabled ab initio MD simulations of the \ce{H2O}/\ce{\gamma-Al2O3} interface,\cite{raybaud_morphology_2001,motta_aimd_2012,ngouana-wakou_atomistic_2017} but they are still computationally demanding compared to reactive force fields.
Compared to linear-scaling DFT, ReaxFF has a much lower computational cost, allowing for large-scale MD simulations of alumina.\cite{Zhang2004, Russo2011, Joshi2013, Sen2014, Joshi2014, Hong2015, Gunkelmann2018, Ramrez2019, Rosandi2020}
The first alumina and water ReaxFF parameters were proposed by Zhang \textit{et al.}, \cite{Zhang2004} and these were later refined and extended by Joshi \textit{et al.}\ for aluminosilicates and water, \cite{Joshi2013, Joshi2014} which is particularly relevant for simulations of alumina-supported catalysts.
However, as shown in the results, the state-of-the-art ReaxFF parameters by Joshi \textit{et al.}\ poorly reproduce DFT reference data for water adsorption on alumina.
This motivated us to demonstrate the relevance of Balanced Loss with a reparameterization of ReaxFF for \ce{H2O}/\ce{\gamma-Al2O3} interactions, using DFT data from the literature. \cite{raybaud_morphology_2001, digne_use_2004, chiche_growth_2009, batista_structure_2023}

The rest of the paper is structured as follows.
Section~\ref{sec:methodology} contains the methodological details of the study: a brief overview of ReaxFF, the generation of the training and validation data sets, the parameter selection and the optimization algorithm.
The Balanced Loss function and workflow are described and motivated in detail in section~\ref{sec:BL}.
Section~\ref{sec:res_discussion} presents the results of the ReaxFF training and validation, and it demonstrates the suitability of the resulting force field for MD simulations.
The last section formulates the main conclusions and gives an outlook on future work.

    \section{Methodology}
    \label{sec:methodology}
\subsection{ReaxFF Reactive Force Fields}

ReaxFF was developed and introduced in 2001 by van Duin \textit{et al.}\ for reactive MD simulations, initially of hydrocarbons\cite{van_duin_reaxff_2001}, and has since been regularly extended to other chemistries. \cite{senftle_reaxff_2016}
Like all force fields, it is a mathematical model of the interactions between atoms in a molecule or a condensed phase, as a function of the Cartesian coordinates of the atomic nuclei.
Unlike most classical force fields, it can describe bond breaking and formation.

The ReaxFF potential energy of an atomistic model is defined as:
\begin{equation}
    \label{eq:full_energy_reaxff}
    \begin{aligned}
        E_\text{system}&=E_\text{bond}+E_\text{over}+E_\text{under}+E_\text{val}+E_\text{tors}\\
        &\hphantom{=} +E_\text{vdW}+E_\text{charge}+E_\text{specific}\\
    \end{aligned}
\end{equation}
where $E_\text{bond}$ describes the energy of an atom pair in all relevant regimes: bonded, in transition states and dissociated. $E_\text{over}$ and $E_\text{under}$ are correction terms for over- and under-coordination, respectively.
$E_\text{val}$ is the valence angle energy and $E_\text{tors}$ is the torsional angle energy between four different particles.
Non-covalent interactions are modeled with $E_\text{charge}$ and $E_\text{vdW}$, the charge and the van der Waals interactions, respectively.
The atomic charges are variable and account for polarization and Coulomb forces.
\cite{mortier_eem_1986, verstraelen_acks2_2013}
In addition to these commonly used energy terms, ReaxFF contains additional contributions for specific use cases, grouped into $E_\text{specific}$, which are not used in this work.

The covalent terms depend on bond orders (BO), which are defined for each pair of atoms and allow ReaxFF to describe bond breaking and formation processes in chemical reactions.
The uncorrected bond order of a pair of atoms consists of three terms, each corresponding to one type of covalent bond, $\sigma$, $\pi$ and $\pi\pi$:
\begin{equation}
    \label{eq:bond_order_uncorrected}
    \begin{aligned}
        \BO^{'}_{ij} &= \BO^{'}_{ij,\sigma} + \BO^{'}_{ij,\pi} + \BO^{'}_{ij,\pi \pi} \\
        &=
        \exp\left(p_{\text{bo},1} \left(\frac{r_{ij}}{r^\sigma_0}\right)^{p_{\text{bo},2}}\right)
        + \exp\left(p_{\text{bo},3} \left(\frac{r_{ij}}{r^\pi_0}\right)^{p_{\text{bo},4}}\right)
        + \exp\left(p_{\text{bo},5} \left(\frac{r_{ij}}{r^{\pi \pi}_0}\right)^{p_{\text{bo},6}}\right) \\
    \end{aligned}
\end{equation}
where $p_\text{bo,\{x\}}$ represent tunable parameters that can be different for each pair of chemical elements.
$r_{ij}$ is the interatomic distance and $r^\sigma_0$, $r^\pi_0$ and $r^{\pi \pi}_0$ are the element-specific $\sigma$, $\pi$ and $\pi\pi$ equilibrium bond lengths, respectively.
The expression for the uncorrected bond orders in Eq.~\eqref{eq:bond_order_uncorrected} features only a small subset of all the adjustable ReaxFF parameters.
ReaxFF has additional equations (with more parameters) to convert uncorrected to corrected bond orders, which are then used in expressions for the covalent energy terms.
A full description can be found in the AMS documentation\cite{ffield_doc} and in the supporting information of Ref.~\citenum{Chenoweth2008}.

ReaxFF has been implemented in several software packages.
The most established ones are the original ``Standalone ReaxFF'' distributed by van Duin,
the commercial implementation in the Amsterdam Modeling Suite (AMS)\cite{ams}
and the open-source version in the LAMMPS package. \cite{plimpton_fast_1995, thompson_lammps_2022}
In this paper, the ReaxFF implementation from AMS (release 2023.101) is used.
The parameter optimization, discussed below, is implemented with ParAMS, \cite{params_doc, komissarov_params_2021} which is a recently developed tool in AMS for the parameterization of approximate potential energy surfaces, such as ReaxFF or Density Functional Tight-Binding (DFTB) models.\cite{Gaus2011, Grimme2017}
In addition, the Atomistic Simulation Environment\cite{ase-paper} was used for processing DFT calculations in the training and validation sets.
Visual Molecular Dynamics (VMD) is used for the 3D visualizations in this work.\cite{Humphrey1996}
\subsection{Training Set Development}
\label{sec:trainingset_dev}

The development of the training set goes through the following steps: (i) the selection of bulk, (hydrated) surface and (hydrated) edge structures, (ii) periodic DFT reference calculations on these structures and (iii) the selection of properties from these calculations as training targets.

\begin{figure}
    \centering
    \includegraphics{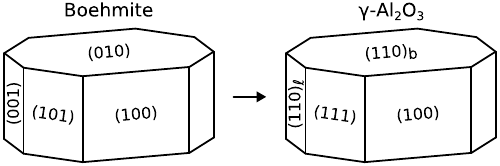}
    \caption{
        Overview of the crystal surfaces of boehmite and \ce{\gamma-Al2O3},
        with the relation between the two as proposed in Ref.~\citenum{pigeon_revisiting_2022}.
    }
    \label{fig:crystsurfaces}
\end{figure}

\paragraph{(i) Structures.}
A training set for optimizing ReaxFF parameters requires reference structures and associated training targets, such as internal coordinates or energies, that ReaxFF should reproduce.
The reference structures were taken from previous publications\cite{raybaud_morphology_2001, digne_use_2004, chiche_growth_2009, batista_structure_2023} and can be divided into five groups, summarized in Table~\ref{tab:structures_summary} and described in more detail below.
For clarity, the relevant crystal surfaces are shown in Figure~\ref{fig:crystsurfaces}.
Note that \ce{\gamma-Al2O3} is industrially the most relevant material, yet other forms of (hydrated) aluminum oxide were included, most notably boehmite, to increase the diversity of the training data.
A complete list of structures is provided in Table S1 of the Supporting Information.
\begin{itemize}
    \item
    The group of \textbf{Bulk Structures} contains 3D-periodic models of boehmite, \ce{\gamma-Al2O3} and \ce{\alpha-Al2O3}. \cite{krokidis_theoretical_2001}

    \item
    The group of \textbf{\ce{\gamma-Al2O3} Surfaces} is based on three different slab models, cut along the (100), (110) or (111) crystal planes.
    In addition to the bare surfaces, structures are included with an increasing number of water molecules adsorbed on the surface. \cite{digne_use_2004}
    All the structures were published before a distinction was made between the lateral (110)$_\ell$ and basal (110)$_\text{b}$ surfaces of \ce{\gamma-Al2O3}, as shown in Figure~\ref{fig:crystsurfaces}. \cite{pigeon_revisiting_2022, batista_structure_2023}
    The (110) \ce{\gamma-Al2O3} surfaces in the training set are in fact all lateral (110)$_\ell$ surfaces.

    \item
    The group of \textbf{Boehmite Surfaces} contains slabs with four surface orientations: (101), (010), (100) or (001).
    Boehmite already contains water in its bulk structure, which is preserved upon cleaving the slabs.
    In addition to the bare slabs, some have additional water molecules adsorbed.
    \cite{chiche_growth_2009, raybaud_morphology_2001}

    \item
    The group of \textbf{\ce{\gamma-Al2O3} Edges} comprises structures that represent the edge between surface orientations (100) and (110).
    In addition to the bare edge structure, six structures with an increasing number of adsorbed water molecules are included. \cite{batista_beyond_2019}

    \item
    The group of \textbf{Small Molecules} contains two structures: a $\gamma$-alumina monomer, \ce{[Al(OH)4]^-H^+}, and water. \cite{kennes_multiscale_2022}

\end{itemize}
\begin{table}
    \centering
    \begin{tabular}{llp{4cm}}
    Group & Structure & Nr.\ of structures with additional water\\
    \hline
    Bulk Structures
    & Bulk \ce{\alpha-Al2O3} & --\\
    & Bulk boehmite & --\\
    & Bulk \ce{\gamma-Al2O3} & --\\
    \hline
    \ce{\gamma-Al2O3} Surfaces
    & \ce{\gamma-Al2O3} (100) & 4\\
    & \ce{\gamma-Al2O3} (110) & 6\\
    & \ce{\gamma-Al2O3} (111) & 3\\
    \hline
    Boehmite Surfaces
    & Boehmite (101) & 1\\
    & Boehmite (010) & 0\\
    & Boehmite (100) & 2\\
    & Boehmite (001) & 1\\
    \hline
    \ce{\gamma-Al2O3} Edges
    & \ce{\gamma-Al2O3} (100-110) & 6\\
    \hline
    Small molecules
    & \ce{[Al(OH)4]^-H^+} monomer & --\\
    & Water & --\\
    \end{tabular}
    \caption{
        Overview of the structures in the training set.
        For surface and edge structures, the number of structures with additional water adsorbed on the surface is mentioned in the last column.
    }
    \label{tab:structures_summary}
\end{table}

\paragraph{(ii) Periodic DFT calculations.}
The geometries of all structures in Table~\ref{tab:structures_summary} are optimized with DFT using periodic boundary conditions and the Perdew, Burke and Ernzerhof (PBE) exchange-correlation functional as implemented in the Vienna Ab initio Simulation Package (VASP) version 5.4. \cite{perdew_rationale_1996, hafner_ab-initio_2008}
The valence interactions are described with the Projected Augmented Wave (PAW) method. \cite{blochl_projector_1994}
An additional \textit{a posteriori} density-dependent dispersion correction {dDsC} is applied. \cite{steinmann_comprehensive_2011}
The plane wave basis set cutoff is \SI{600}{\eV},
which is increased compared to original works from which the structures were taken,
\cite{raybaud_morphology_2001, digne_use_2004, chiche_growth_2009, batista_structure_2023}
to improve the precision of the forces and to facilitate the geometry optimizations.
The k-point spacing is set to \SI{0.5}{\angstrom\tothe{-1}} and a Gaussian smearing is used with a width of \SI{0.05}{\eV}.
The convergence criterion for the self-consistent field calculation is set to \SI{1e-5}{\eV}.
The geometry optimization is performed with the conjugate gradient algorithm and uses a convergence criterion of \SI{0.02}{\eV/\angstrom} on the forces.
The cell parameters were not allowed to change and were taken from previous works.
\cite{raybaud_morphology_2001, digne_use_2004, chiche_growth_2009, batista_structure_2023}

\paragraph{(iii) Property Extraction.}
The selection of internal coordinates, used as training targets, consists of two phases: an analysis of interatomic distances to determine appropriate cutoffs for relevant atom pairs, followed by a classification and enumeration of all relevant distances and angles.

In phase 1, histograms of all interatomic distances up to \SI{5.0}{\angstrom} were constructed per pair of chemical elements, as shown in Figure S1 in the supporting information.
From these histograms, cutoff distances were derived to classify and enumerate all relevant atom pairs.
All \ce{OH} pairs with a distance below \SI{1.2}{\angstrom} are classified as covalent \ce{O-H} bonds.
Remaining \ce{OH} distances below \SI{2.1}{\angstrom} are identified as \ce{O$\cdots$H} hydrogen bonds.
The \ce{AlO} distances below \SI{2.8}{\angstrom} are treated as \ce{Al-O} bonds.
\ce{AlAl} distances below \SI{4.0}{\angstrom} are not directly bonded, but are included because they are relevant for the local structure of alumina.
No other distances were included in the training set.

In phase 2, the distances defined in phase 1 are used to construct the final set of internal coordinates.
In addition to distances, valence angles are constructed by combining all pairs of bonded atom pairs sharing one central atom.
Dihedral angles are not included, because most \ce{O-Al-O-X} quartets, where X can be H or Al, contain nearly co-linear bonds, making the dihedral angle ill-defined.
Furthermore, \ce{H-O-H} angles are not included because, for reasons of backward compatibility explained in the following section, the corresponding valence angle term parameters are kept fixed at the values of Joshi \textit{et al.}\cite{Joshi2013, Joshi2014}
To avoid obvious redundancies in the training set, internal coordinates involving no hydrogen atoms were discarded for structures with additional water molecules, as these internal coordinates already appear in the bare surface and edge structures.
The final set comprises four categories of distances (\ce{Al-O}, \ce{Al-Al}, \ce{O-H} and \ce{O$\cdots$H}) and six categories of valence angles (\ce{O-Al-O}, \ce{Al-O-Al}, \ce{Al-O-H}, \ce{Al-O$\cdots$H}, \ce{H-O$\cdots$H} and \ce{H$\cdots$O$\cdots$H}).
Note that the last two categories are still relevant to include, unlike \ce{H-O-H} angles, because hydrogen bonding angles are sensitive to the structure of the boehmite and alumina surfaces.

Histograms of the final selection of internal coordinates can be found in Figures~S2 (combined) and S3 (per material) of the Supporting Information.
The internal coordinates of boehmite are somewhat similar to those of \ce{\gamma-Al2O3}, with larger deviations related to hydrogen bonds.
Hence, the inclusion of boehmite in the training set increases the diversity of hydrogen bonding information.
The prevalence of each class of internal coordinates per structure is given in Table~S1.

Energies in a ReaxFF training set are conventionally formulated as internal energy differences between reactant and product states.
A full list of reaction energies in the training set is provided in Table~S2 of the Supporting Information.
These energies are grouped into five classes, denoted with three-letter codes: \textbf{BSH}, \textbf{GEH}, \textbf{GSH}, \textbf{SUR} and \textbf{FOR}.
A short summary of the included reactions and their classification is given below.

For a given alumina surface or edge structure \ce{X}, all unique pair of adsorption states are used to construct adsorption energy training data, to avoid bias towards a particular reference state.
Let $m_i$ and $m_j > m_i$ be the number of water molecules adsorbed in states $\ce{X_i}$ and $\ce{X_j}$, respectively, then the corresponding adsorption energy in the training set is defined as:
\begin{equation}
    \label{eq:eads}
    \Delta_r E_{\text{ads},\ce{X_j},\ce{X_i}} = \frac{E_{\ce{X_j}} - E_{\ce{X_j}}}{m_j - m_i} - E_{\ce{H2O}}
\end{equation}
Thus, if there are $N$ states with water adsorbed for a given structure, there are $(N-1)N/2$ corresponding adsorption energies in the training set.
By normalizing the adsorption energies on the number of \ce{H2O} molecules added, they all have approximately the same order of magnitude.
The adsorption energies were grouped into three classes: adsorption on boehmite surfaces (\textbf{BSH}), on \ce{\gamma-Al2O3} edges (\textbf{GEH}) and on \ce{\gamma-Al2O3} surfaces (\textbf{GSH}).

While the water adsorption energies are of primary interest, also other energies were included to diversify the training set:
the transformation from bulk to a slab models (class \textbf{SUR}) and the formation of bulk and slab models from the alumina monomer (class \textbf{FOR}) are included.
Such ``reactions'' do not correspond to specific reactive events, but they do provide useful information for the covalent ReaxFF parameters.

The resulting training set contains 12931 distances, 13409 angles, 63 water adsorption energies and 19 other energies, for a total of 26422 training targets.
Notably, the geometrical features, such as angles and distances, far outnumber the energy entries in the training set.
The categories of internal coordinates and energies will be used in the remainder of the paper for a detailed statistical analysis and in the construction of the Balanced Loss function.

It is worth noting that our training set is considerably larger than those of previous ReaxFF parameterizations, which typically contain hundred to a several thousand training targets. \cite{labrosse_2010, Iype_2013, JaramilloBotero2014, muller_2016, boes_2016, Sengul2021, Komissarov2023, FreitasGustavo2023}
To the best of our knowledge, the only exception is the ReaxFF model by Trnka \textit{et al.}\ for enzyme chemistry, which was trained with 385826 entries. \cite{Trnka2017}
Note that their training set consists of single-point energies and forces, making it computationally less expensive, whereas ReaxFF is normally trained by optimizing geometries at each iteration of the parameter optimization. \cite{kaymak_jaxreaxff_2022}
\subsection{Parameter Selection}
\label{sec:paramsel}

When calibrating ReaxFF parameters, a careful selection of adjustable parameters must be made.
Some parameters can be taken from previous works without refinement, others are not intended to be adjusted, such as the atomic mass, and some parameters are not meaningful for the application of interest.
In addition, simply optimizing all parameters would result in an intractable high-dimensional optimization problem.
The total number of ReaxFF parameters depends on the number of chemical elements in the system.
The parameters are typically grouped into blocks, most of which can be repeated several times for different combinations of chemical elements.
Parameter blocks can be independent of chemical elements (41 general parameters), defined per chemical element (32 atomic parameters), per pair (16 bond and 6 off-diagonal parameters), per triplet (7 angle and 4 hydrogen-bond parameters), or per quartet of elements (7 torsion parameters).

In this work, the selection of parameters follows a top-down approach.
Initially, all parameters from a literature force field are considered,\cite{Joshi2013, Joshi2014} after which several selection criteria are introduced to fix parameters to their literature values, leaving only the remainder to be refined.
The selection process aims for a trade-off between an acceptable dimensionality of the optimization problem and a sufficient model flexibility to obtain a good fit.
In the following paragraphs, we motivate our selection criteria, which may be helpful for future ReaxFF calibrations.

Our starting point is the aluminosilicate force field by Joshi \textit{et al.}\cite{Joshi2013, Joshi2014}
These ReaxFF parameters were calibrated to improve the description of water adsorption at acid sites, \ce{Si-O(H)-Al}, in the H-ZSM-5 zeolite.\cite{kokotailo_structure_1978}
The literature \texttt{ffield} file contains parameters for 13 chemical elements and one dummy element, resulting in a set of 2961 parameters, many of which are irrelevant to aluminosilicates.
For this work, we only consider parameter blocks that contain at least one \ce{Al} element, and otherwise only allow \ce{O} or \ce{H}.
All other parameters are kept fixed, including those related to \ce{Si}, or those describing water.

Ideally, our reparameterization would maintain backward compatibility with the Joshi \textit{et al.}\ model, changing only parameters specific to alumina and irrelevant to acid sites in zeolites.
However, this severely restricts the adjustable parameters to those of atom pairs and valence angles involving at least two \ce{Al} atoms, namely the \ce{Al-Al} pairs, and the \ce{Al-{X}-Al} and \ce{Al-Al-{X}} angles, where \ce{X} can be \ce{Al}, \ce{O} or \ce{H}.
Of this selection, only \ce{Al-O-Al} angles and \ce{Al-Al} pairs ($\ge \SI{2.5}{\angstrom}$) appear in the training set.
As a result of this mismatch, no satisfactory reproduction of our training data was possible when imposing backward compatibility.
To avoid this mismatch, we made a pragmatic selection that includes some parameters related to acid sites in zeolites, but also excludes parameters that are only remotely related to our training data.
This selection includes \ce{Al} atom parameters, \ce{Al-{X}} bond or pair parameters and \ce{Al-H-O}, \ce{Al-O-H}, \ce{Al-O-Al}, \ce{H-Al-O} and \ce{O-Al-O} valence angle parameters.

We further narrow down the parameter selection using the recommendations from the ParAMS documentation.\cite{params_doc, komissarov_params_2021}
In the ParAMS, each parameter is classified with one of the following three labels: ``Standard'', ``Expert'' or ``DoNotOptimize''.
The first label indicates that the parameter is generally safe to optimize.
The second label is used for parameters that should not be changed without a strong motivation.
Parameters with the ``DoNotOptimize' label should never be touched, \textit{e.g.}, because they contain boolean values or atomic data.\cite{params_doc}
In this paper, only bond and angle parameters with the ``Standard'' label are considered for optimization.

Finally, some parameters are (de)activated for very specific reasons:
\begin{itemize}
    \item
    The atomic parameter \texttt{r\_0\^{}sigma}, which is $r_0^\sigma$ in Eq.~\eqref{eq:bond_order_uncorrected}, for Al is deactivated because it can be overruled by corresponding pair parameters for \ce{Al-H} and \ce{Al-O} bonds.
    \item
    Because no \ce{Al-Al} bonds are present, of the \ce{Al-Al} parameters, only \texttt{D\_e\^{}sigma} is optimized.
    This introduces some freedom to tune the weak bonding interactions between pairs of \ce{Al} atoms that are not directly bonded.
    \item
    $\pi$ and $\pi\pi$ bond parameters are deactivated because no such bonds are present in our training data.
    This includes parameters with labels containing any of the following strings: \texttt{pi}, \texttt{p\_bo3}, \texttt{p\_bo4}, \texttt{p\_bo5} or \texttt{p\_bo6}.
    \item
    The following ``Expert'' parameters were activated to improve the angular energy terms: \texttt{p\_val3}, \texttt{p\_val4} and \texttt{p\_val5}.
\end{itemize}

These selection criteria result in a subset of 36 activate parameters.
For each parameter, lower and upper bounds of suitable values are determined and used to restrict the search space during the parameter optimization.
For each parameter, the bounds are set equal to the corresponding range of historical values in the ReaxFF parameter database curated by Software for Chemistry \& Materials B.V.\ (SCM). \cite{ams}
Subsequently the bounds are extended to also include a window of \SI{\pm 20}{\percent} around the values from the Joshi \textit{et al.}\ force field.
Note that such choices are subjective for lack of a better alternative:
there are no established defaults for the parameter bounds.
The list of active parameters and their bounds can be found in Table~S3 of the Supporting Information.
\subsection{Optimization Settings}

ReaxFF parameters are typically calibrated by minimizing a loss function with a numerical optimizer.
We have developed a novel loss function for this work, which will be discussed in section~\ref{sec:BL}.
Here we focus on the details of the numerical optimization algorithm.

Several optimization algorithms have been proposed to refine ReaxFF parameters.
The original method proposed by van Duin was a deterministic algorithm that optimized one parameter at a time with a parabolic extrapolation.\cite{vanDuin1994}
More recent algorithms are stochastic, which makes them more robust to the non-trivial structure of a standard ReaxFF loss function, such as many local minima and discontinuities.\cite{Pahari2011, Deetz2014, JaramilloBotero2014, Dittner2015, Trnka2017, shchygol_reaxff_2019}
These difficulties arise from the small discontinuities in the ReaxFF energy itself, and the noisy sensitivity of geometry optimizations (while training) to the ReaxFF parameters. \cite{shchygol_reaxff_2019, FreitasGustavo2022}
These difficulties are still present in this work, and therefore we use a stochastic derivative-free optimizer that has proven its effectiveness, i.e.\ the Covariance Matrix Adaptation Evolutionary Strategy (CMA-ES). \cite{shchygol_reaxff_2019, hansen_cma_2006, Hansen2003, hansen_cma_2023}

The CMA-ES settings in this work follow the best practices from the literature. \cite{hansen_cma_2006,hansen_cma_2023,Hansen2003}
The algorithm is repeated 40 times, starting from the Joshi's parameters, to reduce the risk of getting stuck in an unfavorable local minimum.
These repetitions are also used to test the robustness of the new loss function proposed in Section~\ref{sec:BL}.
The CMA population size is set to the value recommended by Hansen, $\lfloor 4 + 3\ln N_{\text{par}}\rfloor = 14$, where $N_{\text{par}}=36$ is the number of activated force field parameters. \cite{hansen_cma_2023, hansen_cma_2006, Hansen2003}
ParAMS communicates dimensionless parameters to CMA-ES, by linearly transforming the original parameters so that their bounds all become $[0, 1]$.
The initial CMA-ES step size, in these dimensionless parameters, is set to \SI{0.2}{}.
This is sufficient to let the algorithm randomize the parameters in the first few CMA iterations, after which it starts to converge, thereby guaranteeing an initial exploration of the parameter space.
Each CMA run is terminated after 1000 iterations and the parameters with the lowest loss value are selected for further analysis.

Before evaluating the loss function in each CMA iteration, all structures in the training set are optimized with the parameters generated by CMA.
The maximum number of geometry iterations is set to 500, which is much higher than the default value of 30 in ParAMS.
For the training set in this work, a lower setting, such as 50, 100 or 200, produces force fields that are overfitted to this lower number of geometry steps.
Each CMA run is performed on 18 cores (Intel Xeon Gold 6140), with \texttt{ParallelLevels} \texttt{ParameterVectors=14} and \texttt{Jobs=2}.
This results in a slight over-commitment of the cores, which is normally not recommended, but it improves the overall efficiency in this case, which can be understood as follows.
The CMA-ES algorithm synchronizes after each iteration, resulting in idle time when the members of the population require different CPU times.
This is generally the case for ReaxFF, since the number of required geometry steps depends strongly on the parameters.
By over-committing the cores, the idle time is reduced, resulting in a more efficient use of resources.
\subsection{Validation Set}
\paragraph{(i) Structures}
The validation set is taken from a dataset by Raybaud \textit{et al.}, available on NOMAD, containing alumina structures optimized with VASP, using the same level of theory as the training data, \cite{raybaud_nomad_2021, pigeon_revisiting_2022}
except that a plane-wave cutoff of \SI{400}{\eV} was used.
This set contains 53 new \ce{\gamma-Al2O3} surface structures not used for parameter optimization, with different numbers of adsorbed water molecules.
The surface orientations comprise (001), (111), (110)$_\ell$ and (110)$_\text{b}$ as shown in Fig.~\ref{fig:crystsurfaces}.
The subscripts $\ell$ and b are used to distinguish between lateral and basal surfaces, respectively, which feature different Br{\o}nsted and Lewis acid sites. \cite{pigeon_revisiting_2022, batista_structure_2023}
The set also includes a bulk \ce{\gamma-Al2O3} model and an isolated water molecule was added in this work, using consistent VASP settings.
A complete list of structures is provided in Table~S4 of the Supporting Information.

Recent developments in alumina characterization have revealed an ambiguity in the terminology used in older works.
In particular, earlier spinel models considered the (100), (010) and (001) surfaces to be equivalent, but it has recently been shown from non-spinel models that this is not the case. \cite{pigeon_revisiting_2022}
To remain consistent with published datasets and with the optimizations performed in this work, the notation remains (001) for surfaces with this orientation in the validation set and (100) for surfaces with this orientation in the training set.
However, they are structurally equivalent.

\paragraph{(ii) Property Extraction}
Properties are extracted using the same methodology and classification as described in Section \ref{sec:trainingset_dev}.
The resulting validation set contains 16745 distances, 9513 angles, 101 adsorption energies, and 6 other energies, for a total of 26365 validation targets.
Figures~S3, S5, S6 and S7 in the Supporting Information show the histograms of these data, whereas Table~S5 lists the individual reaction energies.
The final force field in this work and the original one by Joshi \textit{et al.}\cite{Joshi2013, Joshi2014} are validated by comparing these geometrical properties and energies to the VASP reference data.

\section{Balanced Loss function and optimization workflow}
\label{sec:BL}

ReaxFF parameters are conventionally calibrated by minimizing a loss function $L$, which is often a weighted Sum-of-Squares Error (SSE) or Root-Mean-Squared Error (RMSE):
\begin{align}
    L_{\text{SSE}}(\mathbf{x}) &= \sum_{i=1}^N w_i s_i(\mathbf{x})
    \label{eq:sse}\\
    L_{\text{RMSE}}(\mathbf{x}) &= \sqrt{\frac{1}{N}\sum_{i=1}^N w_i s_i(\mathbf{x}) }
    \label{eq:rmse}
\end{align}
where $s_i$ are the squared residuals:
\begin{align}
    s_i(\mathbf{x}) &= r_i^2(\mathbf{x}) = \left(\frac{y_i - \hat{y}_i(\mathbf{x})}{\sigma_i}\right)^2
    \label{eq:residual_x}
\end{align}
and where the sum over $i$ runs over all items in the training set.
In every term, the property value $i$ is calculated with a reference method ($y_i$) and ReaxFF ($\hat{y}_i$).
Through the ReaxFF property values, the loss function depends on a vector of adjustable ReaxFF parameter vector $\mathbf{x}$.
Note that CMA-ES is insensitive to the application of any monotonically increasing transformation of the loss function, so from CMA's perspective, $L_{\text{RMSE}}(\mathbf{x})$ and $L_{\text{SSE}}(\mathbf{x})$ are equivalent.

The constant $\sigma_i$ is a configurable scaling factor with the same unit as the property of $y_i$ to make the residual $r_i$ dimensionless.
In ParAMS, $\sigma_i$ is only used as a reasonable order of magnitude for the corresponding $y_i$.\cite{params_doc}
The weight $w_i$ controls the importance of each training set entry in the total loss function.
In principle, one can absorb $w_i$ into $\sigma_i$ or vice versa.
The main motivation for supporting both factors in ParAMS is to cater to different user groups, some of which may prefer one over the other.
This may seem surprising, since textbook treatments of the least-squares method do not mention the weights $w_i$ and only introduce $\sigma_i$ as a measurement error.
However, a basic assumption of the standard least-squares method does not hold here:
Our data have no measurement errors.
Any discrepancies between the training data and ReaxFF are due to systematic errors, mainly in ReaxFF and, in principle, also in the model used to compute the training data.

At first glance, setting the weights seems straightforward:
the more important an item in the training set, the higher its weight should be.
However, there are different (possibly competing) motivations for adjusting the weights.
A first purpose of the weights is to compensate for an imbalance in the training set.
For example, our training set contains many more distances and angles than energies, while the energies are also important.
This imbalance can be addressed by classifying the data into categories and setting the weight to the inverse of the number of elements in each category.
This strategy is common in the context of machine learning potentials, e.g. when atomic forces are much more abundant than molecular energies. \cite{Wang2018, Unke2019, CoolsCeuppens2022}
A second purpose of the weights is to emphasize the importance of some residuals.
For example, when an initial optimization leads to parameters for which some residuals $r_i$ are perceived to be too large, one may increase the corresponding weights $w_i$ and re-optimize.
Assigning different weights to subsets of data is also known in the field of multi-objective optimization as the ``scalarization'' of multiple objectives into a single loss function. \cite{Giagkiozis2015}
Unlike scalarization methods, multi-objective evolutionary algorithms do not assume any tradeoffs between categories \textit{a priori} and instead find many Pareto-optimal solutions. \cite{Tian2021}

While seemingly intuitive, manual weight adjustment becomes intractable when many weights need to be adjusted differently.
Due to the non-linear response of the residuals to the weights,\cite{verstraelen_computation_2012} multiple combinations of weights must be tried before one reaches the residuals of interest.
If the model has insufficient functional flexibility, it may even be impossible to reach the desired residuals.
In addition, when some residuals of interest decrease, others will inevitably increase.
It is difficult to predict which residuals will increase and by how much, and this can force the operator to keep adjusting the weights.
In practice, this resembles a cat-and-mouse game between weights and residuals.
Manually adjusting the weights also provides little insight into the optimization problem:
If some residuals are large, there is no straightforward way to understand whether their weights should be increased further, or whether the model is simply unable to reproduce the training data.

It is clear that the development of a training set alone is rarely sufficient to find the optimal parameters.
Only if the training data are completely homogeneous, one can simply set all weights equal.
This is typically not the case for ReaxFF training set, which contain different types of data, such as distances, angles and energies in this work.
Assigning weights to the data is therefore an unavoidable and potentially tedious task before and during a ReaxFF parameter optimization. \cite{Furman2018, fedkin_development_2019, brown_reaxff_2022}

To simplify the tedious adjustment of weights, we introduce a new method, hereafter called ``Balanced Loss''.
As will be demonstrated in the results, this method allows for a swift balancing of the training set, and we believe that this methodology will be equally beneficial for other optimization problems facing similar challenges.
The Balanced Loss method introduces a new loss function and an intuitive workflow to balance the data and to gain more insight into how well the model can reproduce subsets of the training data.

Balanced Loss requires a classification of the training data into categories. Technically, the categories are $C$ mutually exclusive and exhaustive subsets: $S_c\,\forall c\in\{1\ldots C\}$.
They should be defined so that residuals within a category respond in roughly the same way to a change in the model parameters.
For example, one might expect that all \ce{O-H} bond lengths in a training set, while not exactly the same, do respond similarly to changes in the ReaxFF parameters.
With this partitioning, the Balanced Loss function is defined as:
\begin{align}
    L_\text{BL} = \tau f^{-1} \left(\sum_{c=1}^C f\left({\frac{R_c}{\tau}}\right)\right)
    \label{eq:bl}
\end{align}
where $R_c$ is the RMSE on the entries in category $c$:
\begin{align}
    R_c = \sqrt{\frac{1}{|S_c|}\sum_{i\in S_c} s_i}
\end{align}
$L_\text{BL}$ is dimensionless by construction.
The function $f$ and its inverse must be monotonically increasing functions, and by default $f(x)=\exp(x)$ is used, which will be denoted as the Log-Sum-Exp (LSE) form, referring to the mathematical operations in Eq.~\eqref{eq:bl}.
To illustrate the benefits of the Log-Sum-Exp form, all parameterizations will be repeated with two other forms of $f$: $f(x)=x^2$, denoted as Root-Sum-Square (RSS) and $f(x)=x$, denoted as Identity-Sum-Identity (ISI).
Note that the RSS form makes Balanced Loss formally equivalent with a standard loss function in Eq.~\eqref{eq:rmse}, with $w_i = \frac{N}{S_c}$, where $c$ is the category to which training set item $i$ belongs.

By default, $L_\text{BL}$ is thus a Log-Sum-Exp function, a well-established smooth approximation of the maximum over multiple inputs.
It is popular in the machine learning context \cite{blanchard_accurately_2021} and it has been used for scalarization of multi-objective problems. \cite{Palaz2016}
Here, the inputs to Log-Sum-Exp are all the $R_c$ values.
The hyperparameter $\tau$, sometimes called the effective temperature, controls the smoothness of the approximation of the maximum.
In the ``cold'' limit $\tau \rightarrow 0$, Log-Sum-Exp loses its smoothness and reduces to the maximum over all $R_c$.
The parameter $\tau$ appears in two places, such that $L_\text{BL}=L_\text{RMSE}$ in the trivial case of one category and $w_i=1\,\forall i$.
Throughout this paper we have used $\tau=1$.

Unlike $L_\text{SSE}$ and $L_\text{RMSE}$,
user-defined weights $w_i$ are missing from $L_\text{BL}$,
which implies that $\sigma_i$ must play a slightly different role.
We propose to set each $\sigma_i$ to the desired accuracy of the corresponding entry $y_i$ in the training set, which resembles its meaning in conventional least-squares methods.
To make the distinction with $\sigma_i$ in other contexts, we call them ``tolerances'' in the context of Balanced Loss,
because they represent the magnitudes of residuals one is willing to tolerate.
At this stage, simply defining tolerances may seem like wishful thinking, but it will become clear later that Balanced Loss helps finding a consistent set of ReaxFF parameters and tolerances.
Our definition of $\sigma_i$ (as the desired accuracy) also facilitates the interpretation of $R_c$:
It expresses, in the RMS sense, the average ratio between the actual and the desired accuracy.
In the ideal case, after optimizing the parameters, one obtains $R_c=1\, \forall c$.

Given the interpretation of $R_c$, the Log-Sum-Exp form of Balanced Loss is easily motivated.
If one of the $R_c$ values is much higher than all others, one finds $L_\text{BL}\approx R_c$, \textit{i.e.}, category $c$ dominates the loss function.
If the optimization algorithm explores a region of the parameter space where category $c$ dominates, it will focus only on reducing $R_c$, with other categories acting at best as a form of regularization.
This is a desirable feature, since category $c$ is then the worst reproduced subset of the training data, and therefore deserves the optimizer's full attention.

One may also understand the effect of Log-Sum-Exp by comparing the gradients of $L_\text{BL}$ and $L_\text{RMSE}$ with respect to the ReaxFF parameters:
\begin{align}
    \frac{\partial L_\text{BL}}{\partial x_k}
        &= \sum_{i=1}^N \frac{\partial L_\text{BL}}{\partial s_i}
           \frac{\partial s_i}{\partial x_k} \\
    \frac{\partial L_\text{RMSE}}{\partial x_k}
        &= \sum_{i=1}^N \frac{\partial L_\text{RMSE}}{\partial s_i}
           \frac{\partial s_i}{\partial x_k}
\end{align}
Both loss gradients are linear combinations of the gradients of squared residuals, $\frac{\partial s_i}{\partial x_k}$,
but they combine them with different ``weights'':
\begin{align}
    \frac{\partial L_\text{BL}}{\partial s_i}
        &= \frac{\exp\left(\frac{R_d}{\tau}\right)}{\sum_{c=1}^C \exp\left(\frac{R_c}{\tau}\right)}
           \frac{\tau}{2|S_d|R_d} &
    \text{with}&&
    i&\in S_d \\
    \frac{\partial L_\text{RMSE}}{\partial s_i}
        &= \frac{w_i}{2 N L_\text{RMSE}}
\end{align}
In the case of $L_\text{RMSE}$, the weight $\frac{\partial L_\text{RMSE}}{\partial s_i}$ is simply proportional to the user-defined weight $w_i$.
For Balanced Loss, however, the weight $\frac{\partial L_\text{BL}}{\partial s_i}$ contains a new and crucial factor:
\begin{align}
    P_d &= \frac{\exp\frac{R_d}{\tau}}{\sum_{c=1}^C \exp\frac{R_c}{\tau}}
\end{align}
with
\begin{align}
    \sum_{d=1}^C P_d &= 1.
\end{align}
This factor is known as SoftMax, a continuous generalization of the ArgMax function, used to identify the position of a maximum in an ordered list. \cite{Nian2020}
This shows how Balanced Loss borrows a strategy from reinforcement learning, known as the Gradient Bandit Algorithm:
At each iteration in the optimization, the most violated subset of the training data determines the action, \cite{Sutton2018} in this context action being the direction in which the parameters must evolve.
The analogy between $P_d$ (in $L_\text{BL}$) and the user-defined weights $w_i$ (in $ L_\text{RMSE}$) also suggests another interpretation.
Instead of a human operator tuning the weights $w_i$, as in the cat-and-mouse metaphor introduced above, Balanced Loss adjusts the weights algorithmically within a single optimization run.

So far, we have assumed that one simply sets the tolerances $\sigma_i$ to the desired accuracy of the corresponding $y_i$.
However, such a choice may be subjective and incompatible with the capabilities of the ReaxFF model to be trained.
In practice, we recommend such ``naive'' tolerances $\sigma_i$ as a first guess.
Parameter optimization can then be used to test these expectations.
To do so, we recommend the following workflow:
\begin{enumerate}
    \item[W1.]
    First gather all the elements of a conventional parameter optimization:
    (i) the model,
    (ii) the training data,
    (iii) a selection of parameters to optimize and their bounds,
    (iv) an initial guess of the parameters, and
    (v) an optimization algorithm.
    In this paper, all these elements are described in Section \ref{sec:methodology}.
    \item[W2.]
    Then define the additional elements needed for Balanced Loss:
    the categories of training data and an initial configuration of the tolerances $\sigma_i$.
    Such choices are domain-specific, but a few general recommendations can be given, in addition to the ones discussed above.
    It is convenient to have data with consistent units within one category $c$ and to assign the same tolerance to all its members, for which the symbol $\sigma_c$ will be used below.
    Furthermore, it is useful to introduce categories for data that deserve special attention, e.g. with key properties for the intended application of the force field, or with properties that are harder to reproduce than others.
    By placing these data in separate categories, their RMSEs are easily monitored and large errors within these categories will be prioritized during the optimization.

    For ReaxFF, one can introduce different categories for distances, angles and energies.
    In this paper, the categories are more fine-grained:
    All internal coordinates are classified by the chemical elements and the bond types involved.
    In fact, we categorize all training data as they were introduced in Section \ref{sec:trainingset_dev}:
    the 4 bond categories are \ce{Al-O}, \ce{Al-Al}, \ce{O-H} and \ce{O$\cdots$H},
    the 6 angle categories are \ce{O-Al-O}, \ce{Al-O-Al}, \ce{Al-O-H}, \ce{Al-O$\cdots$H}, \ce{H-O$\cdots$H} and \ce{H$\cdots$O$\cdots$H},
    and the energy categories are \textbf{BSH}, \textbf{GEH}, \textbf{GSH}, \textbf{SUR} and \textbf{FOR}.
    Note that a regular covalent bond is denoted by minus sign ($-$) and a hydrogen bond by three dots ($\cdots$).
    The tolerances $\sigma_c$ will be described in Section \ref{sec:res_discussion}.
    \item[W3.]
    Finally, minimize the Balanced Loss function and analyze the $R_c$ values of the optimal parameters.
    When one category keeps dominating the loss function throughout the optimization, the only possible explanation is that the corresponding tolerances $\sigma_c$ were set too small.
    There is no way to lower $R_c$ because $L_\text{BL}$ already ignores all other categories.
    The only option left is to accept that the model cannot reproduce items in category $c$ with the desired accuracy, and to adjust one's expectations by increasing the corresponding tolerances $\sigma_c$.
    One can now repeat the parameter optimization and re-evaluate the result, possibly repeating the exercise a few times.
    Unlike tuning the weights in a conventional loss function, these repeated optimizations provide insight:
    They inform the human operator about the capabilities of the model and help manage expectations.
    It may also happen that some $R_c$ end up well below 1, in which case we do not recommend decreasing the corresponding $\sigma_c$.
    Such a fortuitous outcome should not affect the desired accuracy.
\end{enumerate}
Note that steps W2 and W3 in the above workflow involve (possibly subjective) human decisions, and therefore cannot be replaced by an autonomous algorithm.
This is an unavoidable aspect of multi-objective problems:
One has to decide on a compromise between different categories.
The overall goal of Balanced Loss is to facilitate finding suitable compromises.

    \section{Results and Discussion}
    \label{sec:res_discussion}
\subsection{Balanced Loss Optimization Procedure}

This section illustrates how the optimization workflow of Balanced Loss leads to a competitive ReaxFF parameterization, using the alumina training set as a realistic example.
In addition to the final force field, the intermediate steps provide insight into the capabilities of ReaxFF.

\begin{table}
    \centering
    \begin{tabular}{c|c|c|c}
    Category & Unit & Initial $\sigma_c$ & Final $\sigma_c$  \\
    \hline
    \ce{Al-O} & \si{\angstrom} & 0.05 & 0.07 \\
    \ce{Al-Al} & \si{\angstrom} & 0.05 & 0.10 \\
    \ce{O-H} & \si{\angstrom} & 0.05 & 0.05 \\
    \ce{O$\cdots$H} & \si{\angstrom} & 0.05 & 0.12 \\
    \ce{Al-O-Al} & deg & 2.0 & 5.0 \\
    \ce{Al-O-H} & deg & 2.0 & 5.0 \\
    \ce{O-Al-O} & deg & 2.0 & 5.0 \\
    \ce{Al-O$\cdots$H} & deg & 2.0 & 7.0 \\
    \ce{H-O$\cdots$H} & deg & 2.0 & 7.0 \\
    \ce{H$\cdots$O$\cdots$H} & deg & 2.0 & 7.0 \\
    \textbf{BSH} & \si{\kilo\cal \per \mol} & 1.25 & 4.0 \\
    \textbf{GEH} & \si{\kilo\cal \per \mol} & 1.25 & 4.0 \\
    \textbf{GSH} & \si{\kilo\cal \per \mol} & 1.25 & 4.0 \\
    \textbf{SUR} & \si{\kilo\cal \per \mol} & 1.25 & 3.0 \\
    \textbf{FOR} & \si{\kilo\cal \per \mol} & 1.25 & 3.0
    \end{tabular}
    \caption{Tolerances used in the Balanced Loss optimization }
    \label{tab:tolerances}
\end{table}

The data in the training set, described in Section~\ref{sec:trainingset_dev}, have been grouped into categories as described in Section~\ref{sec:BL} and as summarized in the leftmost column of Table~\ref{tab:tolerances}.
The optimizations are carried out in two stages, initial and final, which differ only in the tolerances $\sigma_c$.

\begin{figure}
    \centering
    \includegraphics{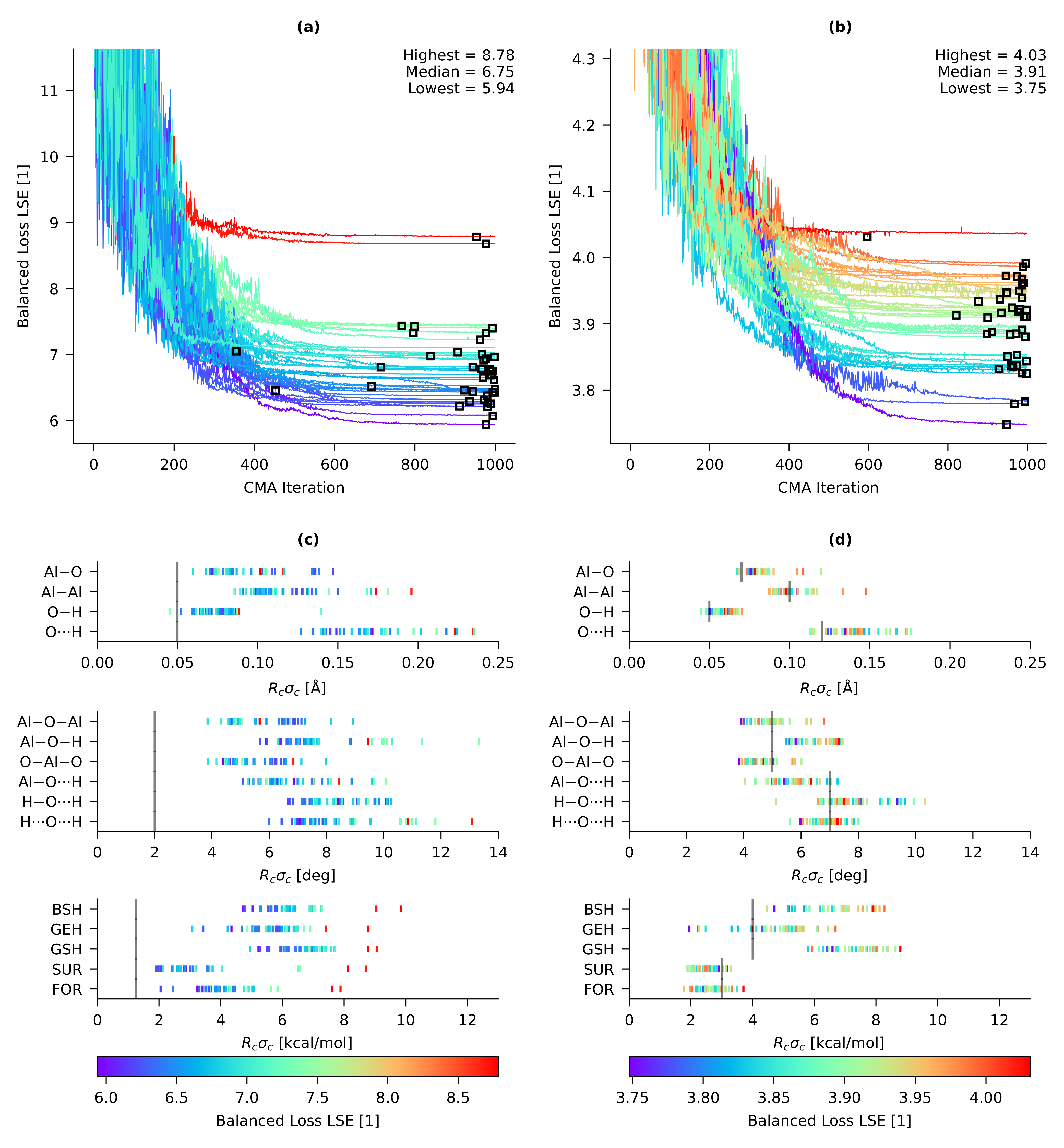}
    \caption{
        The Balanced Loss as a function of CMA iteration (lowest value within the population) of the 40 parameterizations in the initial (a) and final (b) stage.
        The lowest value along each trajectory is indicated by a black square.
        The dimensioned category RMSEs, $R_c\sigma_c$, of the best parameter vector of each of the 40 parameterizations in the initial (c) and final (d) stage.
        All data are color coded by the loss value of the best point along the trajectory.
        The color bar of panel (c)/(d) is also applicable to panel (a)/(b).
        Grey vertical lines in panel (c)/(d) denote the tolerances for the corresponding category.
    }
    \label{fig:loss-stages-lse}
\end{figure}

Each entry is given an initial chemically relevant tolerance, $\sigma_c$, equal to the default sigma value from ParAMS, as shown in Table~\ref{tab:tolerances}.
With these tolerances, the ReaxFF parameters were optimized 40 times using different random seeds to produce independent solutions.
Figure~\ref{fig:loss-stages-lse}(a) shows the evolution of the Balanced Loss during the 40 CMA runs.
Figure~\ref{fig:loss-stages-lse}(c) presents the $R_c \sigma_c$ values (category RMSEs with units) for the 40 optimized parameter vectors.
The curves and data points are colored according to the loss value of the best parameter vector of each run.
Of the 40 CMA runs in the initial stage, two are clearly worse than all others, presumably converging to unfavorable local minima.
All 38 remaining runs produce comparable RMSE values, but are not identical, which is the expected behavior.
ReaxFF loss functions are known to exhibit many local minima and apparent noise due to high sensitivity of the geometry optimizations in the training set to the ReaxFF parameters. \cite{shchygol_reaxff_2019}
Adsorption energies in the categories \textbf{BSH} and \textbf{GSH} are the highest relative to their tolerance, $\sigma_c$.
The performances in all other categories have less effect on the optimized parameters, simply because these errors are closer to their tolerance.
This means that, in this initial stage, the CMA runs train almost exclusively on the adsorption energies.
It is therefore highly unlikely to find ReaxFF parameters that can further lower RMSE on the adsorption energies, let alone reach the tolerance of $\SI{1.25}{\kilo\cal \per \mol}$.
Also, for all other categories, the initial tolerances seem too optimistic, which will be addressed in the next stage.

For the second (and final) stage, the tolerances are revised, as shown in the last column of Table~\ref{tab:tolerances}, to be more consistent with what ReaxFF could achieve in the initial stage.
Without Balanced Loss, these tolerances can only be set by expert judgment, which is greatly facilitated here by the feedback from the initial stage.
Figure~\ref{fig:loss-stages-lse}(b) shows that the revised tolerances result in more consistent loss values across all 40 parameterizations.
This is also reflected in the RMSEs in Figure~\ref{fig:loss-stages-lse}(d), most of which exhibit less scatter.
In other words, for most categories, the 40 parameterizations in the final stage are comparable, with the category \textbf{GEH} being the most notable exception.
Note that the absolute value of the Balanced Loss values cannot be compared between Figures~\ref{fig:loss-stages-lse}(a)~and~\ref{fig:loss-stages-lse}(b), because the two optimization stages use different tolerances.
One could slightly tweak the tolerances further to bring them closer to the errors on the training set, but this would amount to relatively small adjustments that we do not expect to lead to significant improvements.

\begin{figure}
    \centering
    \includegraphics{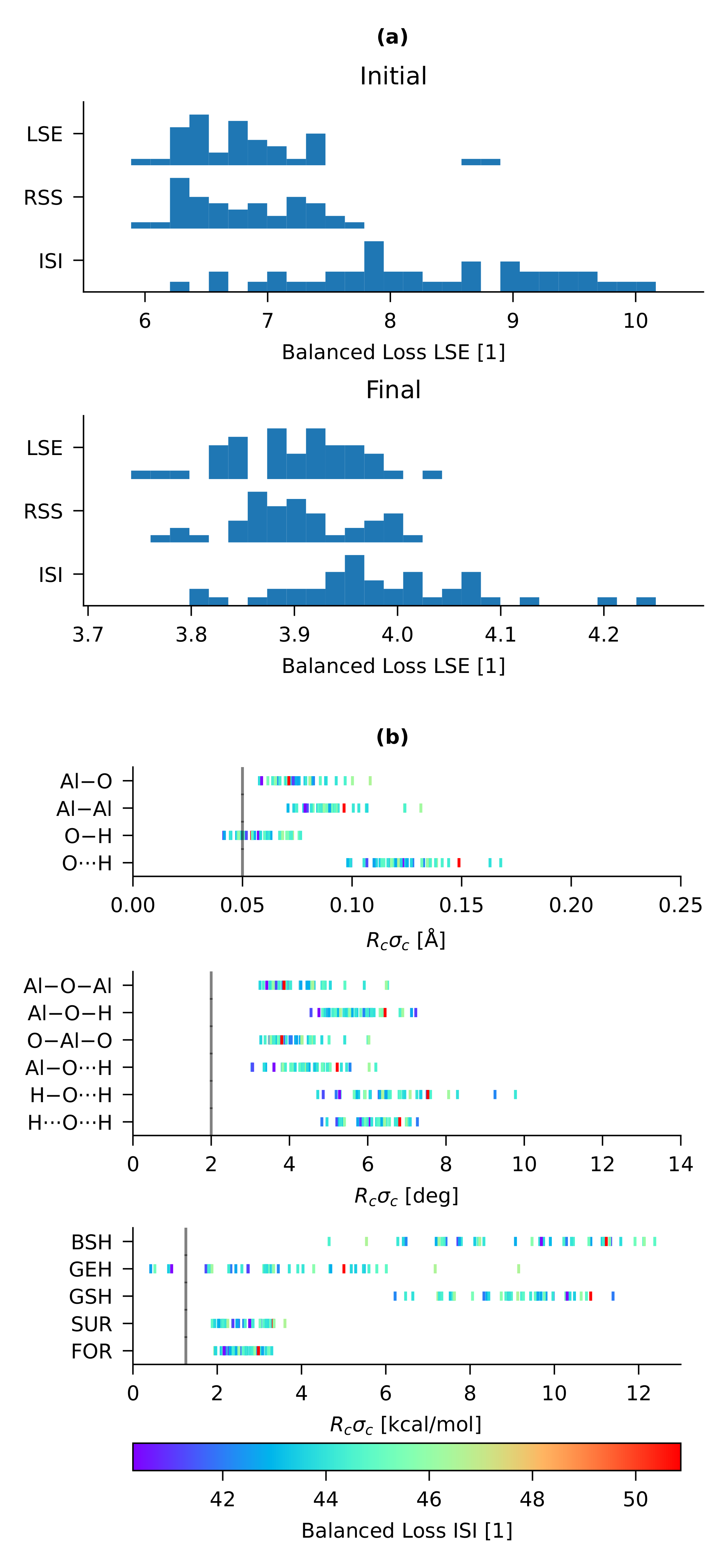}
    \caption{
        (a)
        Distribution of Balanced Loss values, computed with $f(x)=\exp(x)$, for parameter vectors optimized with different functions $f$ in Eq.~\eqref{eq:bl}: $f(x)=\exp(x)$ (LSE), $f(x)=x^2$ (RSS) and $f(x)=x$ (ISI).
        The histograms are computed for both the initial and final tolerances defined in Table~\ref{tab:tolerances}.
        (b)
        The dimensioned category RMSEs, $R_c\sigma_c$, of the best parameter vector of each of the 40 parameterizations in the initial stage using $f(x)=x$ (ISI).
    }
    \label{fig:func-comparison}
\end{figure}

To illustrate the importance of the Log-Sum-Exp form of Balanced Loss, we have performed the 40 CMA optimizations in six different ways: with initial and final tolerances and using different functions $f$: $f(x)=\exp(x)$ (Log-Sum-Exp or LSE, the default, same results as above), $f(x)=x^2$ (Root-Sum-Square or RSS) and $f(x)=x$ (Identity-Sum-Identity or ISI).
For comparison, the LSE form of Balanced Loss is computed for all the 240 optimized parameter vectors, and their distribution is shown in Figure~\ref{fig:func-comparison}(a).
In the initial stage, the function $f$ has a significant influence.
The choice of the function $f$ determines the compromise between the RMSEs of the individual categories:
In the case of RSS and ISI, the parameterization no longer exclusively prioritizes the adsorption energies in the categories \textbf{BSH} and \textbf{GSH}, resulting in higher RMSEs for these categories, as illustrated in Figure~\ref{fig:func-comparison}(b).
The results in this figure do not reveal whether the poor performance of ReaxFF for the categories \textbf{BSH} and \textbf{GSH} can be remedied by giving these categories a higher weight in the loss function, or whether they are high due to intrinsic limitations of the model.
The Log-Sum-Exp form automatically resolves this ambiguity.
Since this form approximates the maximum over all $R_c$, this loss function is dominated by the categories \textbf{BSH} and \textbf{GSH}, resulting in the lowest possible RMSE for these categories.
It is simply impossible to give these categories a higher weight, leaving no other option than to assign more humble tolerances.

Figure~\ref{fig:func-comparison}(a) reveals two additional insights.
First, the results become less sensitive to the choice of the function $f$ in the final stage.
The RMSEs for all categories are close to the final tolerances, meaning that the argument of the function $f$ (for the optimized parameters) is close to one, reducing the importance of the non-linearity of $f$.
Second, with $f(x)=x^2$ (RSS), the loss function is mathematically equivalent to Eq.~\eqref{eq:rmse}, a standard loss functions used for ReaxFF.
This implies that the optimized parameters in this work can also be found with a more conventional loss function, when the weights and sigmas are set consistently with the tolerances in Balanced Loss.
Hence, the added value of Balanced Loss is essentially the feedback provided from the initial stage, which facilitates the configuration of the tolerances.

\begin{figure}
    \centering
    \includegraphics{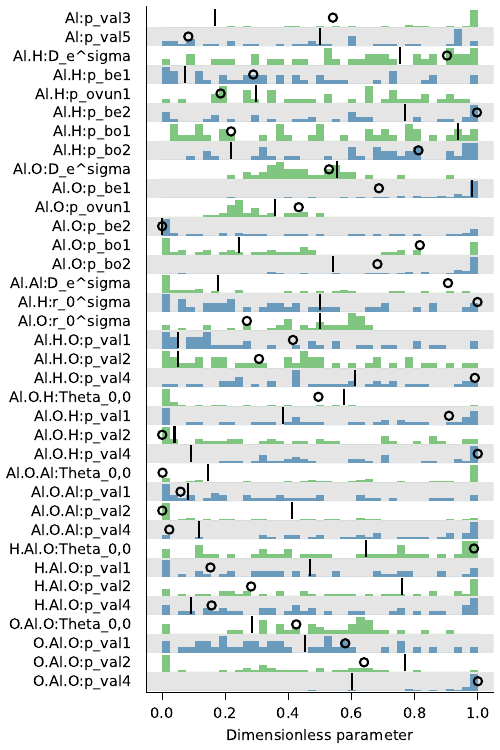}
    \caption{
        Histograms of the 36 components of the 40 optimized dimensionless parameter vectors in the final stage.
        The parameter components are made dimensionless by a linear transformation, such that zero corresponds to the lower bound and one corresponds to the upper bound.
        The bounds are listed in Table S3 in the Supporting Information.
        The initial values are marked with vertical black bars.
        The values corresponding to the lowest loss (over all 40 runs) are marked with circles.
    }
    \label{fig:pardist}
\end{figure}

Figure~\ref{fig:pardist} shows the distribution of the 40 optimized parameter vectors, after transforming them to their dimensionless form.
Although the 40 CMA runs converge to approximately the same Balanced Loss value, the corresponding parameters are not necessarily similar.
Some parameters, such as \texttt{Al.O:D\_e\^{}sigma} or \texttt{Al.O:p\_ovun1}, have a delineated range of optimal values.
However, most parameters can be found across the entire interval of allowed values.
This does not mean that all these parameters are completely random:
They could be correlated, which is not apparent in the individual histograms.
In any case, the optimal parameters are degenerate to some degree, which has also been observed in previous ReaxFF parameterizations on other chemical systems. \cite{Pahari2011, shchygol_reaxff_2019, FreitasGustavo2022, FreitasGustavo2023}

Since ReaxFF is at least partially inspired by physical principles, one might hope that all parameters always converge to the same values.
However, Sethna \textit{et al.} have extensively shown in their work on ``sloppy models'' that broad parameter distributions are virtually always found for models with more than a few parameters, across different scientific disciplines. \cite{brown_2003, Waterfall2006, Gutenkunst_2007, transtrum_2015}
This is a universal pattern, regardless of the degree of physical interpretation that the model parameters may have.
It is observed that some degrees of freedom in the parameter space of complex models are systematically ill-defined, not due to a lack of data, but because nearly the same model predictions (for all possible inputs) are found for different parameter vectors.
As a consequence, predictions on unseen data are robust, despite uncertainties in the parameters.
For the Electronegativity Equalization Method (EEM), which is included in ReaxFF, this parameter degeneracy has been investigated in more detail. \cite{verstraelen_2011}
ReaxFF also has the characteristics of ``a sloppy model'', as illustrated by Figure~S8 in the Supporting Information:
All the optimized parameter vectors of the 40 CMA runs (LSE, final stage) perform reasonably well for the validation set, despite the fact that they represent different local minima in the parameter space.
This in itself is not a limitation when the model is used for simulations, but it obviously makes any direct interpretation of the parameters impossible.

It is also noticeable that several parameters have a high probability of converging close to the bounds.
One might deduce that the parameter bounds are too narrow and the optimizer is trying to move the parameters to an optimum beyond the bounds, but this is not the only possible cause.
Note that the components of the best parameter vector over all 40 runs, the circles in Figure~\ref{fig:pardist}, are not necessarily close to the bounds, even if the remaining near-optimal values cluster near the edges.
Examples of this pattern are ``Al.O:p\_be1'' and ``Al.O.H:Theta\_0,0''.
A deeper investigation, beyond the scope of the current work, is needed to understand why a disproportionate number of near-optimal solutions is found near the bounds.
For example, this could also be related to an optimizer inefficiency when the parameters approach their bounds, and addressing this problem may make the optimization more efficient.
It should also be noted that some initial parameter values start of close to the interval bounds regardless of the boundary extension as explained in Section \ref{sec:paramsel}.
Since some of the initial parameters have values close to zero, the effect of the boundary extension is negligible.
The most prominent examples are ``Al.O:p\_be2'',  ``Al.H.O:p\_val1'', ``Al.H.O:p\_val2'' and ``Al.O.H:p\_val2''.

For the remainder of this work, the best parameter vector from the final stage is used for all calculations, i.e.\ corresponding to the lowest square in Figure~\ref{fig:loss-stages-lse}(b) and the circles in Figure~\ref{fig:pardist}.
The selected parameters are given in the last column of Table~S3 in the Supporting Information.

\subsection{Force field validation}
\label{sec:static}

\begin{figure}
    \centering
    \includegraphics{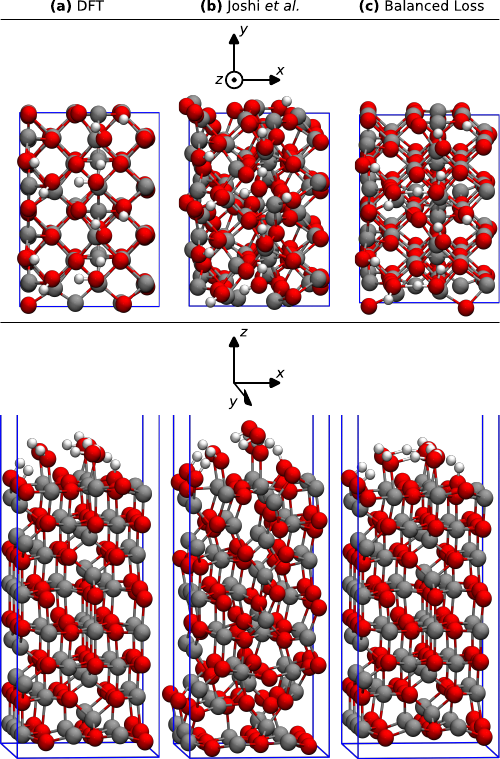}
    \caption{
        Optimized structures of the \ce{\gamma-Al2O3} (001) surface from the validation set at an \ce{OH} coverage of \SI{13.0}{\nano\meter^{-2}}, computed with different models: (a) DFT reference, (b) ReaxFF parameters by Joshi \textit{et al.} and (c) ReaxFF parameters obtained with Balanced Loss.
        Al=gray, O=red, H=white.
    }
    \label{fig:optstructs}
\end{figure}

Figure~\ref{fig:optstructs} offers a first visual impression of how the new force field improves the prediction of dissociative water adsorption on alumina.
It shows a high-hydration structure of the \ce{\gamma-Al2O3} (001) slab model, labeled \texttt{gamma\_surf-001\_03w} in the validation set, geometry-optimized with the reference method (DFT calculation with VASP), the initial force field by Joshi \textit{et al.}\cite{Joshi2014} and the new force field in this work.
The Joshi \textit{et al.}\ force field predicts a severe deformation of the alumina structure, and already desorbs water molecules in this static calculation while all water should be adsorbed according to reference calculation.
In contrast, the new force field in this work predicts a geometry that is visually deviating only slightly from the DFT reference, and the water molecules remain adsorbed on the surface.

\begin{figure}
    \centering
    \includegraphics{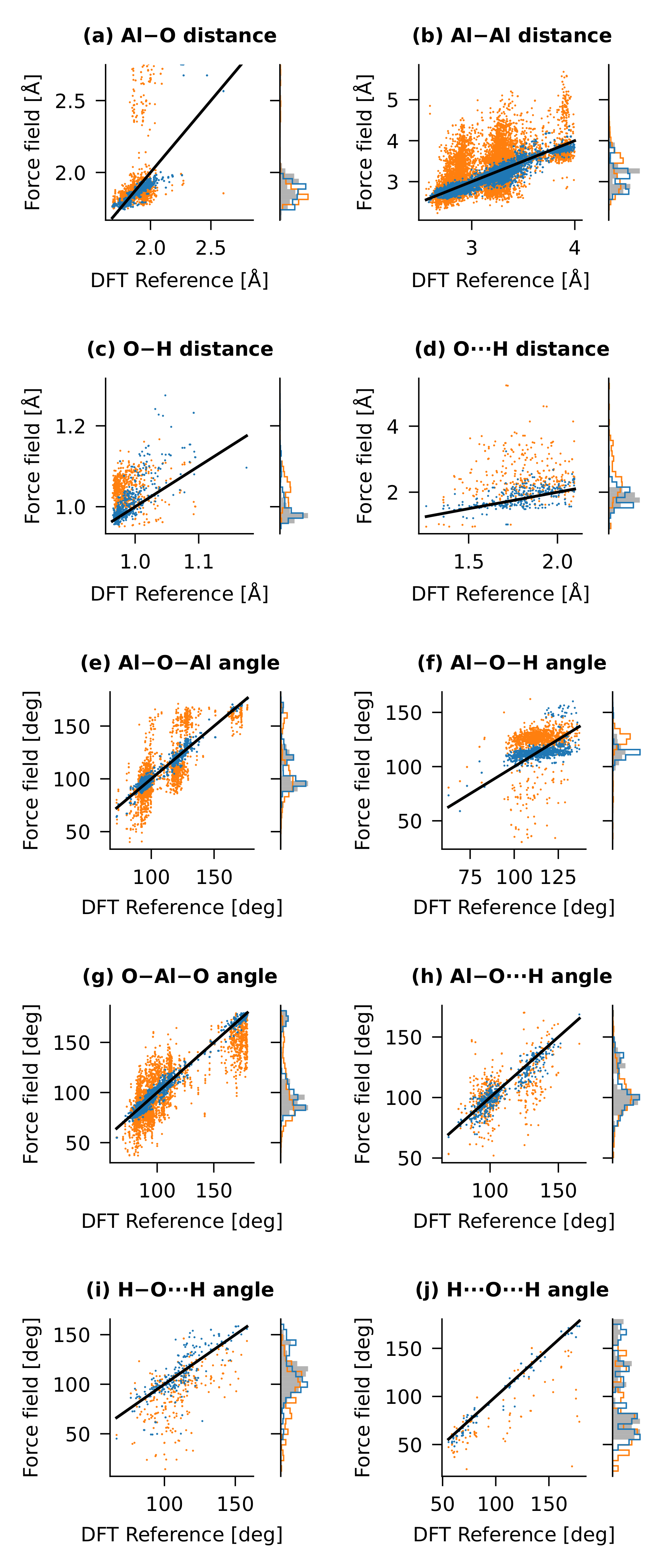}
    \caption{
        Parity plots of the reference internal coordinates from the validation set versus the force field predictions:
        initial parameters by Joshi \textit{et al.}\cite{Joshi2013, Joshi2014} (orange)
        and optimal parameters obtained with Balanced loss (blue).
        Distances: (a) \ce{Al-O}, (b) \ce{Al-Al}, (c) \ce{O-H} and (d) \ce{O$\cdots$H}.
        Angles: (e) \ce{Al-O-Al}, (f) \ce{Al-O-H}, (g) \ce{O-Al-O}, (h) \ce{Al-O$\cdots$H}, (i) \ce{Al-O$\cdots$H} and (j) \ce{Al$\cdots$O$\cdots$H}.
        The parity line is plotted as a black solid line.
        In panels (a) and (c) the vertical axis is manually limited to only show bonded distances.
        The initial parameters from Joshi result in many broken bonds, which are omitted for the sake of clarity.
    }
    \label{fig:parity-combined}
\end{figure}

\begin{figure}
    \centering
    \includegraphics{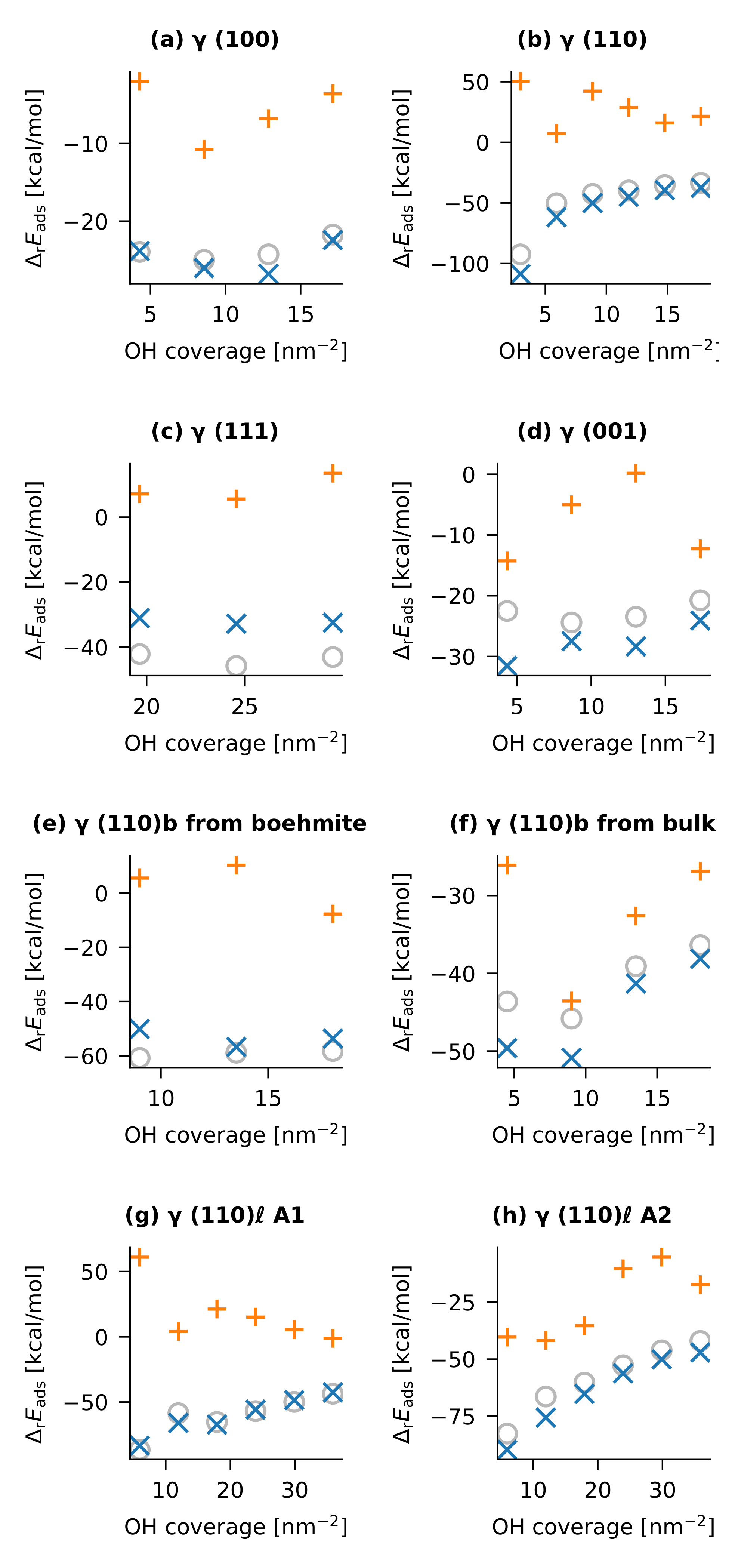}
    \caption{
        Water adsorption energy per water molecule on \ce{\gamma-Al2O3} surfaces as a function of the \ce{OH} coverage, using the bare surface as the reference, i.e.\, using $m_i=0$ in Eq.~\eqref{eq:eads}.
        Results are computed with DFT (gray circle), Joshi FF (orange plus), FF from this work (blue cross).
        Panels (a) to (c) are adsorption energies from the training set, whereas (d) to (h) are derived from the validation set.
        The nomenclature of the surfaces in the validation set is described in Ref.~\citenum{pigeon_revisiting_2022}.
    }
    \label{fig:adsorption-combined}
\end{figure}

The performance of the new force field can be evaluated more in detail by analyzing the distributions of categorized data in the training and validation sets, and their deviations from the DFT reference.
Table~\ref{tab:rmses} shows the RMSEs between force field and reference data, for each category, for the training and validation set, and for the initial force field by Joshi \textit{et al.}\cite{Joshi2013, Joshi2014} and the one optimized in this work.
To make the RMSEs directly comparable, only data related to \ce{\gamma-Al2O3} were taken from the training set since the validation set also contains only \ce{\gamma-Al2O3}.
In addition, only data categories (rows) are included that exist in both data sets.
For reference, also the standard deviations on the reference data per category are reported.
Figure~\ref{fig:parity-combined} shows parity plots for all categories of internal coordinates in the validation set.
In addition, water adsorption energies on \ce{\gamma-Al2O3} surfaces in the training and validation sets are shown in Figure~\ref{fig:adsorption-combined}.
Not all data categories from the training set exist for the validation set, because the validation set is focused on adsorption on \ce{\gamma-Al2O3} only.
This is reflected in the Table~\ref{tab:rmses}, Figure~\ref{fig:parity-combined} and Figure~\ref{fig:adsorption-combined} by only considering \ce{\gamma-Al2O3} structures.
Table S6 and Figure S3 in the Supporting Information contain the results omitted here, i.e.\ not involving \ce{\gamma-Al2O3} surfaces, for which a direct comparison to the validation set is not possible.

\begin{table}
    \centering
    \begin{adjustbox}{width=\textwidth}
    \begin{tabular}{
        c|c|
        S[table-format=3.2]
        S[table-format=3.2]
        S[table-format=3.2]
        S[table-format=4.0]
        |
        S[table-format=3.2]
        S[table-format=3.2]
        S[table-format=3.2]
        S[table-format=4.0]
    }
        Category &
        Unit &
        \multicolumn{4}{c|}{Training} &
        \multicolumn{4}{c}{Validation}
        \\
        & &
        \multicolumn{1}{c}{Ref.} &
        \multicolumn{1}{c}{Joshi} &
        \multicolumn{1}{c}{BL} &
        &
        \multicolumn{1}{c}{Ref.} &
        \multicolumn{1}{c}{Joshi} &
        \multicolumn{1}{c}{BL} &
        \\
        & &
        \multicolumn{1}{c}{SD} &
        \multicolumn{1}{c}{RMSE} &
        \multicolumn{1}{c}{RMSE} &
        \# &
        \multicolumn{1}{c}{SD} &
        \multicolumn{1}{c}{RMSE} &
        \multicolumn{1}{c}{RMSE} &
        \#
        \\
        \hline
\ce{Al-O}  & \si{\angstrom} & 0.16       & 0.45       & 0.09       & 1405       & 0.10       & 0.40       & 0.06       & 2102 \\
\ce{Al-Al} & \si{\angstrom} & 0.30       & 0.40       & 0.11       & 6759       & 0.31       & 0.40       & 0.11       & 13588 \\
\ce{O-H}   & \si{\angstrom} & 0.03       & 0.31       & 0.07       & 352        & 0.02       & 0.18       & 0.06       & 704 \\
\ce{O$\cdots$H} & \si{\angstrom} & 0.18       & 0.85       & 0.17       & 203        & 0.18       & 0.90       & 0.25       & 351 \\
\ce{Al-O-Al} & deg        & 21.2       & 17.0       & 4.4        & 1756       & 20.0       & 16.6       & 3.8        & 2744 \\
\ce{Al-O-H} & deg        & 12.2       & 20.4       & 6.9        & 598        & 9.3        & 18.7       & 8.3        & 1217 \\
\ce{O-Al-O} & deg        & 29.4       & 16.7       & 4.5        & 2943       & 30.9       & 17.4       & 3.9        & 4656 \\
\ce{Al-O$\cdots$H} & deg        & 18.4       & 16.9       & 6.5        & 301        & 17.6       & 17.6       & 6.6        & 554 \\
\ce{H-O$\cdots$H} & deg        & 17.5       & 27.4       & 12.7       & 142        & 17.3       & 29.4       & 14.7       & 254 \\
\ce{H$\cdots$O$\cdots$H} & deg        & 31.1       & 28.9       & 4.4        & 65         & 38.0       & 33.7       & 7.7        & 88 \\
\textbf{GSH} & \si{\kilo\cal \per \mol} & 15.0       & 55.7       & 6.6        & 37         & 16.4       & 54.4       & 8.3        & 101 \\
\textbf{SUR} & \si{\kilo\cal \per \mol} & 6.1        & 11.3       & 2.9        & 8          & 1.4        & 4.7        & 1.4        & 6
        \\ \hline
    \end{tabular}
    \end{adjustbox}
    \caption{
        Comparison of root-mean-square-errors (RMSEs) of the initial force field by Joshi \textit{et al.}\cite{Joshi2013, Joshi2014} and the force field optimized in this work (BL).
        The RMSEs are computed for training and validation sets and are grouped per data category.
        For reference, the standard deviation (SD) on the reference data is included.
            }
    \label{tab:rmses}
\end{table}

The RMSEs in Table~\ref{tab:rmses} show that new force field significantly reduces the errors on the bond lengths compared to the Joshi \textit{et al.}\ force field.\cite{Joshi2013, Joshi2014}
A subset of the bonds is broken after geometry optimization with the force fields, which is not fully visible in Figures~\ref{fig:parity-combined}(a)~and~\ref{fig:parity-combined}(c), because this would require an impractical scale for the vertical axes.
In the validation set, \SI{14.3}{\percent} of the \ce{Al-O} bonds and \SI{1.1}{\percent} of the \ce{O-H} bonds are broken with the Joshi \textit{et al.}\ force field.
With the new force field proposed here, these percentages reduce to \SI{0.0}{\percent} and \SI{0.3}{\percent}, respectively.
These percentages are consistent with the visualization in Figure~\ref{fig:optstructs}(b) and confirm that the Joshi \ce{\gamma-Al2O3} surfaces force field cannot preserve the structural integrity of the (hydrated) \ce{\gamma-Al2O3} slabs.
By consequence, this force field also performs poorly for other categories (angles and energies), for which a correct bonding topology is required.

The parity plots of the distances in Figures~\ref{fig:parity-combined}(a),~\ref{fig:parity-combined}(c)~and~\ref{fig:parity-combined}(d) also reveal that even the new force field captures the variations in bond lengths only approximately.
This is also confirmed by the fact that the RMSEs of the distances in Table~\ref{tab:rmses} are of the same order as the standard deviation on the distances in the reference data.
For the \ce{O-H} and \ce{O$\cdots$H} distances, this was to be expected, because the corresponding bond parameters in ReaxFF were not re-optimized for the sake of compatibility with the silicate parameters in the Joshi \textit{et al.}\ force field.
For the \ce{Al-O} distances, the performance is slightly better, which is consistent with the fact that several \ce{Al-O} parameters were re-optimized.

The new force field also improves upon the Joshi \textit{et al.}\ force field in terms of valence angles, again with a somewhat better performance when no hydrogen atoms are involved.
It is remarkable that the multimodal distributions of the \ce{Al-O-Al}, \ce{O-Al-O} and \ce{Al-O$\cdots$H} angles are reproduced well by the new force field, despite only having a single energy term for these angles in the ReaxFF force field.
Also the improvements of the \ce{H-O$\cdots$H} and \ce{H$\cdots$O$\cdots$H} angles, compared to the initial force field, are remarkable, because no corresponding valence angle of hydrogen bonding terms were reparameterized.

In line with the previous categories, also the RMSEs on the energies are significantly smaller with the new force field.
Before discussing the adsorption energies, it should be noted that the category \textbf{SUR}
was mainly introduced to improve the diversity of the training set. (It comprises reaction energies between bulk and slab models, normalized on the number of \ce{Al} atoms.)
A few data points in the same category can be derived from the validation structures and are included here for the sake of completeness.
The error on these data points has also decreased compared to the initial force field, confirming the ability of Balanced Loss to account for underrepresented categories in the training set.

The new force field reduces the error on the adsorption energies (category \textbf{GSH}) by more than a factor 6 in the validation set.
The improvements are also immediately clear in Figure \ref{fig:adsorption-combined}, which shows the adsorption energy on \ce{\gamma-Al2O3} surfaces as a function of \ce{OH} coverage.
In most cases, the new force field predicts the correct trends in the adsorption energy, with some exceptions at low \ce{OH} coverage in Figures \ref{fig:adsorption-combined}(d) and \ref{fig:adsorption-combined}(f).
For the (001) and the (110)$_\text{b}$ slabs in the validation set, it is unclear why these surfaces exhibit larger errors in adsorption energy at low coverage.
The Joshi \textit{et al.} force field incorrectly predicts water desorption for 12 out of 53
surface structures in the validation set, which hampers reliable energy predictions.
With, the new force field, this problem is far less prevalent:
Only one water molecule (out of six) from only one surface structure spontaneously desorbs.
The new force field can also predict the magnitude of the adsorption energy, with an RMSE of \SI{8.3}{\kilo\cal \per \mol}, compared to a standard deviation of \SI{16.4}{\kilo\cal \per \mol} of the adsorption energies in the validation set.

For all categories of data discussed above, the errors on static calculations are very similar for the training and the validation sets, indicating that the significant improvements of the new ReaxFF force field generalize to structures not used for training.
The errors on distances, valence angles and energy differences are also comparable to those reported for previous ReaxFF models. \cite{rahaman_2010, Goken2011, muller_2016, Hu_2017, Trnka2017, Bertels2020}

Because the training set only includes equilibrium geometries, it should be tested to what extent the new parameters can also reproduce non-equilibrium energies.
To this end, a constant-temperature DFT molecular dynamics run was performed on structure \texttt{gamma\_surf-110l\_A1\_06w} from the validation set, using the same level of theory as the training data.
An elevated temperature of \SI{1000}{\kelvin} stimulates the desorption of water, which is observed during the first \SI{200}{\femto\second}.
Section S2 of the Supporting Information presents a detailed comparison of the DFT and ReaxFF energies computed for snapshots from this trajectory.
In summary, the instantaneous DFT adsorption energy computed with Eq.~\eqref{eq:eads} as a function of time is reproduced qualitatively by the ReaxFF parameters obtained with Balanced Loss: The relative error of about \SI{25}{\percent} over the first \SI{200}{\femto\second} is comparable to the RMSE on the training set for GSH category.
Our new parameters also show a clear improvement compared to the ReaxFF energies obtained with the parameters of Joshi \textit{et al.}
The thermal energy fluctuations due to vibrations within the alumina slab are not well reproduced, which is expected, since no corresponding data was used for training.

Despite our improvements, it remains interesting to explore further refinements, e.g.\ to further reduce errors in adsorption energies or to improve the vibrational states of alumina.
One avenue is to activate more parameters during the training that are now fixed for the sake of backward compatibility.
Giving up backward compatibility would only be useful when extending the chemical space of the training set to aluminosilicates and water, such that all parameters of the Joshi \textit{et al.}\ force field can be re-optimized.
However, this would be a daunting enterprise, because the current training set size is already computationally demanding:
A single CMA run in this work already took more than 24 hours.
In addition to the increased cost of the training set, more active parameters also imply more local minima and a slower convergence of CMA-ES, further exacerbating the computational cost.
This avenue is therefore only feasible when one can drastically speed up the training of ReaxFF parameters.
It is encouraging that efficiency gains were reported in recent publications, e.g.\ by parallel optimization management, \cite{FreitasGustavo2022} or by machine learning surrogates of the loss function.\cite{Sengul2021, Daksha2021, Yeon2021}
One may also reduce the dimensionality of the parameter space through sensitivity analysis to speed up the CMA runs.\cite{FreitasGustavo2023}

\begin{figure}
    \centering
    \includegraphics{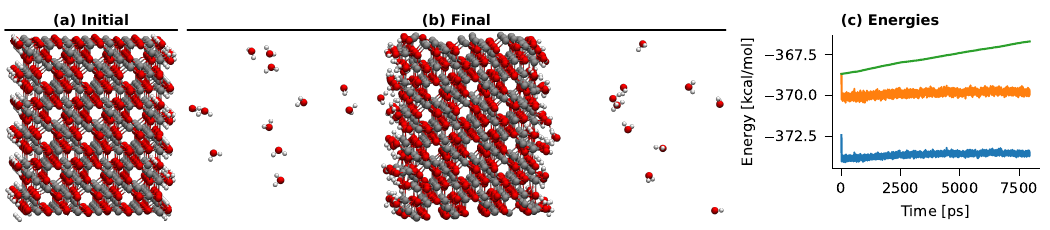}
    \caption{
        The initial (a) and final (b) states of a constant-temperature MD trajectory of a (110)$_\ell$ \ce{\gamma-Al2O3} with water initially adsorbed on the surface.
        Al=gray, O=red, H=white.
        See text for details.
        (c) The kinetic energy (blue), total energy (kinetic + potential, orange) and the conserved quantity (green).
    }
    \label{fig:md}
\end{figure}

As a final check of the new force field, a molecular dynamics (MD) simulations is performed on (110)$_\ell$ \ce{\gamma-Al2O3} slab with a cross section of \SI{6.0}{\nano\meter\tothe{2}}, a thickness of \SI{1.8}{\nano\meter} and surrounded by a vacuum layer of \SI{5.7}{\nano\meter} wide.
In the initial structure, the maximal number of water molecules is dissociatively adsorbed on both sides of the slab, such that the chemical formula is \ce{Al_{504}O_{864}H_216}.
The MD simulation follows a \textit{NVT}-ensemble and uses a Nosé-Hoover thermostat\cite{evans_nosehoover_1985} with a time step of \SI{0.2}{\femto\second}, a temperature of \SI{500}{\kelvin}, a time-constant of \SI{500}{\pico\second} and a fixed periodic box size.
Figure~\ref{fig:md} shows the initial and final states of the MD trajectory, as well as the evolution of the kinetic energy, total energy (kinetic + potential) and the conserved quantity.
The conserved quantity (green) exhibits a slow linear increase, which is acceptable for long ReaxFF MD simulations at a constant temperature.
ReaxFF forces are imperfect due to numerical convergence of the variable charges and small discontinuities in the ReaxFF energy surface.
Such small force errors are practically tiny random kicks on the nulcei, which slowly pump energy into the system, but this is easily compensated for by the thermostat and results in a slowly increasing conserved quantity.
The 3D visualizations show that some of the water molecules desorb, as expected at a temperature of \SI{500}{\kelvin}.
This test shows that the new force field can also be used for MD simulations, even though it is only trained on optimized geometries.
A complete study of water adsorption, with larger slabs, different alumina surfaces and temperatures, goes beyond the scope of this work.
We expect that the new force field will make such simulations possible, at time and length scales that are infeasible for DFT methods.

\section{Conclusion and Outlook}
\label{sec:conclusion}

This work addresses the difficulty of assigning fitting weights (or their inverses, often called sigmas) in a conventional ReaxFF loss function.
Balanced Loss is proposed as a new cost function as well as a workflow to reformulate the weight assignment in terms of more manageable concepts.
One starts by classifying the training data into meaningful categories with a corresponding tolerance, which is the root-mean-square error (between ReaxFF predictions and reference data in that category) that one is willing to tolerate.
When the error on one category exceeds the corresponding tolerance more than other categories, the Log-Sum-Exp form of Balanced Loss guarantees that this error will completely dominate the loss function, effectively forcing the optimizer to reduce this error first.
If one or more of such dominating categories remain after the parameters converge, it is guaranteed that these errors cannot be reduced further at the expense of making larger errors in other categories.
As a result, the parameter optimization also assesses whether ReaxFF can meet the expectations defined by the tolerances.
If necessary, the expectations can be adjusted, followed by a new parameter optimization.
The methodology is applied to a realistic and challenging reparameterization of ReaxFF for water adsorption on alumina surfaces.
This not only results in a competitive force field, but also provides insight into the performance that can be expected from ReaxFF for each category of training data.

The new force field derived in this work is a refinement of the alumina parameters in the aluminosilicate force field of Joshi \textit{et al.}\cite{Joshi2013, Joshi2014}
The training and validation data consisted of geometry and energy data from previous Density Functional Theory (DFT) studies of water adsorption on boehmite and \ce{\gamma-Al2O3} surfaces.
While \ce{\gamma-Al2O3} is industrially the most relevant, boehmite structures were included in the training set to improve the data diversity.
Parameter selection focused on maintaining backward compatibility with the ReaxFF model of Joshi \textit{et al.}\ as much as possible, while still activating sufficient parameters to reproduce the training data.
Covariance Matrix Adaptation (CMA) is used to minimize Balanced Loss as a function of the selected parameters.
40 independent CMA runs were performed to test the robustness of the optimized parameters.
Of all these runs, the result with the lowest error on the training set is used for validation.
Static calculations confirm that the optimized force field produces very similar errors on the \ce{\gamma-Al2O3} properties present in the training and validation sets.
The force field can be used for MD simulations, and we expect it to be applicable to extensive simulations of water adsorption on alumina surfaces, at time and length scales inaccessible to DFT methods.

This study also revealed several challenges and new avenues for future method development.
Obviously, it should be validated whether the Balanced Loss workflow is equally helpful for other ReaxFF parameterizations.
Even beyond the scope of ReaxFF, Balanced Loss may facilitate optimization problems involving multiple (possibly competing) categories of training data.
In addition, this study confirmed known pitfalls of ReaxFF parameterization and suggested new ones.
Despite our careful and relatively small selection of parameters, the minimum of the loss function is still degenerate, suggesting that the number of active parameters can be reduced further with a sensitivity analysis.
Finally, it was observed that parameters often converge to near-optimal values close to the bounds.
A better understanding of this phenomenon may help to speed up the convergence to better parameters.
\section{Data and Code availability}
\label{sec:code_availability}

A data set as made available at \url{https://doi.org/10.5281/zenodo.10491516} comprising: the training and validation data, the results of the 240 CMA optimizations, and scripts used to select parameters, convert datasets and perform the training and validation.

    \begin{acknowledgement}

        The authors thank the company Software for Chemistry \& Materials B.V.\ (SCM) for providing an AMS developer license at no cost.
        In particular, we are grateful to Matti Hellstr\"{o}m and Tom{\'{a}}{\v{s}} Trnka from SCM for their technical support.
        The authors also thank members of the Center for Molecular Modelling at Ghent University, especially Tom Braeckevelt, Leonid Komissarov and Michael Gustavo as well as Pascal Raybaud from IFP Energies nouvelles for the insightful discussions.
        Computational resources and services used were provided by Ghent University (Stevin and
Hortense Supercomputer Infrastructures), the VSC (Flemish Supercomputer Center), funded by
the Research Foundation-Flanders (FWO), and IFP Energies nouvelles (ENER440).
        Funding was provided by IFPEN and UGent (BOF/24J/2023/121).
        T.V.\ Acknowledges the Special Research Fund (BOF) of Ghent University for its continuous support.
    \end{acknowledgement}

    \begin{suppinfo}
        The Supporting Information contains two files:
        \begin{itemize}
            \item
            A PDF document with
            (i) additional figures and tables describing the training and validation sets,
            (ii) the ReaxFF parameters selected for optimization, and
            (iii) additional performance metrics on the training set that have no counterpart in the validation set.
            \item
            A ZIP file with the optimized parameters used for the force field validation,
            and example input files for the MD simulations in this work.
        \end{itemize}
    \end{suppinfo}

    \bibliography{references.bib}

\end{document}


\tableofcontents
    \input{1-additional-display-items.tex}
    \input{2-non-equilibrium-energies.tex}
    \bibliography{references.bib}

%% file: 1-additional-display-items.tex
\begin{landscape}
\section{Additional Display Items}
\scriptsize
\renewcommand*{\arraystretch}{0.6}
\begin{longtable}{
    >{\ttfamily}r|
    r|
    S[table-format=2.0]|
    S[table-format=2.0]|
    S[table-format=2.0]|
    S[table-format=2.0]|
    S[table-format=2.0]|
    S[table-format=2.0]|
    S[table-format=2.0]|
    S[table-format=2.0]|
    S[table-format=2.0]|
    S[table-format=2.0]
}
    \caption{
        Structures in the training set and contributions to each category of data.
        Bonds are represented by pairs of chemical elements, angles by triplets.
        The symbol $\cdots$ denotes a hydrogen bond.
        Internal coordinates with oxygen not bound to aluminum are discarded.
        For hydrated alumina surfaces and edges, and for boehmite surfaces, the internal coordinates without hydrogen are not considered.
    }
    \label{tab:training_set_overview} \tabularnewline
    \multicolumn{1}{c|}{Structure} &
    \multicolumn{1}{c|}{Chem.~Form.} &
    \multicolumn{1}{c|}{\ce{O-H}} &
    \multicolumn{1}{c|}{\ce{O$\cdots$H}} &
    \multicolumn{1}{c|}{\ce{Al-O}} &
    \multicolumn{1}{c|}{\ce{Al-Al}} &
    \multicolumn{1}{c|}{\ce{O-Al-O}} &
    \multicolumn{1}{c|}{\ce{Al-O-Al}} &
    \multicolumn{1}{c|}{\ce{Al-O-H}} &
    \multicolumn{1}{c|}{\ce{Al-O$\cdots$H}} &
    \multicolumn{1}{c|}{\ce{H-O$\cdots$H}} &
    \multicolumn{1}{c|}{\ce{H$\cdots$O$\cdots$H}} \\
    \hline
\input{training_set_overview_rows.tex}
\end{longtable}
\end{landscape}

\newpage
\begin{figure}[H]
    \centering
    \includegraphics{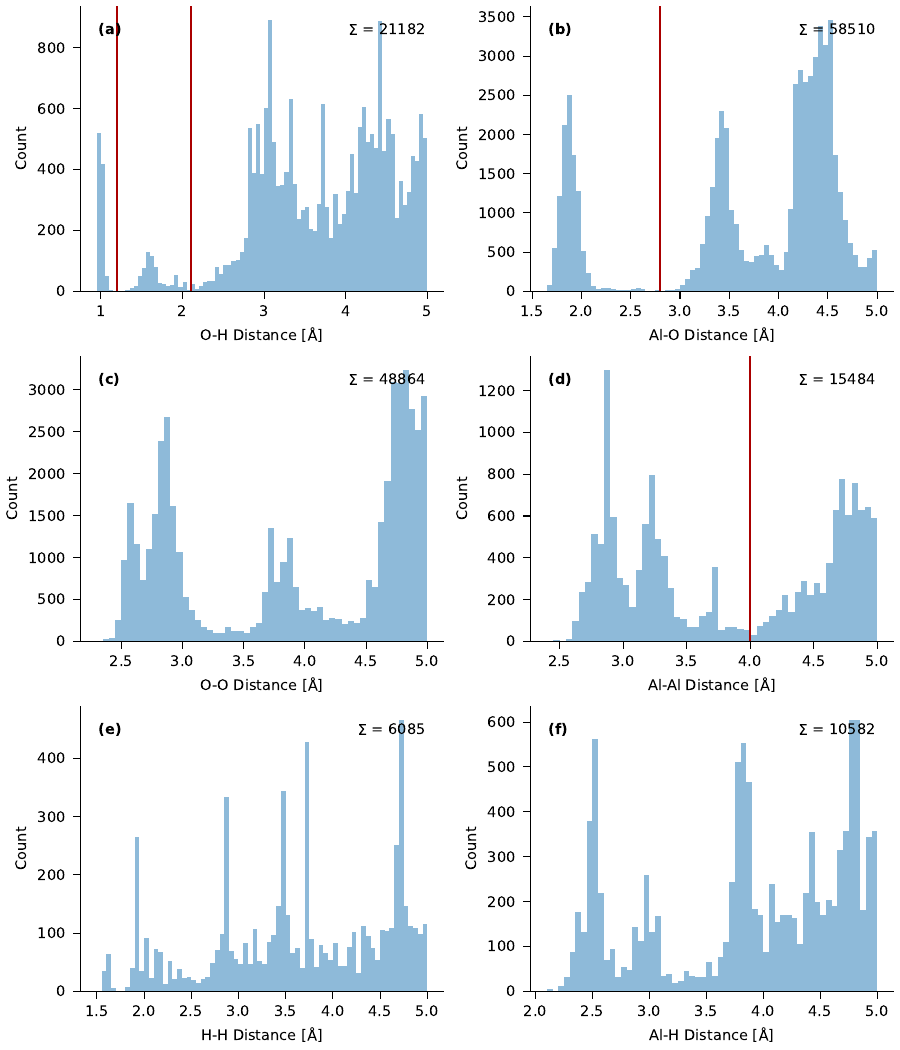}
    \caption{
        Histograms of all interatomic distances in the training set up to \SI{5}{\angstrom},
        grouped per pair of chemical elements.
        Cutoffs for \ce{OH} and \ce{AlO} pairs depicted as vertical red lines:
        \SI{1.2}{\angstrom} for \ce{O-H} bonds,
        \SI{2.1}{\angstrom} for hydrogen bonds and
        \SI{2.8}{\angstrom} for \ce{Al-O} bonds.
        See main text for a more detailed description.
    }
    \label{fig:training_set_ics_phase1}
\end{figure}

\newpage
\begin{figure}[H]
    \centering
    \includegraphics{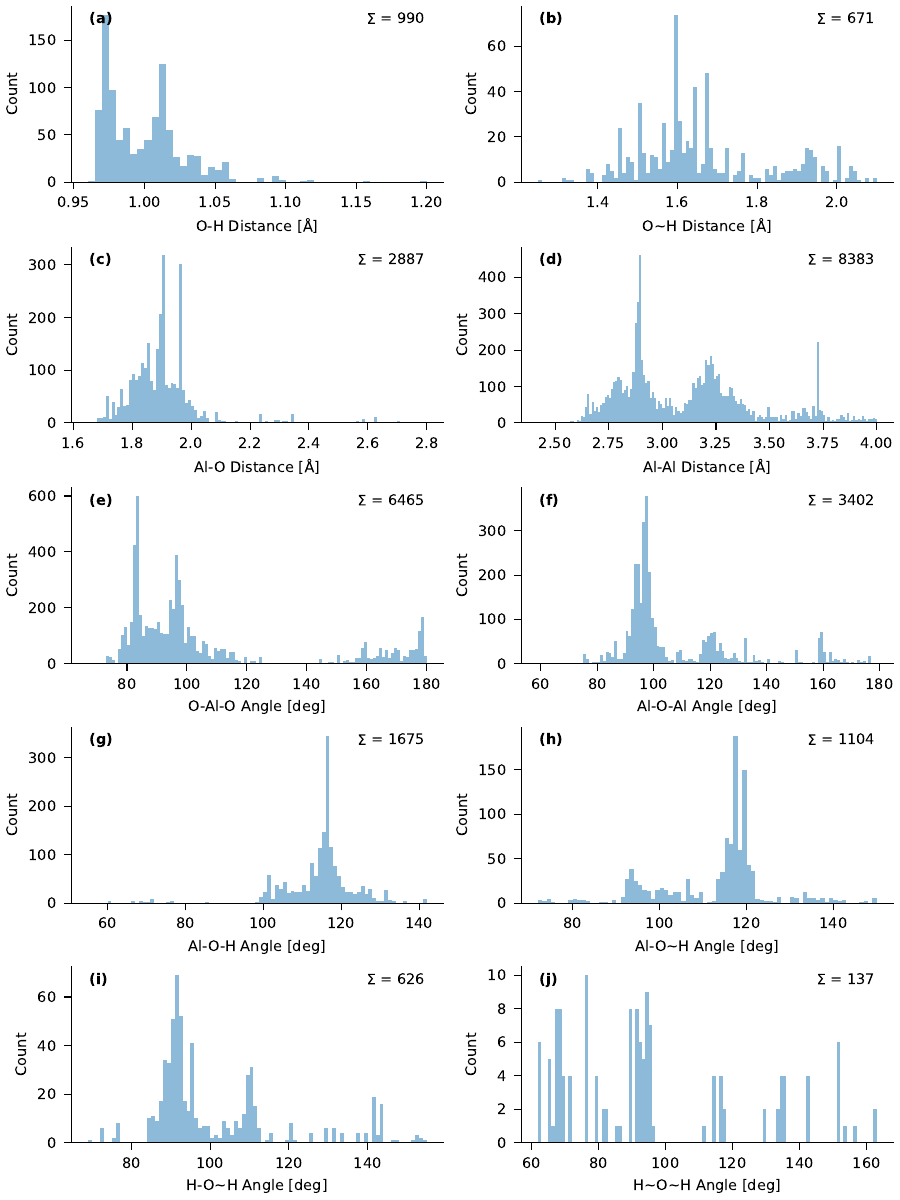}
    \caption{
        Histograms of internal coordinates in the training set.
        In the labels on the horizontal axis, a dash represents a regular bond and a tilde represents a hydrogen bond.
    }
    \label{fig:training_set_ics_phase2}
\end{figure}

\newpage
\begin{figure}[H]
    \centering
    \includegraphics{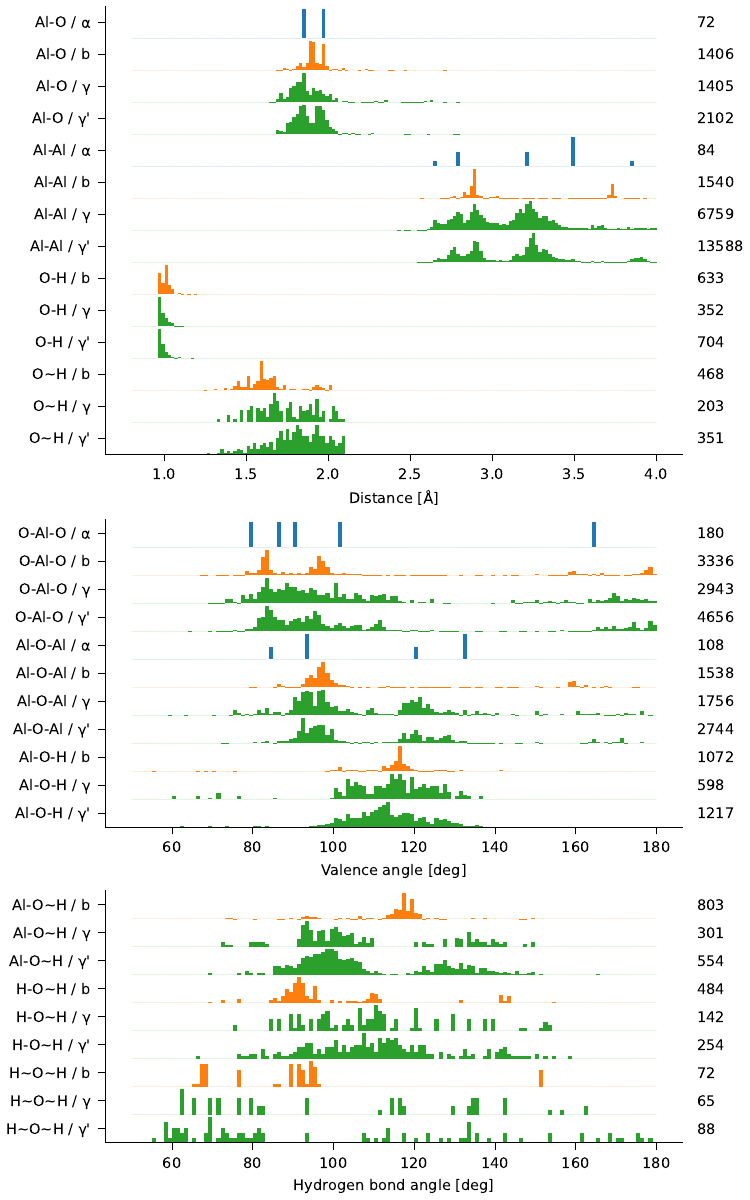}
    \caption{
        Histograms of internal coordinates, using the notation of Figure~\ref{fig:training_set_ics_phase2}, per material.
        Histograms for the training set: $\alpha$\,$=$\,\ce{\alpha-Al2O3} (blue), b\,$=$\,boehmite (orange), $\gamma$\,$=$\,\ce{\gamma-Al2O3} (green).
        Histograms for the validation set: $\gamma'$\,$=$\,\ce{\gamma-Al2O3} (green).
        The number of internal coordinates within each class is shown to the right of the corresponding histogram.
    }
    \label{fig:materials-ics}
\end{figure}

\newpage
\begin{center}
    \renewcommand*{\arraystretch}{0.6}
    \begin{longtable}{
        l|
        r
        >{\ttfamily}l
        l|
        r
        S[table-format=2.0]
    }
        \caption{
            Overview of all chemical equations in the training set.
            Reactants are given negative coefficients.
            For each reaction, three reaction energies are in \si{\kilo\cal\per\mole}:
            the reference DFT result (R), the prediction with the Joshi force field (J) and the prediction with the new force field in this work (B).
            Water adsorption energies are normalized to the number of water molecules.
            All other reaction energies are normalized on the number of \ce{Al} atoms.
            The categories are defined in the main text.
        }
        \label{tab:training_set_energies} \tabularnewline
        \multicolumn{1}{c|}{Category} &
        \multicolumn{1}{c}{Coeff.} &
        \multicolumn{1}{c}{Structure} &
        \multicolumn{1}{c|}{Chem. Form.} &
        \multicolumn{2}{c}{Reaction energy} \\
        \hline
\input{training_set_energies_rows.tex}
    \end{longtable}
\end{center}

\newpage
\begin{figure}[H]
    \centering
    \includegraphics{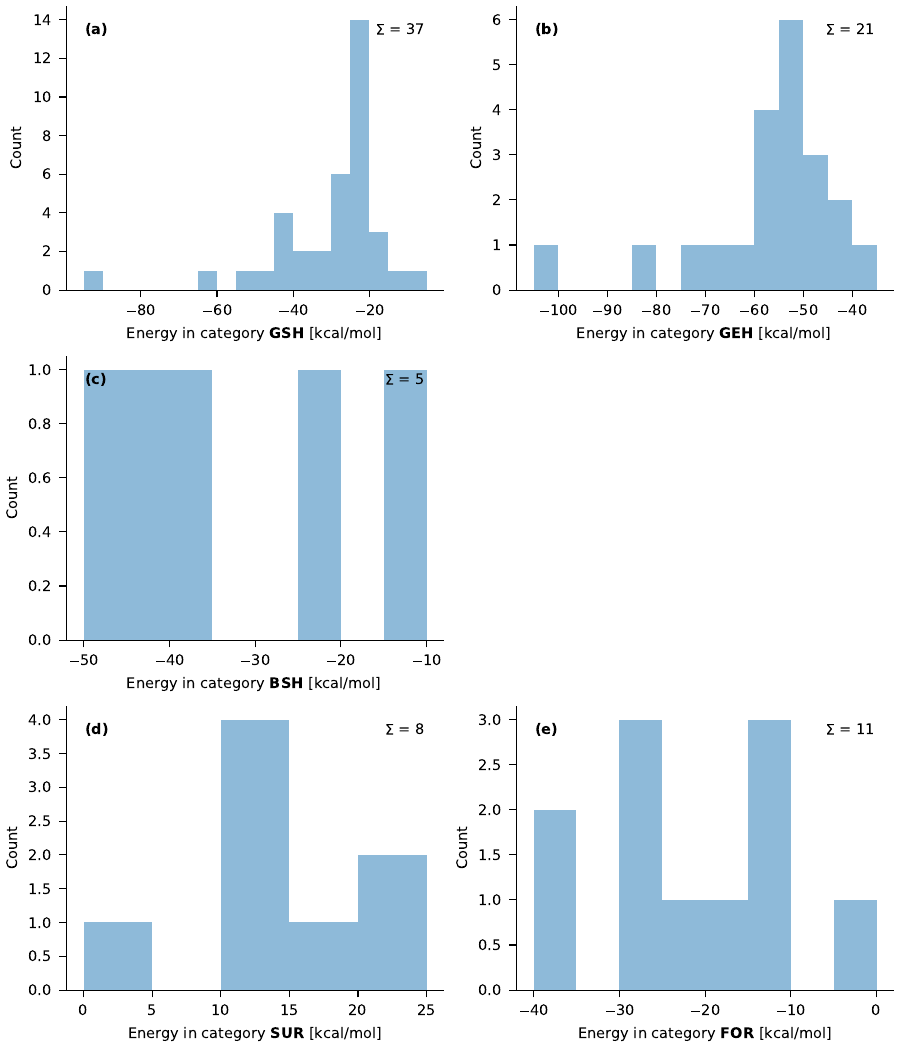}
    \caption{
        Histograms of reaction energies in the training set.
    }
    \label{fig:training_set_energies}
\end{figure}

\newpage
\begin{center}
    \renewcommand*{\arraystretch}{0.6}
    \begin{longtable}{
        >{\ttfamily}l|
        c|
        >{\ttfamily}l|
        >{\ttfamily}l|
        S[table-format=3.1]|
        S[table-format=3.1]|
        S[table-format=3.1]|
        S[table-format=3.1]
    }
        \caption{List of activated parameters and their corresponding block, initial values from \citenum{joshi_reactive_2014} and bounds.}
        \label{tab:activated_parameters} \tabularnewline
        \multicolumn{1}{c|}{Name} &
        Unit &
        \multicolumn{1}{c|}{Atoms} &
        \multicolumn{1}{c|}{Block} &
        \multicolumn{1}{c|}{Joshi 2014} &
        \multicolumn{1}{c|}{Lower bound} &
        \multicolumn{1}{c|}{Upper bound} &
        \multicolumn{1}{c}{This work} \\
            \hline
\input{parameters_active.tex}
    \end{longtable}
\end{center}

\newpage
\begin{landscape}
\begin{center}
    \scriptsize
    \renewcommand*{\arraystretch}{0.6}
    \begin{longtable}{
        >{\ttfamily}r|
        r|
        S[table-format=2.0]|
        S[table-format=2.0]|
        S[table-format=2.0]|
        S[table-format=2.0]|
        S[table-format=2.0]|
        S[table-format=2.0]|
        S[table-format=2.0]|
        S[table-format=2.0]|
        S[table-format=2.0]|
        S[table-format=2.0]
    }
        \caption{
            Structures in the validation set and contributions to each category of data.
            Bonds are represented by pairs of chemical elements, angles by triplets.
            The symbol $\cdots$ denotes a hydrogen bond.
            Internal coordinates with oxygen not bound to aluminum are discarded.
            For hydrated alumina surfaces, the internal coordinates without hydrogen are not considered.
        }
        \label{tab:validation_set_overview} \tabularnewline
        \multicolumn{1}{c|}{Structure} &
        \multicolumn{1}{c|}{Chem.~Form.} &
        \multicolumn{1}{c|}{\ce{O-H}} &
        \multicolumn{1}{c|}{\ce{O$\cdots$H}} &
        \multicolumn{1}{c|}{\ce{Al-O}} &
        \multicolumn{1}{c|}{\ce{Al-Al}} &
        \multicolumn{1}{c|}{\ce{O-Al-O}} &
        \multicolumn{1}{c|}{\ce{Al-O-Al}} &
        \multicolumn{1}{c|}{\ce{Al-O-H}} &
        \multicolumn{1}{c|}{\ce{Al-O$\cdots$H}} &
        \multicolumn{1}{c|}{\ce{H-O$\cdots$H}} &
        \multicolumn{1}{c|}{\ce{H$\cdots$O$\cdots$H}} \\
        \hline
\input{validation_set_overview_rows.tex}
    \end{longtable}
\end{center}
\end{landscape}

\newpage
\begin{figure}[H]
    \centering
    \includegraphics{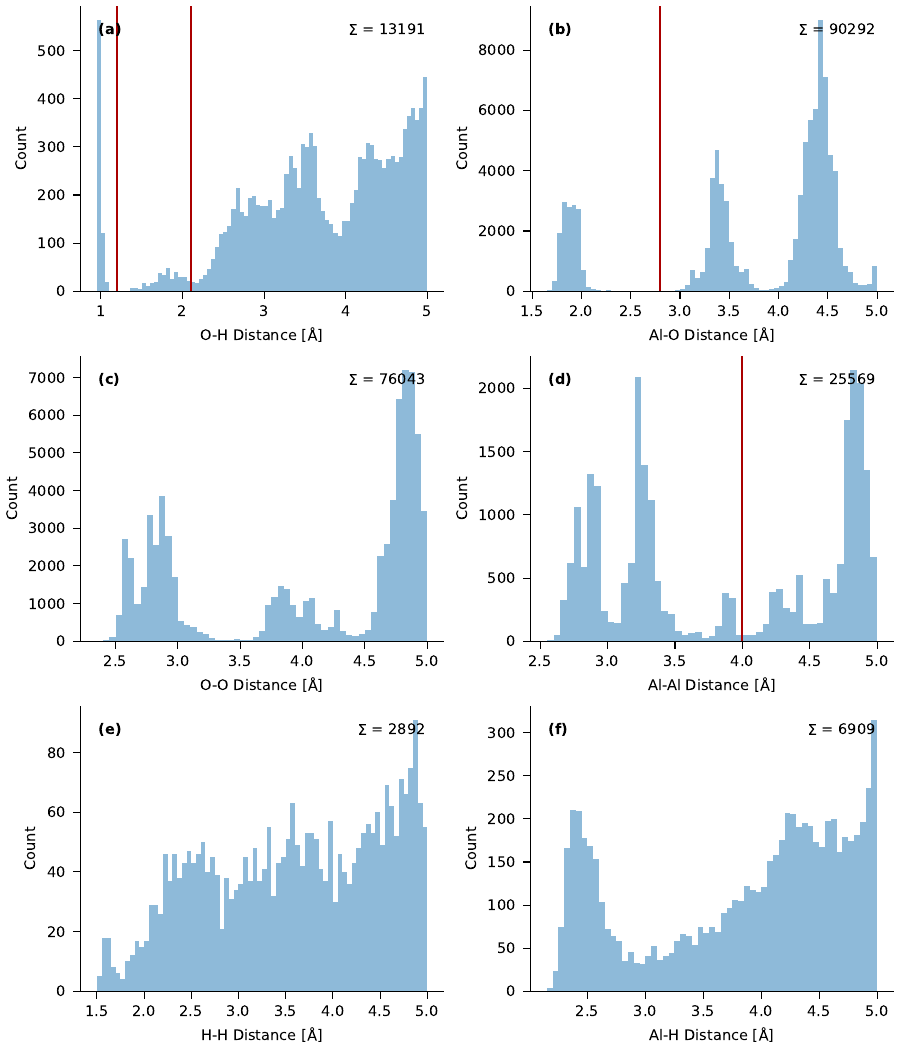}
    \caption{
        Histograms of all interatomic distances in the validation set up to \SI{5}{\angstrom},
        grouped per pair of chemical elements.
        Cutoffs for \ce{OH} and \ce{AlO} pairs depicted as vertical red lines:
        \SI{1.2}{\angstrom} for \ce{O-H} bonds,
        \SI{2.1}{\angstrom} for hydrogen bonds and
        \SI{2.8}{\angstrom} for \ce{Al-O} bonds.
        See main text for a more detailed description.
    }
    \label{fig:validation_set_ics_phase1}
\end{figure}

\newpage
\begin{figure}[H]
    \centering
    \includegraphics{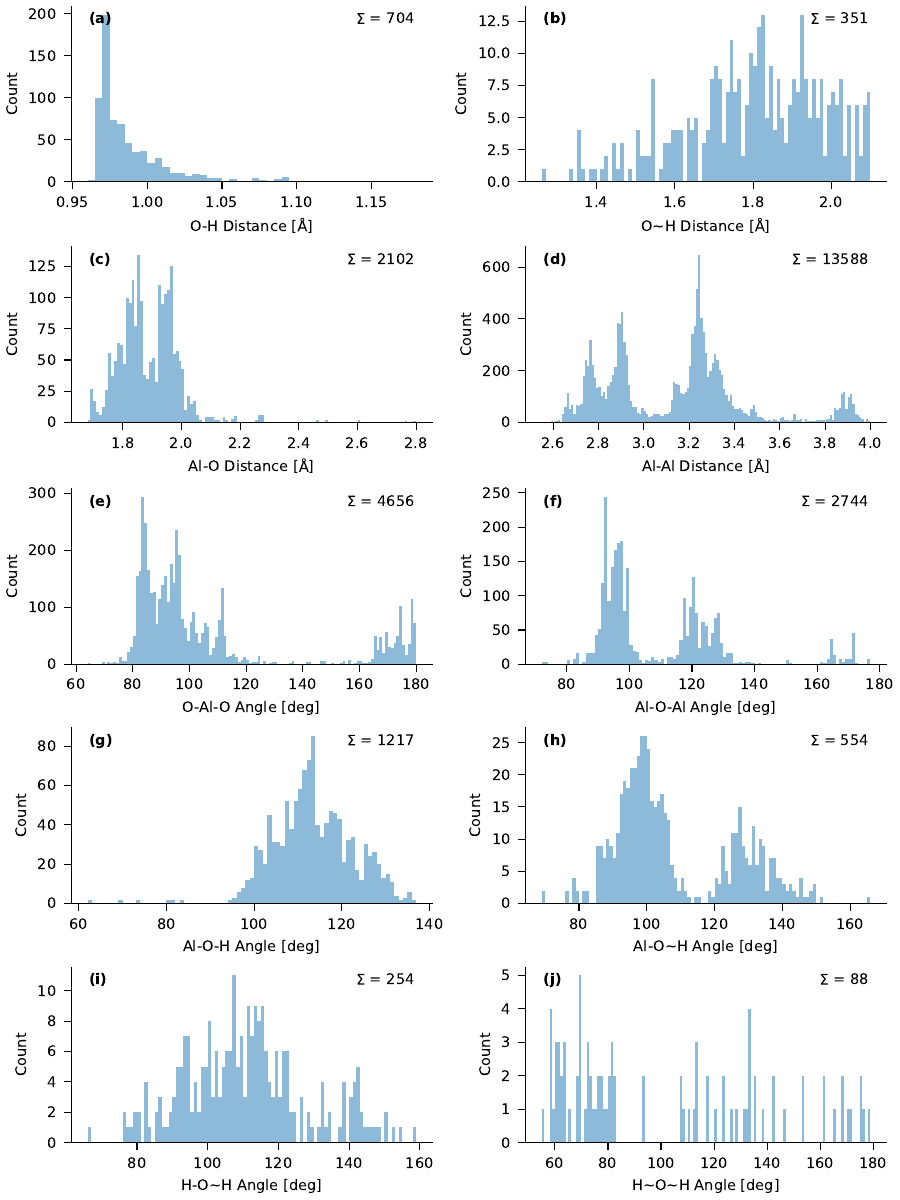}
    \caption{
        Histograms of internal coordinates in the validation set.
        In the labels on the horizontal axis, a dash represents a regular bond and a tilde represents a hydrogen bond.
        }
    \label{fig:validation_set_ics_phase2}
\end{figure}

\newpage
\begin{center}
    \renewcommand*{\arraystretch}{0.6}
    \begin{longtable}{
        l|
        r
        >{\ttfamily}l
        l|
        r
        S[table-format=2.0]
    }
        \caption{
            Overview of all chemical equations in the validation set.
            Reactants are given negative coefficients.
            For each reaction, three reaction energies are in \si{\kilo\cal\per\mole}:
            the reference DFT result (R), the prediction with the Joshi force field (J) and the prediction with the new force field in this work (B).
            Water adsorption energies are normalized to the number of water molecules.
            All other reaction energies are normalized on the number of \ce{Al} atoms.
            The categories are defined in the main text.
        }
        \label{tab:validation_set_energies} \tabularnewline
        \multicolumn{1}{c|}{Category} &
        \multicolumn{1}{c}{Coeff.} &
        \multicolumn{1}{c}{Structure} &
        \multicolumn{1}{c|}{Chem. Form.} &
        \multicolumn{2}{c}{Reaction energy} \\
        \hline
\input{validation_set_energies_rows.tex}
    \end{longtable}
\end{center}

\newpage
\begin{figure}[H]
    \centering
    \includegraphics{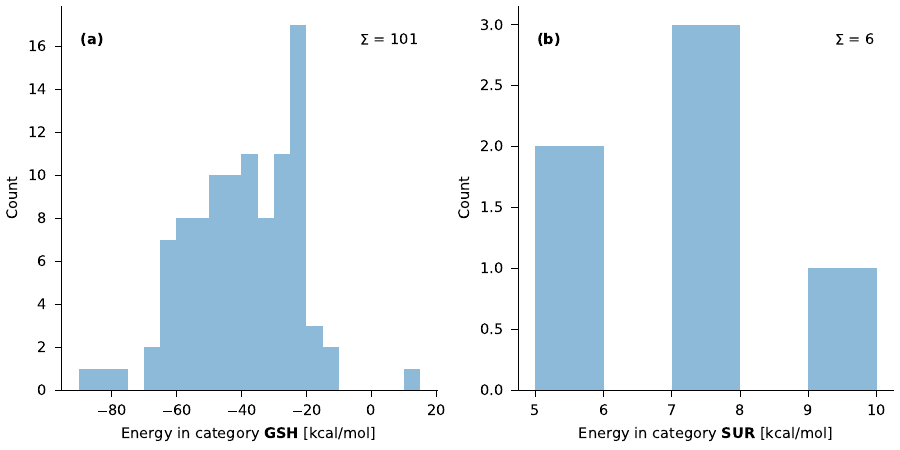}
    \caption{
        Histograms of reaction energies in the validation set.
    }
    \label{fig:validation_set_energies}
\end{figure}

\newpage
\begin{figure}[H]
    \centering
    \includegraphics{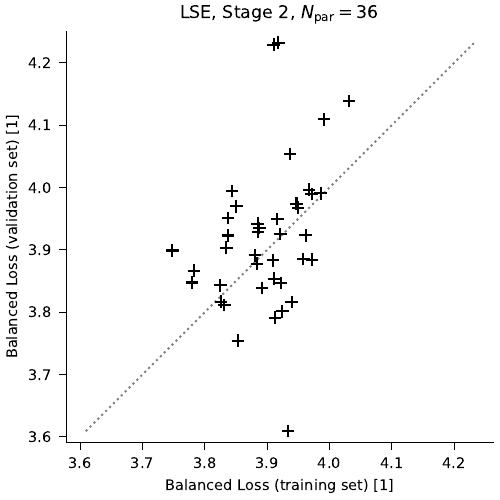}
    \caption{
        Parity plot comparing the value of the Balanced Loss function, for the training and validation sets, for the best solution from all 40 CMA runs (LSE form, optimization stage 2, see main text for details).
        Despite that the 40 optimized parameter vectors different significantly, their performance for the training and validation sets is similar.
    }
    \label{fig:sloppiness}
\end{figure}

\begin{table}[H]
    \centering
    \begin{tabular}{
        c|c|
        S[table-format=3.2]
        S[table-format=3.2]
        S[table-format=3.2]
        S[table-format=4.0]
    }
        Category &
        Unit &
        \multicolumn{1}{c}{Ref.} &
        \multicolumn{1}{c}{Joshi} &
        \multicolumn{1}{c}{BL} &
        \\
        & &
        \multicolumn{1}{c}{SD} &
        \multicolumn{1}{c}{RMSE} &
        \multicolumn{1}{c}{RMSE} &
        \#
        \\
        \hline
\input{rmses.tex}
        \\ \hline
    \end{tabular}
    \caption{
        Comparison of root-mean-square-errors (RMSEs) of the initial force field by Joshi \textit{et al.}\cite{joshi_reactive_2014} and the force field optimized in this work (BL).
        The RMSEs are computed for categories of training data and structures for which there is no counterpart in the validation set, i.e.\ not related to \ce{\gamma-Al2O3} surfaces.
        For reference, the standard deviation (SD) on the reference data is included.
            }
    \label{tab:rmses}
\end{table}

\begin{figure}[H]
    \centering
    \includegraphics{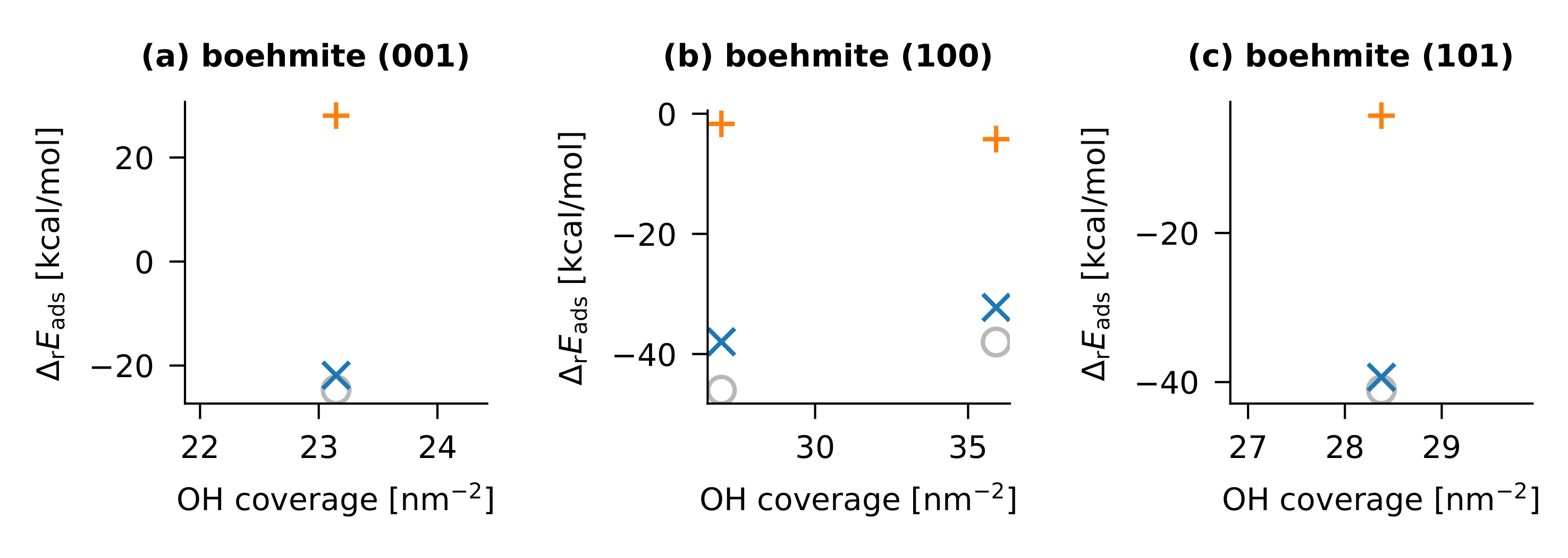}
    \caption{
        Adsorption energies, as defined in Eq.~(3) in the main text, on boehmite surfaces as a function of the \ce{OH} coverage, computed with different models: DFT (gray circle), Joshi (orange plus), this work (blue cross).
        All structures are derived from the training set.
    }
    \label{fig:adsorption-combined}
\end{figure}

%% file: training_set_overview_rows.tex
alpha\_bulk & \ce{Al_12 O_18} &  &  & 72 & 84 & 180 & 108 &  &  &  &  \\
boehm\_bulk & \ce{Al_32 O_64 H_32} & 32 & 32 & 192 & 128 & 480 & 224 & 64 & 64 & 32 &  \\
boehm\_surf-001\_00w & \ce{Al_48 O_96 H_48} & 48 & 32 & 272 & 160 & 640 & 288 & 96 & 64 & 32 &  \\
boehm\_surf-001\_08w & \ce{Al_48 O_112 H_80} & 80 & 40 &  & 160 &  &  & 160 & 80 & 40 &  \\
boehm\_surf-010\_00w & \ce{Al_72 O_144 H_72} & 72 & 48 & 432 & 288 & 1080 & 504 & 144 & 96 & 48 &  \\
boehm\_surf-100\_00w & \ce{Al_48 O_96 H_48} & 48 & 48 & 256 & 160 & 576 & 272 & 80 & 80 & 48 &  \\
boehm\_surf-100\_12w & \ce{Al_48 O_120 H_96} & 96 & 80 &  & 160 &  &  & 128 & 112 & 96 & 24 \\
boehm\_surf-100\_16w & \ce{Al_48 O_128 H_112} & 112 & 89 &  & 160 &  &  & 144 & 137 & 98 & 24 \\
boehm\_surf-101\_00w & \ce{Al_48 O_96 H_48} & 49 & 33 & 254 & 168 & 560 & 250 & 100 & 68 & 18 & 6 \\
boehm\_surf-101\_12w & \ce{Al_48 O_120 H_96} & 96 & 66 &  & 156 &  &  & 156 & 102 & 72 & 18 \\
gamma\_bulk & \ce{Al_16 O_24} &  &  & 88 & 88 & 204 & 120 &  &  &  &  \\
gamma\_edge-100-110\_00w & \ce{Al_96 O_144} &  &  & 481 & 424 & 1003 & 598 &  &  &  &  \\
gamma\_edge-100-110\_01w & \ce{Al_96 O_145 H_2} & 2 &  &  & 425 &  &  & 3 &  &  &  \\
gamma\_edge-100-110\_02w & \ce{Al_96 O_146 H_4} & 4 & 1 &  & 426 &  &  & 6 & 2 &  &  \\
gamma\_edge-100-110\_03w & \ce{Al_96 O_147 H_6} & 6 & 2 &  & 422 &  &  & 10 & 4 &  &  \\
gamma\_edge-100-110\_04w & \ce{Al_96 O_148 H_8} & 8 & 3 &  & 426 &  &  & 13 & 6 &  & 1 \\
gamma\_edge-100-110\_05w & \ce{Al_96 O_149 H_10} & 10 & 4 &  & 424 &  &  & 17 & 8 &  & 2 \\
gamma\_edge-100-110\_07w & \ce{Al_96 O_151 H_14} & 14 & 7 &  & 424 &  &  & 23 & 13 & 2 & 2 \\
gamma\_surf-100\_00w & \ce{Al_64 O_96} &  &  & 320 & 280 & 656 & 384 &  &  &  &  \\
gamma\_surf-100\_01w & \ce{Al_64 O_100 H_8} & 8 & 4 &  & 284 &  &  & 16 & 4 & 4 &  \\
gamma\_surf-100\_02w & \ce{Al_64 O_104 H_16} & 16 & 8 &  & 282 &  &  & 24 & 8 & 8 & 4 \\
gamma\_surf-100\_03w & \ce{Al_64 O_108 H_24} & 24 & 16 &  & 280 &  &  & 40 & 16 & 16 & 8 \\
gamma\_surf-100\_04w & \ce{Al_64 O_112 H_32} & 32 & 20 &  & 280 &  &  & 52 & 28 & 20 & 8 \\
gamma\_surf-110\_00w & \ce{Al_64 O_96} &  &  & 304 & 240 & 600 & 368 &  &  &  &  \\
gamma\_surf-110\_01w & \ce{Al_64 O_100 H_8} & 8 & 4 &  & 232 &  &  & 16 & 8 &  &  \\
gamma\_surf-110\_02w & \ce{Al_64 O_104 H_16} & 16 & 8 &  & 240 &  &  & 24 & 16 &  &  \\
gamma\_surf-110\_03w & \ce{Al_64 O_108 H_24} & 24 & 12 &  & 240 &  &  & 44 & 16 & 8 & 4 \\
gamma\_surf-110\_04w & \ce{Al_64 O_112 H_32} & 32 & 24 &  & 240 &  &  & 52 & 44 & 12 & 8 \\
gamma\_surf-110\_05w & \ce{Al_64 O_116 H_40} & 40 & 28 &  & 240 &  &  & 60 & 52 & 12 & 12 \\
gamma\_surf-110\_06w & \ce{Al_64 O_120 H_48} & 48 & 32 &  & 236 &  &  & 88 & 32 & 32 & 12 \\
gamma\_surf-111\_00w & \ce{Al_40 O_60} &  &  & 212 & 190 & 480 & 286 &  &  &  &  \\
gamma\_surf-111\_04w & \ce{Al_32 O_56 H_16} & 16 & 8 &  & 144 &  &  & 30 & 14 & 6 &  \\
gamma\_surf-111\_05w & \ce{Al_32 O_58 H_20} & 20 & 10 &  & 146 &  &  & 36 & 14 & 10 & 2 \\
gamma\_surf-111\_06w & \ce{Al_32 O_60 H_24} & 24 & 12 &  & 146 &  &  & 44 & 16 & 12 & 2 \\
monomer & \ce{Al O_4 H_5} & 5 &  & 4 &  & 6 &  & 5 &  &  &  \\
water & \ce{O H_2} &  &  &  &  &  &  &  &  &  &  \\
\hline total &  & 990 & 671 & 2887 & 8383 & 6465 & 3402 & 1675 & 1104 & 626 & 137

%% file: training_set_energies_rows.tex
\textbf{BSH} & $-1 / 16 \times$ & boehm\_surf-001\_00w & \ce{Al_48 O_96 H_48} & R & -24.7 \\
 & $-1 \times$ & water & \ce{O H_2} & J & 28.0 \\
 & $1 / 16 \times$ & boehm\_surf-001\_08w & \ce{Al_48 O_112 H_80} & B & -21.9 \\
\hline
\textbf{BSH} & $-1 / 24 \times$ & boehm\_surf-100\_00w & \ce{Al_48 O_96 H_48} & R & -46.0 \\
 & $-1 \times$ & water & \ce{O H_2} & J & -1.7 \\
 & $1 / 24 \times$ & boehm\_surf-100\_12w & \ce{Al_48 O_120 H_96} & B & -38.0 \\
\hline
\textbf{BSH} & $-1 / 32 \times$ & boehm\_surf-100\_00w & \ce{Al_48 O_96 H_48} & R & -38.0 \\
 & $-1 \times$ & water & \ce{O H_2} & J & -4.2 \\
 & $1 / 32 \times$ & boehm\_surf-100\_16w & \ce{Al_48 O_128 H_112} & B & -32.2 \\
\hline
\textbf{BSH} & $-1 / 8 \times$ & boehm\_surf-100\_12w & \ce{Al_48 O_120 H_96} & R & -14.0 \\
 & $-1 \times$ & water & \ce{O H_2} & J & -11.9 \\
 & $1 / 8 \times$ & boehm\_surf-100\_16w & \ce{Al_48 O_128 H_112} & B & -15.1 \\
\hline
\textbf{BSH} & $-1 / 24 \times$ & boehm\_surf-101\_00w & \ce{Al_48 O_96 H_48} & R & -41.0 \\
 & $-1 \times$ & water & \ce{O H_2} & J & -4.3 \\
 & $1 / 24 \times$ & boehm\_surf-101\_12w & \ce{Al_48 O_120 H_96} & B & -39.3 \\
\hline
\textbf{GEH} & $-1 \times$ & gamma\_edge-100-110\_00w & \ce{Al_96 O_144} & R & -104.1 \\
 & $-1 \times$ & water & \ce{O H_2} & J & -101.6 \\
 & $1 \times$ & gamma\_edge-100-110\_01w & \ce{Al_96 O_145 H_2} & B & -103.0 \\
\hline
\textbf{GEH} & $-1 / 2 \times$ & gamma\_edge-100-110\_00w & \ce{Al_96 O_144} & R & -81.2 \\
 & $-1 \times$ & water & \ce{O H_2} & J & -86.6 \\
 & $1 / 2 \times$ & gamma\_edge-100-110\_02w & \ce{Al_96 O_146 H_4} & B & -80.3 \\
\hline
\textbf{GEH} & $-1 \times$ & gamma\_edge-100-110\_01w & \ce{Al_96 O_145 H_2} & R & -58.3 \\
 & $-1 \times$ & water & \ce{O H_2} & J & -71.5 \\
 & $1 \times$ & gamma\_edge-100-110\_02w & \ce{Al_96 O_146 H_4} & B & -57.5 \\
\hline
\textbf{GEH} & $-1 / 3 \times$ & gamma\_edge-100-110\_00w & \ce{Al_96 O_144} & R & -72.1 \\
 & $-1 \times$ & water & \ce{O H_2} & J & -18.0 \\
 & $1 / 3 \times$ & gamma\_edge-100-110\_03w & \ce{Al_96 O_147 H_6} & B & -71.2 \\
\hline
\textbf{GEH} & $-1 / 2 \times$ & gamma\_edge-100-110\_01w & \ce{Al_96 O_145 H_2} & R & -56.1 \\
 & $-1 \times$ & water & \ce{O H_2} & J & 23.8 \\
 & $1 / 2 \times$ & gamma\_edge-100-110\_03w & \ce{Al_96 O_147 H_6} & B & -55.3 \\
\hline
\textbf{GEH} & $-1 \times$ & gamma\_edge-100-110\_02w & \ce{Al_96 O_146 H_4} & R & -53.9 \\
 & $-1 \times$ & water & \ce{O H_2} & J & 119.1 \\
 & $1 \times$ & gamma\_edge-100-110\_03w & \ce{Al_96 O_147 H_6} & B & -53.0 \\
\hline
\textbf{GEH} & $-1 / 4 \times$ & gamma\_edge-100-110\_00w & \ce{Al_96 O_144} & R & -67.5 \\
 & $-1 \times$ & water & \ce{O H_2} & J & -71.0 \\
 & $1 / 4 \times$ & gamma\_edge-100-110\_04w & \ce{Al_96 O_148 H_8} & B & -67.0 \\
\hline
\textbf{GEH} & $-1 / 3 \times$ & gamma\_edge-100-110\_01w & \ce{Al_96 O_145 H_2} & R & -55.3 \\
 & $-1 \times$ & water & \ce{O H_2} & J & -60.8 \\
 & $1 / 3 \times$ & gamma\_edge-100-110\_04w & \ce{Al_96 O_148 H_8} & B & -55.0 \\
\hline
\textbf{GEH} & $-1 / 2 \times$ & gamma\_edge-100-110\_02w & \ce{Al_96 O_146 H_4} & R & -53.9 \\
 & $-1 \times$ & water & \ce{O H_2} & J & -55.4 \\
 & $1 / 2 \times$ & gamma\_edge-100-110\_04w & \ce{Al_96 O_148 H_8} & B & -53.7 \\
\hline
\textbf{GEH} & $-1 \times$ & gamma\_edge-100-110\_03w & \ce{Al_96 O_147 H_6} & R & -53.8 \\
 & $-1 \times$ & water & \ce{O H_2} & J & -230.0 \\
 & $1 \times$ & gamma\_edge-100-110\_04w & \ce{Al_96 O_148 H_8} & B & -54.5 \\
\hline
\textbf{GEH} & $-1 / 5 \times$ & gamma\_edge-100-110\_00w & \ce{Al_96 O_144} & R & -63.3 \\
 & $-1 \times$ & water & \ce{O H_2} & J & -22.9 \\
 & $1 / 5 \times$ & gamma\_edge-100-110\_05w & \ce{Al_96 O_149 H_10} & B & -63.8 \\
\hline
\textbf{GEH} & $-1 / 4 \times$ & gamma\_edge-100-110\_01w & \ce{Al_96 O_145 H_2} & R & -53.1 \\
 & $-1 \times$ & water & \ce{O H_2} & J & -3.2 \\
 & $1 / 4 \times$ & gamma\_edge-100-110\_05w & \ce{Al_96 O_149 H_10} & B & -54.0 \\
\hline
\textbf{GEH} & $-1 / 3 \times$ & gamma\_edge-100-110\_02w & \ce{Al_96 O_146 H_4} & R & -51.4 \\
 & $-1 \times$ & water & \ce{O H_2} & J & 19.5 \\
 & $1 / 3 \times$ & gamma\_edge-100-110\_05w & \ce{Al_96 O_149 H_10} & B & -52.8 \\
\hline
\textbf{GEH} & $-1 / 2 \times$ & gamma\_edge-100-110\_03w & \ce{Al_96 O_147 H_6} & R & -50.2 \\
 & $-1 \times$ & water & \ce{O H_2} & J & -30.3 \\
 & $1 / 2 \times$ & gamma\_edge-100-110\_05w & \ce{Al_96 O_149 H_10} & B & -52.7 \\
\hline
\textbf{GEH} & $-1 \times$ & gamma\_edge-100-110\_04w & \ce{Al_96 O_148 H_8} & R & -46.6 \\
 & $-1 \times$ & water & \ce{O H_2} & J & 169.4 \\
 & $1 \times$ & gamma\_edge-100-110\_05w & \ce{Al_96 O_149 H_10} & B & -51.0 \\
\hline
\textbf{GEH} & $-1 / 7 \times$ & gamma\_edge-100-110\_00w & \ce{Al_96 O_144} & R & -55.8 \\
 & $-1 \times$ & water & \ce{O H_2} & J & -12.6 \\
 & $1 / 7 \times$ & gamma\_edge-100-110\_07w & \ce{Al_96 O_151 H_14} & B & -57.1 \\
\hline
\textbf{GEH} & $-1 / 6 \times$ & gamma\_edge-100-110\_01w & \ce{Al_96 O_145 H_2} & R & -47.8 \\
 & $-1 \times$ & water & \ce{O H_2} & J & 2.2 \\
 & $1 / 6 \times$ & gamma\_edge-100-110\_07w & \ce{Al_96 O_151 H_14} & B & -49.5 \\
\hline
\textbf{GEH} & $-1 / 5 \times$ & gamma\_edge-100-110\_02w & \ce{Al_96 O_146 H_4} & R & -45.7 \\
 & $-1 \times$ & water & \ce{O H_2} & J & 16.9 \\
 & $1 / 5 \times$ & gamma\_edge-100-110\_07w & \ce{Al_96 O_151 H_14} & B & -47.9 \\
\hline
\textbf{GEH} & $-1 / 4 \times$ & gamma\_edge-100-110\_03w & \ce{Al_96 O_147 H_6} & R & -43.6 \\
 & $-1 \times$ & water & \ce{O H_2} & J & -8.6 \\
 & $1 / 4 \times$ & gamma\_edge-100-110\_07w & \ce{Al_96 O_151 H_14} & B & -46.6 \\
\hline
\textbf{GEH} & $-1 / 3 \times$ & gamma\_edge-100-110\_04w & \ce{Al_96 O_148 H_8} & R & -40.2 \\
 & $-1 \times$ & water & \ce{O H_2} & J & 65.1 \\
 & $1 / 3 \times$ & gamma\_edge-100-110\_07w & \ce{Al_96 O_151 H_14} & B & -43.9 \\
\hline
\textbf{GEH} & $-1 / 2 \times$ & gamma\_edge-100-110\_05w & \ce{Al_96 O_149 H_10} & R & -37.0 \\
 & $-1 \times$ & water & \ce{O H_2} & J & 13.0 \\
 & $1 / 2 \times$ & gamma\_edge-100-110\_07w & \ce{Al_96 O_151 H_14} & B & -40.4 \\
\hline
\textbf{GSH} & $-1 / 4 \times$ & gamma\_surf-100\_00w & \ce{Al_64 O_96} & R & -24.0 \\
 & $-1 \times$ & water & \ce{O H_2} & J & -2.0 \\
 & $1 / 4 \times$ & gamma\_surf-100\_01w & \ce{Al_64 O_100 H_8} & B & -23.8 \\
\hline
\textbf{GSH} & $-1 / 8 \times$ & gamma\_surf-100\_00w & \ce{Al_64 O_96} & R & -25.0 \\
 & $-1 \times$ & water & \ce{O H_2} & J & -10.7 \\
 & $1 / 8 \times$ & gamma\_surf-100\_02w & \ce{Al_64 O_104 H_16} & B & -26.0 \\
\hline
\textbf{GSH} & $-1 / 4 \times$ & gamma\_surf-100\_01w & \ce{Al_64 O_100 H_8} & R & -26.0 \\
 & $-1 \times$ & water & \ce{O H_2} & J & -19.5 \\
 & $1 / 4 \times$ & gamma\_surf-100\_02w & \ce{Al_64 O_104 H_16} & B & -28.2 \\
\hline
\textbf{GSH} & $-1 / 12 \times$ & gamma\_surf-100\_00w & \ce{Al_64 O_96} & R & -24.3 \\
 & $-1 \times$ & water & \ce{O H_2} & J & -6.8 \\
 & $1 / 12 \times$ & gamma\_surf-100\_03w & \ce{Al_64 O_108 H_24} & B & -26.8 \\
\hline
\textbf{GSH} & $-1 / 8 \times$ & gamma\_surf-100\_01w & \ce{Al_64 O_100 H_8} & R & -24.4 \\
 & $-1 \times$ & water & \ce{O H_2} & J & -9.2 \\
 & $1 / 8 \times$ & gamma\_surf-100\_03w & \ce{Al_64 O_108 H_24} & B & -28.3 \\
\hline
\textbf{GSH} & $-1 / 4 \times$ & gamma\_surf-100\_02w & \ce{Al_64 O_104 H_16} & R & -22.8 \\
 & $-1 \times$ & water & \ce{O H_2} & J & 1.1 \\
 & $1 / 4 \times$ & gamma\_surf-100\_03w & \ce{Al_64 O_108 H_24} & B & -28.3 \\
\hline
\textbf{GSH} & $-1 / 16 \times$ & gamma\_surf-100\_00w & \ce{Al_64 O_96} & R & -21.7 \\
 & $-1 \times$ & water & \ce{O H_2} & J & -3.6 \\
 & $1 / 16 \times$ & gamma\_surf-100\_04w & \ce{Al_64 O_112 H_32} & B & -22.4 \\
\hline
\textbf{GSH} & $-1 / 12 \times$ & gamma\_surf-100\_01w & \ce{Al_64 O_100 H_8} & R & -21.0 \\
 & $-1 \times$ & water & \ce{O H_2} & J & -4.1 \\
 & $1 / 12 \times$ & gamma\_surf-100\_04w & \ce{Al_64 O_112 H_32} & B & -22.0 \\
\hline
\textbf{GSH} & $-1 / 8 \times$ & gamma\_surf-100\_02w & \ce{Al_64 O_104 H_16} & R & -18.5 \\
 & $-1 \times$ & water & \ce{O H_2} & J & 3.5 \\
 & $1 / 8 \times$ & gamma\_surf-100\_04w & \ce{Al_64 O_112 H_32} & B & -18.8 \\
\hline
\textbf{GSH} & $-1 / 4 \times$ & gamma\_surf-100\_03w & \ce{Al_64 O_108 H_24} & R & -14.1 \\
 & $-1 \times$ & water & \ce{O H_2} & J & 6.0 \\
 & $1 / 4 \times$ & gamma\_surf-100\_04w & \ce{Al_64 O_112 H_32} & B & -9.3 \\
\hline
\textbf{GSH} & $-1 / 4 \times$ & gamma\_surf-110\_00w & \ce{Al_64 O_96} & R & -92.4 \\
 & $-1 \times$ & water & \ce{O H_2} & J & 50.4 \\
 & $1 / 4 \times$ & gamma\_surf-110\_01w & \ce{Al_64 O_100 H_8} & B & -108.6 \\
\hline
\textbf{GSH} & $-1 / 8 \times$ & gamma\_surf-110\_00w & \ce{Al_64 O_96} & R & -50.2 \\
 & $-1 \times$ & water & \ce{O H_2} & J & 7.3 \\
 & $1 / 8 \times$ & gamma\_surf-110\_02w & \ce{Al_64 O_104 H_16} & B & -61.8 \\
\hline
\textbf{GSH} & $-1 / 4 \times$ & gamma\_surf-110\_01w & \ce{Al_64 O_100 H_8} & R & -8.0 \\
 & $-1 \times$ & water & \ce{O H_2} & J & -35.8 \\
 & $1 / 4 \times$ & gamma\_surf-110\_02w & \ce{Al_64 O_104 H_16} & B & -15.0 \\
\hline
\textbf{GSH} & $-1 / 12 \times$ & gamma\_surf-110\_00w & \ce{Al_64 O_96} & R & -42.5 \\
 & $-1 \times$ & water & \ce{O H_2} & J & 42.3 \\
 & $1 / 12 \times$ & gamma\_surf-110\_03w & \ce{Al_64 O_108 H_24} & B & -50.1 \\
\hline
\textbf{GSH} & $-1 / 8 \times$ & gamma\_surf-110\_01w & \ce{Al_64 O_100 H_8} & R & -17.6 \\
 & $-1 \times$ & water & \ce{O H_2} & J & 38.2 \\
 & $1 / 8 \times$ & gamma\_surf-110\_03w & \ce{Al_64 O_108 H_24} & B & -20.8 \\
\hline
\textbf{GSH} & $-1 / 4 \times$ & gamma\_surf-110\_02w & \ce{Al_64 O_104 H_16} & R & -27.2 \\
 & $-1 \times$ & water & \ce{O H_2} & J & 112.3 \\
 & $1 / 4 \times$ & gamma\_surf-110\_03w & \ce{Al_64 O_108 H_24} & B & -26.7 \\
\hline
\textbf{GSH} & $-1 / 16 \times$ & gamma\_surf-110\_00w & \ce{Al_64 O_96} & R & -39.6 \\
 & $-1 \times$ & water & \ce{O H_2} & J & 28.9 \\
 & $1 / 16 \times$ & gamma\_surf-110\_04w & \ce{Al_64 O_112 H_32} & B & -44.9 \\
\hline
\textbf{GSH} & $-1 / 12 \times$ & gamma\_surf-110\_01w & \ce{Al_64 O_100 H_8} & R & -22.1 \\
 & $-1 \times$ & water & \ce{O H_2} & J & 21.7 \\
 & $1 / 12 \times$ & gamma\_surf-110\_04w & \ce{Al_64 O_112 H_32} & B & -23.7 \\
\hline
\textbf{GSH} & $-1 / 8 \times$ & gamma\_surf-110\_02w & \ce{Al_64 O_104 H_16} & R & -29.1 \\
 & $-1 \times$ & water & \ce{O H_2} & J & 50.5 \\
 & $1 / 8 \times$ & gamma\_surf-110\_04w & \ce{Al_64 O_112 H_32} & B & -28.1 \\
\hline
\textbf{GSH} & $-1 / 4 \times$ & gamma\_surf-110\_03w & \ce{Al_64 O_108 H_24} & R & -31.0 \\
 & $-1 \times$ & water & \ce{O H_2} & J & -11.3 \\
 & $1 / 4 \times$ & gamma\_surf-110\_04w & \ce{Al_64 O_112 H_32} & B & -29.5 \\
\hline
\textbf{GSH} & $-1 / 20 \times$ & gamma\_surf-110\_00w & \ce{Al_64 O_96} & R & -35.2 \\
 & $-1 \times$ & water & \ce{O H_2} & J & 16.0 \\
 & $1 / 20 \times$ & gamma\_surf-110\_05w & \ce{Al_64 O_116 H_40} & B & -39.4 \\
\hline
\textbf{GSH} & $-1 / 16 \times$ & gamma\_surf-110\_01w & \ce{Al_64 O_100 H_8} & R & -20.9 \\
 & $-1 \times$ & water & \ce{O H_2} & J & 7.4 \\
 & $1 / 16 \times$ & gamma\_surf-110\_05w & \ce{Al_64 O_116 H_40} & B & -22.1 \\
\hline
\textbf{GSH} & $-1 / 12 \times$ & gamma\_surf-110\_02w & \ce{Al_64 O_104 H_16} & R & -25.1 \\
 & $-1 \times$ & water & \ce{O H_2} & J & 21.9 \\
 & $1 / 12 \times$ & gamma\_surf-110\_05w & \ce{Al_64 O_116 H_40} & B & -24.5 \\
\hline
\textbf{GSH} & $-1 / 8 \times$ & gamma\_surf-110\_03w & \ce{Al_64 O_108 H_24} & R & -24.1 \\
 & $-1 \times$ & water & \ce{O H_2} & J & -23.4 \\
 & $1 / 8 \times$ & gamma\_surf-110\_05w & \ce{Al_64 O_116 H_40} & B & -23.4 \\
\hline
\textbf{GSH} & $-1 / 4 \times$ & gamma\_surf-110\_04w & \ce{Al_64 O_112 H_32} & R & -17.3 \\
 & $-1 \times$ & water & \ce{O H_2} & J & -35.4 \\
 & $1 / 4 \times$ & gamma\_surf-110\_05w & \ce{Al_64 O_116 H_40} & B & -17.2 \\
\hline
\textbf{GSH} & $-1 / 24 \times$ & gamma\_surf-110\_00w & \ce{Al_64 O_96} & R & -33.5 \\
 & $-1 \times$ & water & \ce{O H_2} & J & 21.5 \\
 & $1 / 24 \times$ & gamma\_surf-110\_06w & \ce{Al_64 O_120 H_48} & B & -37.5 \\
\hline
\textbf{GSH} & $-1 / 20 \times$ & gamma\_surf-110\_01w & \ce{Al_64 O_100 H_8} & R & -21.7 \\
 & $-1 \times$ & water & \ce{O H_2} & J & 15.8 \\
 & $1 / 20 \times$ & gamma\_surf-110\_06w & \ce{Al_64 O_120 H_48} & B & -23.3 \\
\hline
\textbf{GSH} & $-1 / 16 \times$ & gamma\_surf-110\_02w & \ce{Al_64 O_104 H_16} & R & -25.1 \\
 & $-1 \times$ & water & \ce{O H_2} & J & 28.7 \\
 & $1 / 16 \times$ & gamma\_surf-110\_06w & \ce{Al_64 O_120 H_48} & B & -25.3 \\
\hline
\textbf{GSH} & $-1 / 12 \times$ & gamma\_surf-110\_03w & \ce{Al_64 O_108 H_24} & R & -24.4 \\
 & $-1 \times$ & water & \ce{O H_2} & J & 0.8 \\
 & $1 / 12 \times$ & gamma\_surf-110\_06w & \ce{Al_64 O_120 H_48} & B & -24.9 \\
\hline
\textbf{GSH} & $-1 / 8 \times$ & gamma\_surf-110\_04w & \ce{Al_64 O_112 H_32} & R & -21.1 \\
 & $-1 \times$ & water & \ce{O H_2} & J & 6.8 \\
 & $1 / 8 \times$ & gamma\_surf-110\_06w & \ce{Al_64 O_120 H_48} & B & -22.6 \\
\hline
\textbf{GSH} & $-1 / 4 \times$ & gamma\_surf-110\_05w & \ce{Al_64 O_116 H_40} & R & -24.9 \\
 & $-1 \times$ & water & \ce{O H_2} & J & 49.1 \\
 & $1 / 4 \times$ & gamma\_surf-110\_06w & \ce{Al_64 O_120 H_48} & B & -27.9 \\
\hline
\textbf{GSH} & $-1 / 10 \times$ & gamma\_surf-111\_00w & \ce{Al_40 O_60} & R & -42.1 \\
 & $-1 \times$ & water & \ce{O H_2} & J & 7.2 \\
 & $1 / 8 \times$ & gamma\_surf-111\_04w & \ce{Al_32 O_56 H_16} & B & -31.1 \\
\hline
\textbf{GSH} & $-2 / 25 \times$ & gamma\_surf-111\_00w & \ce{Al_40 O_60} & R & -45.8 \\
 & $-1 \times$ & water & \ce{O H_2} & J & 5.6 \\
 & $1 / 10 \times$ & gamma\_surf-111\_05w & \ce{Al_32 O_58 H_20} & B & -32.8 \\
\hline
\textbf{GSH} & $-1 / 2 \times$ & gamma\_surf-111\_04w & \ce{Al_32 O_56 H_16} & R & -60.5 \\
 & $-1 \times$ & water & \ce{O H_2} & J & -0.7 \\
 & $1 / 2 \times$ & gamma\_surf-111\_05w & \ce{Al_32 O_58 H_20} & B & -39.8 \\
\hline
\textbf{GSH} & $-1 / 15 \times$ & gamma\_surf-111\_00w & \ce{Al_40 O_60} & R & -43.1 \\
 & $-1 \times$ & water & \ce{O H_2} & J & 13.5 \\
 & $1 / 12 \times$ & gamma\_surf-111\_06w & \ce{Al_32 O_60 H_24} & B & -32.5 \\
\hline
\textbf{GSH} & $-1 / 4 \times$ & gamma\_surf-111\_04w & \ce{Al_32 O_56 H_16} & R & -45.0 \\
 & $-1 \times$ & water & \ce{O H_2} & J & 26.2 \\
 & $1 / 4 \times$ & gamma\_surf-111\_06w & \ce{Al_32 O_60 H_24} & B & -35.3 \\
\hline
\textbf{GSH} & $-1 / 2 \times$ & gamma\_surf-111\_05w & \ce{Al_32 O_58 H_20} & R & -29.4 \\
 & $-1 \times$ & water & \ce{O H_2} & J & 53.2 \\
 & $1 / 2 \times$ & gamma\_surf-111\_06w & \ce{Al_32 O_60 H_24} & B & -30.9 \\
\hline
\textbf{SUR} & $-1 / 32 \times$ & boehm\_bulk & \ce{Al_32 O_64 H_32} & R & 11.2 \\
 & $1 / 48 \times$ & boehm\_surf-001\_00w & \ce{Al_48 O_96 H_48} & J & 5.5 \\
 & & & & B & 9.0 \\
\hline
\textbf{SUR} & $-1 / 32 \times$ & boehm\_bulk & \ce{Al_32 O_64 H_32} & R & 2.9 \\
 & $1 / 72 \times$ & boehm\_surf-010\_00w & \ce{Al_72 O_144 H_72} & J & 2.1 \\
 & & & & B & 3.1 \\
\hline
\textbf{SUR} & $-1 / 32 \times$ & boehm\_bulk & \ce{Al_32 O_64 H_32} & R & 23.0 \\
 & $1 / 48 \times$ & boehm\_surf-100\_00w & \ce{Al_48 O_96 H_48} & J & -0.4 \\
 & & & & B & 19.1 \\
\hline
\textbf{SUR} & $-1 / 32 \times$ & boehm\_bulk & \ce{Al_32 O_64 H_32} & R & 18.6 \\
 & $1 / 48 \times$ & boehm\_surf-101\_00w & \ce{Al_48 O_96 H_48} & J & 7.5 \\
 & & & & B & 16.4 \\
\hline
\textbf{SUR} & $-1 / 16 \times$ & gamma\_bulk & \ce{Al_16 O_24} & R & 13.5 \\
 & $1 / 96 \times$ & gamma\_edge-100-110\_00w & \ce{Al_96 O_144} & J & 7.0 \\
 & & & & B & 14.0 \\
\hline
\textbf{SUR} & $-1 / 16 \times$ & gamma\_bulk & \ce{Al_16 O_24} & R & 10.4 \\
 & $1 / 64 \times$ & gamma\_surf-100\_00w & \ce{Al_64 O_96} & J & 6.6 \\
 & & & & B & 13.0 \\
\hline
\textbf{SUR} & $-1 / 16 \times$ & gamma\_bulk & \ce{Al_16 O_24} & R & 21.2 \\
 & $1 / 64 \times$ & gamma\_surf-110\_00w & \ce{Al_64 O_96} & J & 7.4 \\
 & & & & B & 26.6 \\
\hline
\textbf{SUR} & $-1 / 16 \times$ & gamma\_bulk & \ce{Al_16 O_24} & R & 13.5 \\
 & $1 / 40 \times$ & gamma\_surf-111\_00w & \ce{Al_40 O_60} & J & 5.6 \\
 & & & & B & 11.1 \\
\hline
\textbf{FOR} & $-1 \times$ & monomer & \ce{Al O_4 H_5} & R & -27.8 \\
 & $1 / 12 \times$ & alpha\_bulk & \ce{Al_12 O_18} & J & -44.9 \\
 & $5 / 2 \times$ & water & \ce{O H_2} & B & -31.9 \\
\hline
\textbf{FOR} & $-1 \times$ & monomer & \ce{Al O_4 H_5} & R & -38.8 \\
 & $1 / 32 \times$ & boehm\_bulk & \ce{Al_32 O_64 H_32} & J & -38.6 \\
 & $2 \times$ & water & \ce{O H_2} & B & -36.5 \\
\hline
\textbf{FOR} & $-1 \times$ & monomer & \ce{Al O_4 H_5} & R & -27.6 \\
 & $1 / 48 \times$ & boehm\_surf-001\_00w & \ce{Al_48 O_96 H_48} & J & -33.1 \\
 & $2 \times$ & water & \ce{O H_2} & B & -27.4 \\
\hline
\textbf{FOR} & $-1 \times$ & monomer & \ce{Al O_4 H_5} & R & -35.9 \\
 & $1 / 72 \times$ & boehm\_surf-010\_00w & \ce{Al_72 O_144 H_72} & J & -36.5 \\
 & $2 \times$ & water & \ce{O H_2} & B & -33.4 \\
\hline
\textbf{FOR} & $-1 \times$ & monomer & \ce{Al O_4 H_5} & R & -15.8 \\
 & $1 / 48 \times$ & boehm\_surf-100\_00w & \ce{Al_48 O_96 H_48} & J & -39.0 \\
 & $2 \times$ & water & \ce{O H_2} & B & -17.4 \\
\hline
\textbf{FOR} & $-1 \times$ & monomer & \ce{Al O_4 H_5} & R & -20.2 \\
 & $1 / 48 \times$ & boehm\_surf-101\_00w & \ce{Al_48 O_96 H_48} & J & -31.1 \\
 & $2 \times$ & water & \ce{O H_2} & B & -20.1 \\
\hline
\textbf{FOR} & $-1 \times$ & monomer & \ce{Al O_4 H_5} & R & -25.3 \\
 & $1 / 16 \times$ & gamma\_bulk & \ce{Al_16 O_24} & J & -48.6 \\
 & $5 / 2 \times$ & water & \ce{O H_2} & B & -26.4 \\
\hline
\textbf{FOR} & $-1 \times$ & monomer & \ce{Al O_4 H_5} & R & -11.7 \\
 & $1 / 96 \times$ & gamma\_edge-100-110\_00w & \ce{Al_96 O_144} & J & -41.6 \\
 & $5 / 2 \times$ & water & \ce{O H_2} & B & -12.4 \\
\hline
\textbf{FOR} & $-1 \times$ & monomer & \ce{Al O_4 H_5} & R & -14.8 \\
 & $1 / 64 \times$ & gamma\_surf-100\_00w & \ce{Al_64 O_96} & J & -42.0 \\
 & $5 / 2 \times$ & water & \ce{O H_2} & B & -13.4 \\
\hline
\textbf{FOR} & $-1 \times$ & monomer & \ce{Al O_4 H_5} & R & -4.1 \\
 & $1 / 64 \times$ & gamma\_surf-110\_00w & \ce{Al_64 O_96} & J & -41.2 \\
 & $5 / 2 \times$ & water & \ce{O H_2} & B & 0.2 \\
\hline
\textbf{FOR} & $-1 \times$ & monomer & \ce{Al O_4 H_5} & R & -11.7 \\
 & $1 / 40 \times$ & gamma\_surf-111\_00w & \ce{Al_40 O_60} & J & -43.0 \\
 & $5 / 2 \times$ & water & \ce{O H_2} & B & -15.3 \\
\hline

%% file: parameters_active.tex
p\_val3 & 1 & Al & ATM & 1.5000 & 1.2000 & 3.0000 & 2.9993 \\
p\_val5 & 1 & Al & ATM & 2.5791 & 2.0633 & 3.0949 & 2.4980 \\
D\_e\^{}sigma & \si{\kilo\cal\per\mol} & Al-H & BND & 92.8579 & 0.8579 & 122.7844 & 77.1779 \\
p\_be1 & 1 & Al-H & BND & -0.6528 & -0.7834 & 1.0000 & -0.7823 \\
p\_ovun1 & 1 & Al-H & BND & 0.1551 & 0.0100 & 0.5000 & 0.3089 \\
p\_be2 & 1 & Al-H & BND & 10.0663 & 0.2281 & 13.0000 & 2.0985 \\
p\_bo1 & 1 & Al-H & BND & -0.0842 & -0.3320 & -0.0674 & -0.1398 \\
p\_bo2 & 1 & Al-H & BND & 7.1758 & 5.0015 & 15.0000 & 11.7930 \\
D\_e\^{}sigma & \si{\kilo\cal\per\mol} & Al-O & BND & 182.0654 & 118.9203 & 232.7313 & 166.4329 \\
p\_be1 & 1 & Al-O & BND & -0.0920 & -1.0000 & -0.0736 & -0.0963 \\
p\_ovun1 & 1 & Al-O & BND & 0.1688 & 0.0100 & 0.4562 & 0.1457 \\
p\_be2 & 1 & Al-O & BND & 0.0010 & 0.0008 & 1.5477 & 0.7025 \\
p\_bo1 & 1 & Al-O & BND & -0.1959 & -0.2351 & -0.0740 & -0.2053 \\
p\_bo2 & 1 & Al-O & BND & 6.0894 & 4.6533 & 7.3073 & 7.1635 \\
D\_e\^{}sigma & \si{\kilo\cal\per\mol} & Al-Al & BND & 34.0777 & 27.2622 & 65.7742 & 31.9434 \\
r\_0\^{}sigma & \si{\angstrom} & Al-H & OFD & 1.7276 & 1.3821 & 2.0731 & 1.4893 \\
r\_0\^{}sigma & \si{\angstrom} & Al-O & OFD & 1.5646 & 1.2517 & 1.8775 & 1.6172 \\
p\_val1 & 1 & Al-H-O & ANG & 4.2750 & 3.4200 & 20.0000 & 19.7283 \\
p\_val2 & 1 & Al-H-O & ANG & 1.0250 & 0.8200 & 4.8339 & 3.9208 \\
p\_val4 & 1 & Al-H-O & ANG & 1.4750 & 1.0100 & 1.7700 & 1.0931 \\
Theta\_0,0 & \si{\deg} & Al-O-H & ANG & 88.6163 & 64.6197 & 106.3396 & 66.0975 \\
p\_val1 & 1 & Al-O-H & ANG & 10.1310 & 4.2037 & 19.7491 & 16.7556 \\
p\_val2 & 1 & Al-O-H & ANG & 1.6896 & 1.3517 & 10.0000 & 2.4332 \\
p\_val4 & 1 & Al-O-H & ANG & 1.0000 & 0.8000 & 3.0000 & 1.4778 \\
Theta\_0,0 & \si{\deg} & Al-O-Al & ANG & 13.8580 & 5.2474 & 64.5513 & 64.5486 \\
p\_val1 & 1 & Al-O-Al & ANG & 12.3669 & 9.8935 & 40.0000 & 12.8201 \\
p\_val2 & 1 & Al-O-Al & ANG & 4.4355 & 0.5527 & 9.9945 & 7.0734 \\
p\_val4 & 1 & Al-O-Al & ANG & 1.1908 & 0.9526 & 3.0000 & 2.9786 \\
Theta\_0,0 & \si{\deg} & H-Al-O & ANG & 41.8108 & 0.0000 & 64.8437 & 9.7233 \\
p\_val1 & 1 & H-Al-O & ANG & 17.3800 & 5.4547 & 30.9495 & 23.4162 \\
p\_val2 & 1 & H-Al-O & ANG & 2.6618 & 0.9702 & 3.1942 & 3.0390 \\
p\_val4 & 1 & H-Al-O & ANG & 1.0100 & 0.8080 & 3.0000 & 1.0218 \\
Theta\_0,0 & \si{\deg} & O-Al-O & ANG & 55.4358 & 43.7395 & 84.7469 & 61.5592 \\
p\_val1 & 1 & O-Al-O & ANG & 22.1089 & 7.3926 & 40.0000 & 24.2337 \\
p\_val2 & 1 & O-Al-O & ANG & 3.7402 & 1.2450 & 4.4882 & 1.2622 \\
p\_val4 & 1 & O-Al-O & ANG & 2.2064 & 1.0123 & 3.0000 & 2.9975

%% file: validation_set_overview_rows.tex
gamma\_bulk & \ce{Al_32 O_48} &  &  & 176 & 176 & 408 & 240 &  &  &  &  \\
gamma\_surf-001\_00w & \ce{Al_64 O_96} &  &  & 336 & 316 & 736 & 432 &  &  &  &  \\
gamma\_surf-001\_01w & \ce{Al_64 O_98 H_4} & 4 & 2 &  & 316 &  &  & 8 & 2 & 2 &  \\
gamma\_surf-001\_02w & \ce{Al_64 O_100 H_8} & 8 & 4 &  & 318 &  &  & 12 & 4 & 4 & 2 \\
gamma\_surf-001\_03w & \ce{Al_64 O_102 H_12} & 12 & 8 &  & 316 &  &  & 20 & 8 & 8 & 4 \\
gamma\_surf-001\_04w & \ce{Al_64 O_104 H_16} & 16 & 10 &  & 316 &  &  & 26 & 14 & 10 & 4 \\
gamma\_surf-110b\_from\_boehm\_00w & \ce{Al_80 O_120} &  &  & 414 & 416 & 902 & 532 &  &  &  &  \\
gamma\_surf-110b\_from\_boehm\_02w & \ce{Al_80 O_124 H_8} & 8 &  &  & 436 &  &  & 16 &  &  &  \\
gamma\_surf-110b\_from\_boehm\_03w & \ce{Al_80 O_126 H_12} & 12 &  &  & 436 &  &  & 22 &  &  &  \\
gamma\_surf-110b\_from\_boehm\_04w & \ce{Al_80 O_128 H_16} & 16 & 8 &  & 436 &  &  & 24 & 8 & 8 &  \\
gamma\_surf-110b\_from\_bulk\_00w & \ce{Al_64 O_96} &  &  & 336 & 324 & 744 & 440 &  &  &  &  \\
gamma\_surf-110b\_from\_bulk\_02w & \ce{Al_64 O_98 H_4} & 4 & 2 &  & 324 &  &  & 6 & 4 &  &  \\
gamma\_surf-110b\_from\_bulk\_04w & \ce{Al_64 O_100 H_8} & 8 & 4 &  & 324 &  &  & 12 & 4 & 4 &  \\
gamma\_surf-110b\_from\_bulk\_06w & \ce{Al_64 O_102 H_12} & 12 & 6 &  & 324 &  &  & 16 & 10 & 4 &  \\
gamma\_surf-110b\_from\_bulk\_08w & \ce{Al_64 O_104 H_16} & 16 & 10 &  & 324 &  &  & 20 & 18 & 6 & 2 \\
gamma\_surf-110l\_A1\_00w & \ce{Al_48 O_72} &  &  & 250 & 238 & 552 & 324 &  &  &  &  \\
gamma\_surf-110l\_A1\_01w & \ce{Al_48 O_74 H_4} & 4 & 2 &  & 234 &  &  & 8 & 4 &  &  \\
gamma\_surf-110l\_A1\_02w & \ce{Al_48 O_76 H_8} & 8 & 4 &  & 240 &  &  & 14 & 6 & 2 &  \\
gamma\_surf-110l\_A1\_03w & \ce{Al_48 O_78 H_12} & 12 & 4 &  & 240 &  &  & 22 & 6 & 4 &  \\
gamma\_surf-110l\_A1\_04w & \ce{Al_48 O_80 H_16} & 16 & 12 &  & 236 &  &  & 26 & 22 & 6 & 4 \\
gamma\_surf-110l\_A1\_05w & \ce{Al_48 O_82 H_20} & 20 & 14 &  & 236 &  &  & 30 & 26 & 8 & 4 \\
gamma\_surf-110l\_A1\_06w & \ce{Al_48 O_84 H_24} & 24 & 17 &  & 236 &  &  & 30 & 37 & 4 & 9 \\
gamma\_surf-110l\_A2\_00w & \ce{Al_56 O_84} &  &  & 296 & 276 & 660 & 392 &  &  &  &  \\
gamma\_surf-110l\_A2\_01w & \ce{Al_56 O_86 H_4} & 4 & 2 &  & 274 &  &  & 8 & 4 &  &  \\
gamma\_surf-110l\_A2\_02w & \ce{Al_56 O_88 H_8} & 8 & 4 &  & 282 &  &  & 14 & 8 &  & 2 \\
gamma\_surf-110l\_A2\_03w & \ce{Al_56 O_90 H_12} & 12 & 7 &  & 284 &  &  & 20 & 11 & 5 & 1 \\
gamma\_surf-110l\_A2\_04w & \ce{Al_56 O_92 H_16} & 16 & 10 &  & 284 &  &  & 24 & 18 & 8 &  \\
gamma\_surf-110l\_A2\_05w & \ce{Al_56 O_94 H_20} & 20 & 14 &  & 282 &  &  & 26 & 30 & 4 & 6 \\
gamma\_surf-110l\_A2\_06w & \ce{Al_56 O_96 H_24} & 24 & 18 &  & 282 &  &  & 30 & 38 & 6 & 8 \\
gamma\_surf-110l\_L1\_04w & \ce{Al_48 O_80 H_16} & 16 & 6 &  & 240 &  &  & 32 & 12 & 4 &  \\
gamma\_surf-110l\_L2\_00w & \ce{Al_56 O_84} &  &  & 294 & 286 & 654 & 384 &  &  &  &  \\
gamma\_surf-110l\_L2\_01w & \ce{Al_56 O_86 H_4} & 4 & 2 &  & 276 &  &  & 8 & 4 &  &  \\
gamma\_surf-110l\_L2\_02w & \ce{Al_56 O_88 H_8} & 8 & 4 &  & 282 &  &  & 14 & 8 &  & 2 \\
gamma\_surf-110l\_L2\_03w & \ce{Al_56 O_90 H_12} & 12 & 4 &  & 282 &  &  & 22 & 10 & 2 &  \\
gamma\_surf-110l\_L2\_04w & \ce{Al_56 O_92 H_16} & 16 & 6 &  & 282 &  &  & 32 & 16 & 2 &  \\
gamma\_surf-111\_D1\_03w & \ce{Al_32 O_54 H_12} & 12 & 5 &  & 144 &  &  & 22 & 6 & 4 & 2 \\
gamma\_surf-111\_D1\_04w & \ce{Al_32 O_56 H_16} & 16 & 5 &  & 146 &  &  & 30 & 6 & 5 & 2 \\
gamma\_surf-111\_D1\_05w & \ce{Al_32 O_58 H_20} & 20 & 10 &  & 146 &  &  & 36 & 14 & 10 & 2 \\
gamma\_surf-111\_D1\_06w & \ce{Al_32 O_60 H_24} & 24 & 14 &  & 146 &  &  & 44 & 17 & 14 & 5 \\
gamma\_surf-111\_D2\_03w & \ce{Al_32 O_54 H_12} & 12 & 8 &  & 144 &  &  & 22 & 12 & 4 & 2 \\
gamma\_surf-111\_D2\_04w & \ce{Al_32 O_56 H_16} & 16 & 9 &  & 146 &  &  & 30 & 14 & 7 & 2 \\
gamma\_surf-111\_D2\_05w & \ce{Al_32 O_58 H_20} & 20 & 12 &  & 146 &  &  & 36 & 15 & 12 & 6 \\
gamma\_surf-111\_D2\_06w & \ce{Al_32 O_60 H_24} & 24 & 15 &  & 146 &  &  & 44 & 18 & 15 & 6 \\
gamma\_surf-111\_P1\_1\_05w & \ce{Al_40 O_70 H_20} & 20 & 3 &  & 184 &  &  & 43 & 6 & 3 &  \\
gamma\_surf-111\_P1\_1\_06w & \ce{Al_40 O_72 H_24} & 24 & 4 &  & 178 &  &  & 54 & 8 & 4 &  \\
gamma\_surf-111\_P1\_2\_03w & \ce{Al_40 O_66 H_12} & 12 & 2 &  & 178 &  &  & 24 & 4 & 2 &  \\
gamma\_surf-111\_P1\_2\_04w & \ce{Al_40 O_68 H_16} & 16 & 4 &  & 178 &  &  & 34 & 8 & 4 &  \\
gamma\_surf-111\_P1\_2\_05w & \ce{Al_40 O_70 H_20} & 20 & 5 &  & 178 &  &  & 44 & 10 & 5 &  \\
gamma\_surf-111\_P1\_2\_06w & \ce{Al_40 O_72 H_24} & 24 & 2 &  & 182 &  &  & 54 & 4 & 2 &  \\
gamma\_surf-111\_P2\_1\_05w & \ce{Al_40 O_70 H_20} & 20 & 14 &  & 196 &  &  & 27 & 15 & 13 & 4 \\
gamma\_surf-111\_P2\_1\_06w & \ce{Al_40 O_72 H_24} & 24 & 16 &  & 200 &  &  & 38 & 16 & 16 & 4 \\
gamma\_surf-111\_P2\_2\_04w & \ce{Al_40 O_68 H_16} & 16 & 10 &  & 194 &  &  & 26 & 16 & 6 &  \\
gamma\_surf-111\_P2\_2\_05w & \ce{Al_40 O_70 H_20} & 20 & 14 &  & 190 &  &  & 32 & 18 & 12 & 3 \\
gamma\_surf-111\_P2\_2\_06w & \ce{Al_40 O_72 H_24} & 24 & 15 &  & 192 &  &  & 35 & 15 & 15 & 2 \\
water & \ce{O H_2} &  &  &  &  &  &  &  &  &  &  \\
\hline total &  & 704 & 351 & 2102 & 13588 & 4656 & 2744 & 1217 & 554 & 254 & 88

%% file: validation_set_energies_rows.tex
\textbf{GSH} & $-1 / 2 \times$ & gamma\_surf-001\_00w & \ce{Al_64 O_96} & R & -22.5 \\
 & $-1 \times$ & water & \ce{O H_2} & J & -14.3 \\
 & $1 / 2 \times$ & gamma\_surf-001\_01w & \ce{Al_64 O_98 H_4} & B & -31.6 \\
\hline
\textbf{GSH} & $-1 / 4 \times$ & gamma\_surf-001\_00w & \ce{Al_64 O_96} & R & -24.4 \\
 & $-1 \times$ & water & \ce{O H_2} & J & -5.0 \\
 & $1 / 4 \times$ & gamma\_surf-001\_02w & \ce{Al_64 O_100 H_8} & B & -27.5 \\
\hline
\textbf{GSH} & $-1 / 2 \times$ & gamma\_surf-001\_01w & \ce{Al_64 O_98 H_4} & R & -26.3 \\
 & $-1 \times$ & water & \ce{O H_2} & J & 4.2 \\
 & $1 / 2 \times$ & gamma\_surf-001\_02w & \ce{Al_64 O_100 H_8} & B & -23.4 \\
\hline
\textbf{GSH} & $-1 / 6 \times$ & gamma\_surf-001\_00w & \ce{Al_64 O_96} & R & -23.5 \\
 & $-1 \times$ & water & \ce{O H_2} & J & 0.2 \\
 & $1 / 6 \times$ & gamma\_surf-001\_03w & \ce{Al_64 O_102 H_12} & B & -28.4 \\
\hline
\textbf{GSH} & $-1 / 4 \times$ & gamma\_surf-001\_01w & \ce{Al_64 O_98 H_4} & R & -24.0 \\
 & $-1 \times$ & water & \ce{O H_2} & J & 7.4 \\
 & $1 / 4 \times$ & gamma\_surf-001\_03w & \ce{Al_64 O_102 H_12} & B & -26.8 \\
\hline
\textbf{GSH} & $-1 / 2 \times$ & gamma\_surf-001\_02w & \ce{Al_64 O_100 H_8} & R & -21.6 \\
 & $-1 \times$ & water & \ce{O H_2} & J & 10.6 \\
 & $1 / 2 \times$ & gamma\_surf-001\_03w & \ce{Al_64 O_102 H_12} & B & -30.1 \\
\hline
\textbf{GSH} & $-1 / 8 \times$ & gamma\_surf-001\_00w & \ce{Al_64 O_96} & R & -20.8 \\
 & $-1 \times$ & water & \ce{O H_2} & J & -12.3 \\
 & $1 / 8 \times$ & gamma\_surf-001\_04w & \ce{Al_64 O_104 H_16} & B & -24.1 \\
\hline
\textbf{GSH} & $-1 / 6 \times$ & gamma\_surf-001\_01w & \ce{Al_64 O_98 H_4} & R & -20.2 \\
 & $-1 \times$ & water & \ce{O H_2} & J & -11.6 \\
 & $1 / 6 \times$ & gamma\_surf-001\_04w & \ce{Al_64 O_104 H_16} & B & -21.6 \\
\hline
\textbf{GSH} & $-1 / 4 \times$ & gamma\_surf-001\_02w & \ce{Al_64 O_100 H_8} & R & -17.1 \\
 & $-1 \times$ & water & \ce{O H_2} & J & -19.5 \\
 & $1 / 4 \times$ & gamma\_surf-001\_04w & \ce{Al_64 O_104 H_16} & B & -20.7 \\
\hline
\textbf{GSH} & $-1 / 2 \times$ & gamma\_surf-001\_03w & \ce{Al_64 O_102 H_12} & R & -12.6 \\
 & $-1 \times$ & water & \ce{O H_2} & J & -49.7 \\
 & $1 / 2 \times$ & gamma\_surf-001\_04w & \ce{Al_64 O_104 H_16} & B & -11.3 \\
\hline
\textbf{GSH} & $-1 / 4 \times$ & gamma\_surf-110b\_from\_boehm\_00w & \ce{Al_80 O_120} & R & -60.7 \\
 & $-1 \times$ & water & \ce{O H_2} & J & 5.5 \\
 & $1 / 4 \times$ & gamma\_surf-110b\_from\_boehm\_02w & \ce{Al_80 O_124 H_8} & B & -50.1 \\
\hline
\textbf{GSH} & $-1 / 6 \times$ & gamma\_surf-110b\_from\_boehm\_00w & \ce{Al_80 O_120} & R & -58.9 \\
 & $-1 \times$ & water & \ce{O H_2} & J & 10.3 \\
 & $1 / 6 \times$ & gamma\_surf-110b\_from\_boehm\_03w & \ce{Al_80 O_126 H_12} & B & -56.7 \\
\hline
\textbf{GSH} & $-1 / 2 \times$ & gamma\_surf-110b\_from\_boehm\_02w & \ce{Al_80 O_124 H_8} & R & -55.1 \\
 & $-1 \times$ & water & \ce{O H_2} & J & 19.7 \\
 & $1 / 2 \times$ & gamma\_surf-110b\_from\_boehm\_03w & \ce{Al_80 O_126 H_12} & B & -70.0 \\
\hline
\textbf{GSH} & $-1 / 8 \times$ & gamma\_surf-110b\_from\_boehm\_00w & \ce{Al_80 O_120} & R & -58.2 \\
 & $-1 \times$ & water & \ce{O H_2} & J & -7.7 \\
 & $1 / 8 \times$ & gamma\_surf-110b\_from\_boehm\_04w & \ce{Al_80 O_128 H_16} & B & -53.6 \\
\hline
\textbf{GSH} & $-1 / 4 \times$ & gamma\_surf-110b\_from\_boehm\_02w & \ce{Al_80 O_124 H_8} & R & -55.7 \\
 & $-1 \times$ & water & \ce{O H_2} & J & -21.0 \\
 & $1 / 4 \times$ & gamma\_surf-110b\_from\_boehm\_04w & \ce{Al_80 O_128 H_16} & B & -57.2 \\
\hline
\textbf{GSH} & $-1 / 2 \times$ & gamma\_surf-110b\_from\_boehm\_03w & \ce{Al_80 O_126 H_12} & R & -56.3 \\
 & $-1 \times$ & water & \ce{O H_2} & J & -61.7 \\
 & $1 / 2 \times$ & gamma\_surf-110b\_from\_boehm\_04w & \ce{Al_80 O_128 H_16} & B & -44.4 \\
\hline
\textbf{GSH} & $-1 / 2 \times$ & gamma\_surf-110b\_from\_bulk\_00w & \ce{Al_64 O_96} & R & -43.6 \\
 & $-1 \times$ & water & \ce{O H_2} & J & -26.1 \\
 & $1 / 2 \times$ & gamma\_surf-110b\_from\_bulk\_02w & \ce{Al_64 O_98 H_4} & B & -49.6 \\
\hline
\textbf{GSH} & $-1 / 4 \times$ & gamma\_surf-110b\_from\_bulk\_00w & \ce{Al_64 O_96} & R & -45.8 \\
 & $-1 \times$ & water & \ce{O H_2} & J & -43.6 \\
 & $1 / 4 \times$ & gamma\_surf-110b\_from\_bulk\_04w & \ce{Al_64 O_100 H_8} & B & -50.9 \\
\hline
\textbf{GSH} & $-1 / 2 \times$ & gamma\_surf-110b\_from\_bulk\_02w & \ce{Al_64 O_98 H_4} & R & -48.0 \\
 & $-1 \times$ & water & \ce{O H_2} & J & -61.0 \\
 & $1 / 2 \times$ & gamma\_surf-110b\_from\_bulk\_04w & \ce{Al_64 O_100 H_8} & B & -52.1 \\
\hline
\textbf{GSH} & $-1 / 6 \times$ & gamma\_surf-110b\_from\_bulk\_00w & \ce{Al_64 O_96} & R & -39.1 \\
 & $-1 \times$ & water & \ce{O H_2} & J & -32.6 \\
 & $1 / 6 \times$ & gamma\_surf-110b\_from\_bulk\_06w & \ce{Al_64 O_102 H_12} & B & -41.3 \\
\hline
\textbf{GSH} & $-1 / 4 \times$ & gamma\_surf-110b\_from\_bulk\_02w & \ce{Al_64 O_98 H_4} & R & -36.8 \\
 & $-1 \times$ & water & \ce{O H_2} & J & -35.9 \\
 & $1 / 4 \times$ & gamma\_surf-110b\_from\_bulk\_06w & \ce{Al_64 O_102 H_12} & B & -37.2 \\
\hline
\textbf{GSH} & $-1 / 2 \times$ & gamma\_surf-110b\_from\_bulk\_04w & \ce{Al_64 O_100 H_8} & R & -25.6 \\
 & $-1 \times$ & water & \ce{O H_2} & J & -10.7 \\
 & $1 / 2 \times$ & gamma\_surf-110b\_from\_bulk\_06w & \ce{Al_64 O_102 H_12} & B & -22.2 \\
\hline
\textbf{GSH} & $-1 / 8 \times$ & gamma\_surf-110b\_from\_bulk\_00w & \ce{Al_64 O_96} & R & -36.4 \\
 & $-1 \times$ & water & \ce{O H_2} & J & -26.9 \\
 & $1 / 8 \times$ & gamma\_surf-110b\_from\_bulk\_08w & \ce{Al_64 O_104 H_16} & B & -38.1 \\
\hline
\textbf{GSH} & $-1 / 6 \times$ & gamma\_surf-110b\_from\_bulk\_02w & \ce{Al_64 O_98 H_4} & R & -34.0 \\
 & $-1 \times$ & water & \ce{O H_2} & J & -27.1 \\
 & $1 / 6 \times$ & gamma\_surf-110b\_from\_bulk\_08w & \ce{Al_64 O_104 H_16} & B & -34.3 \\
\hline
\textbf{GSH} & $-1 / 4 \times$ & gamma\_surf-110b\_from\_bulk\_04w & \ce{Al_64 O_100 H_8} & R & -26.9 \\
 & $-1 \times$ & water & \ce{O H_2} & J & -10.2 \\
 & $1 / 4 \times$ & gamma\_surf-110b\_from\_bulk\_08w & \ce{Al_64 O_104 H_16} & B & -25.4 \\
\hline
\textbf{GSH} & $-1 / 2 \times$ & gamma\_surf-110b\_from\_bulk\_06w & \ce{Al_64 O_102 H_12} & R & -28.2 \\
 & $-1 \times$ & water & \ce{O H_2} & J & -9.7 \\
 & $1 / 2 \times$ & gamma\_surf-110b\_from\_bulk\_08w & \ce{Al_64 O_104 H_16} & B & -28.6 \\
\hline
\textbf{GSH} & $-1 / 2 \times$ & gamma\_surf-110l\_A1\_00w & \ce{Al_48 O_72} & R & -86.5 \\
 & $-1 \times$ & water & \ce{O H_2} & J & 61.0 \\
 & $1 / 2 \times$ & gamma\_surf-110l\_A1\_01w & \ce{Al_48 O_74 H_4} & B & -83.3 \\
\hline
\textbf{GSH} & $-1 / 4 \times$ & gamma\_surf-110l\_A1\_00w & \ce{Al_48 O_72} & R & -58.5 \\
 & $-1 \times$ & water & \ce{O H_2} & J & 4.1 \\
 & $1 / 4 \times$ & gamma\_surf-110l\_A1\_02w & \ce{Al_48 O_76 H_8} & B & -65.9 \\
\hline
\textbf{GSH} & $-1 / 2 \times$ & gamma\_surf-110l\_A1\_01w & \ce{Al_48 O_74 H_4} & R & -30.5 \\
 & $-1 \times$ & water & \ce{O H_2} & J & -52.8 \\
 & $1 / 2 \times$ & gamma\_surf-110l\_A1\_02w & \ce{Al_48 O_76 H_8} & B & -48.5 \\
\hline
\textbf{GSH} & $-1 / 6 \times$ & gamma\_surf-110l\_A1\_00w & \ce{Al_48 O_72} & R & -65.3 \\
 & $-1 \times$ & water & \ce{O H_2} & J & 21.2 \\
 & $1 / 6 \times$ & gamma\_surf-110l\_A1\_03w & \ce{Al_48 O_78 H_12} & B & -67.2 \\
\hline
\textbf{GSH} & $-1 / 4 \times$ & gamma\_surf-110l\_A1\_01w & \ce{Al_48 O_74 H_4} & R & -54.6 \\
 & $-1 \times$ & water & \ce{O H_2} & J & 1.4 \\
 & $1 / 4 \times$ & gamma\_surf-110l\_A1\_03w & \ce{Al_48 O_78 H_12} & B & -59.2 \\
\hline
\textbf{GSH} & $-1 / 2 \times$ & gamma\_surf-110l\_A1\_02w & \ce{Al_48 O_76 H_8} & R & -78.8 \\
 & $-1 \times$ & water & \ce{O H_2} & J & 55.5 \\
 & $1 / 2 \times$ & gamma\_surf-110l\_A1\_03w & \ce{Al_48 O_78 H_12} & B & -69.8 \\
\hline
\textbf{GSH} & $-1 / 8 \times$ & gamma\_surf-110l\_A1\_00w & \ce{Al_48 O_72} & R & -57.0 \\
 & $-1 \times$ & water & \ce{O H_2} & J & 15.1 \\
 & $1 / 8 \times$ & gamma\_surf-110l\_A1\_04w & \ce{Al_48 O_80 H_16} & B & -55.8 \\
\hline
\textbf{GSH} & $-1 / 6 \times$ & gamma\_surf-110l\_A1\_01w & \ce{Al_48 O_74 H_4} & R & -47.1 \\
 & $-1 \times$ & water & \ce{O H_2} & J & -0.2 \\
 & $1 / 6 \times$ & gamma\_surf-110l\_A1\_04w & \ce{Al_48 O_80 H_16} & B & -46.6 \\
\hline
\textbf{GSH} & $-1 / 4 \times$ & gamma\_surf-110l\_A1\_02w & \ce{Al_48 O_76 H_8} & R & -55.4 \\
 & $-1 \times$ & water & \ce{O H_2} & J & 26.1 \\
 & $1 / 4 \times$ & gamma\_surf-110l\_A1\_04w & \ce{Al_48 O_80 H_16} & B & -45.6 \\
\hline
\textbf{GSH} & $-1 / 2 \times$ & gamma\_surf-110l\_A1\_03w & \ce{Al_48 O_78 H_12} & R & -32.0 \\
 & $-1 \times$ & water & \ce{O H_2} & J & -3.4 \\
 & $1 / 2 \times$ & gamma\_surf-110l\_A1\_04w & \ce{Al_48 O_80 H_16} & B & -21.4 \\
\hline
\textbf{GSH} & $-1 / 10 \times$ & gamma\_surf-110l\_A1\_00w & \ce{Al_48 O_72} & R & -49.8 \\
 & $-1 \times$ & water & \ce{O H_2} & J & 5.5 \\
 & $1 / 10 \times$ & gamma\_surf-110l\_A1\_05w & \ce{Al_48 O_82 H_20} & B & -48.5 \\
\hline
\textbf{GSH} & $-1 / 8 \times$ & gamma\_surf-110l\_A1\_01w & \ce{Al_48 O_74 H_4} & R & -40.7 \\
 & $-1 \times$ & water & \ce{O H_2} & J & -8.4 \\
 & $1 / 8 \times$ & gamma\_surf-110l\_A1\_05w & \ce{Al_48 O_82 H_20} & B & -39.8 \\
\hline
\textbf{GSH} & $-1 / 6 \times$ & gamma\_surf-110l\_A1\_02w & \ce{Al_48 O_76 H_8} & R & -44.1 \\
 & $-1 \times$ & water & \ce{O H_2} & J & 6.4 \\
 & $1 / 6 \times$ & gamma\_surf-110l\_A1\_05w & \ce{Al_48 O_82 H_20} & B & -36.9 \\
\hline
\textbf{GSH} & $-1 / 4 \times$ & gamma\_surf-110l\_A1\_03w & \ce{Al_48 O_78 H_12} & R & -26.7 \\
 & $-1 \times$ & water & \ce{O H_2} & J & -18.1 \\
 & $1 / 4 \times$ & gamma\_surf-110l\_A1\_05w & \ce{Al_48 O_82 H_20} & B & -20.4 \\
\hline
\textbf{GSH} & $-1 / 2 \times$ & gamma\_surf-110l\_A1\_04w & \ce{Al_48 O_80 H_16} & R & -21.4 \\
 & $-1 \times$ & water & \ce{O H_2} & J & -32.8 \\
 & $1 / 2 \times$ & gamma\_surf-110l\_A1\_05w & \ce{Al_48 O_82 H_20} & B & -19.4 \\
\hline
\textbf{GSH} & $-1 / 12 \times$ & gamma\_surf-110l\_A1\_00w & \ce{Al_48 O_72} & R & -43.6 \\
 & $-1 \times$ & water & \ce{O H_2} & J & -1.2 \\
 & $1 / 12 \times$ & gamma\_surf-110l\_A1\_06w & \ce{Al_48 O_84 H_24} & B & -42.5 \\
\hline
\textbf{GSH} & $-1 / 10 \times$ & gamma\_surf-110l\_A1\_01w & \ce{Al_48 O_74 H_4} & R & -35.0 \\
 & $-1 \times$ & water & \ce{O H_2} & J & -13.6 \\
 & $1 / 10 \times$ & gamma\_surf-110l\_A1\_06w & \ce{Al_48 O_84 H_24} & B & -34.4 \\
\hline
\textbf{GSH} & $-1 / 8 \times$ & gamma\_surf-110l\_A1\_02w & \ce{Al_48 O_76 H_8} & R & -36.2 \\
 & $-1 \times$ & water & \ce{O H_2} & J & -3.8 \\
 & $1 / 8 \times$ & gamma\_surf-110l\_A1\_06w & \ce{Al_48 O_84 H_24} & B & -30.8 \\
\hline
\textbf{GSH} & $-1 / 6 \times$ & gamma\_surf-110l\_A1\_03w & \ce{Al_48 O_78 H_12} & R & -22.0 \\
 & $-1 \times$ & water & \ce{O H_2} & J & -23.6 \\
 & $1 / 6 \times$ & gamma\_surf-110l\_A1\_06w & \ce{Al_48 O_84 H_24} & B & -17.8 \\
\hline
\textbf{GSH} & $-1 / 4 \times$ & gamma\_surf-110l\_A1\_04w & \ce{Al_48 O_80 H_16} & R & -17.0 \\
 & $-1 \times$ & water & \ce{O H_2} & J & -33.7 \\
 & $1 / 4 \times$ & gamma\_surf-110l\_A1\_06w & \ce{Al_48 O_84 H_24} & B & -16.0 \\
\hline
\textbf{GSH} & $-1 / 2 \times$ & gamma\_surf-110l\_A1\_05w & \ce{Al_48 O_82 H_20} & R & -12.5 \\
 & $-1 \times$ & water & \ce{O H_2} & J & -34.6 \\
 & $1 / 2 \times$ & gamma\_surf-110l\_A1\_06w & \ce{Al_48 O_84 H_24} & B & -12.7 \\
\hline
\textbf{GSH} & $-1 / 2 \times$ & gamma\_surf-110l\_A2\_00w & \ce{Al_56 O_84} & R & -82.6 \\
 & $-1 \times$ & water & \ce{O H_2} & J & -40.3 \\
 & $1 / 2 \times$ & gamma\_surf-110l\_A2\_01w & \ce{Al_56 O_86 H_4} & B & -89.7 \\
\hline
\textbf{GSH} & $-1 / 4 \times$ & gamma\_surf-110l\_A2\_00w & \ce{Al_56 O_84} & R & -66.4 \\
 & $-1 \times$ & water & \ce{O H_2} & J & -41.8 \\
 & $1 / 4 \times$ & gamma\_surf-110l\_A2\_02w & \ce{Al_56 O_88 H_8} & B & -75.6 \\
\hline
\textbf{GSH} & $-1 / 2 \times$ & gamma\_surf-110l\_A2\_01w & \ce{Al_56 O_86 H_4} & R & -50.2 \\
 & $-1 \times$ & water & \ce{O H_2} & J & -43.2 \\
 & $1 / 2 \times$ & gamma\_surf-110l\_A2\_02w & \ce{Al_56 O_88 H_8} & B & -61.5 \\
\hline
\textbf{GSH} & $-1 / 6 \times$ & gamma\_surf-110l\_A2\_00w & \ce{Al_56 O_84} & R & -60.3 \\
 & $-1 \times$ & water & \ce{O H_2} & J & -35.4 \\
 & $1 / 6 \times$ & gamma\_surf-110l\_A2\_03w & \ce{Al_56 O_90 H_12} & B & -65.2 \\
\hline
\textbf{GSH} & $-1 / 4 \times$ & gamma\_surf-110l\_A2\_01w & \ce{Al_56 O_86 H_4} & R & -49.1 \\
 & $-1 \times$ & water & \ce{O H_2} & J & -32.9 \\
 & $1 / 4 \times$ & gamma\_surf-110l\_A2\_03w & \ce{Al_56 O_90 H_12} & B & -52.9 \\
\hline
\textbf{GSH} & $-1 / 2 \times$ & gamma\_surf-110l\_A2\_02w & \ce{Al_56 O_88 H_8} & R & -48.1 \\
 & $-1 \times$ & water & \ce{O H_2} & J & -22.5 \\
 & $1 / 2 \times$ & gamma\_surf-110l\_A2\_03w & \ce{Al_56 O_90 H_12} & B & -44.3 \\
\hline
\textbf{GSH} & $-1 / 8 \times$ & gamma\_surf-110l\_A2\_00w & \ce{Al_56 O_84} & R & -52.7 \\
 & $-1 \times$ & water & \ce{O H_2} & J & -10.4 \\
 & $1 / 8 \times$ & gamma\_surf-110l\_A2\_04w & \ce{Al_56 O_92 H_16} & B & -56.2 \\
\hline
\textbf{GSH} & $-1 / 6 \times$ & gamma\_surf-110l\_A2\_01w & \ce{Al_56 O_86 H_4} & R & -42.7 \\
 & $-1 \times$ & water & \ce{O H_2} & J & -0.4 \\
 & $1 / 6 \times$ & gamma\_surf-110l\_A2\_04w & \ce{Al_56 O_92 H_16} & B & -45.1 \\
\hline
\textbf{GSH} & $-1 / 4 \times$ & gamma\_surf-110l\_A2\_02w & \ce{Al_56 O_88 H_8} & R & -38.9 \\
 & $-1 \times$ & water & \ce{O H_2} & J & 21.0 \\
 & $1 / 4 \times$ & gamma\_surf-110l\_A2\_04w & \ce{Al_56 O_92 H_16} & B & -36.8 \\
\hline
\textbf{GSH} & $-1 / 2 \times$ & gamma\_surf-110l\_A2\_03w & \ce{Al_56 O_90 H_12} & R & -29.8 \\
 & $-1 \times$ & water & \ce{O H_2} & J & 64.5 \\
 & $1 / 2 \times$ & gamma\_surf-110l\_A2\_04w & \ce{Al_56 O_92 H_16} & B & -29.3 \\
\hline
\textbf{GSH} & $-1 / 10 \times$ & gamma\_surf-110l\_A2\_00w & \ce{Al_56 O_84} & R & -46.1 \\
 & $-1 \times$ & water & \ce{O H_2} & J & -5.3 \\
 & $1 / 10 \times$ & gamma\_surf-110l\_A2\_05w & \ce{Al_56 O_94 H_20} & B & -50.1 \\
\hline
\textbf{GSH} & $-1 / 8 \times$ & gamma\_surf-110l\_A2\_01w & \ce{Al_56 O_86 H_4} & R & -37.0 \\
 & $-1 \times$ & water & \ce{O H_2} & J & 3.5 \\
 & $1 / 8 \times$ & gamma\_surf-110l\_A2\_05w & \ce{Al_56 O_94 H_20} & B & -40.2 \\
\hline
\textbf{GSH} & $-1 / 6 \times$ & gamma\_surf-110l\_A2\_02w & \ce{Al_56 O_88 H_8} & R & -32.6 \\
 & $-1 \times$ & water & \ce{O H_2} & J & 19.0 \\
 & $1 / 6 \times$ & gamma\_surf-110l\_A2\_05w & \ce{Al_56 O_94 H_20} & B & -33.0 \\
\hline
\textbf{GSH} & $-1 / 4 \times$ & gamma\_surf-110l\_A2\_03w & \ce{Al_56 O_90 H_12} & R & -24.9 \\
 & $-1 \times$ & water & \ce{O H_2} & J & 39.8 \\
 & $1 / 4 \times$ & gamma\_surf-110l\_A2\_05w & \ce{Al_56 O_94 H_20} & B & -27.4 \\
\hline
\textbf{GSH} & $-1 / 2 \times$ & gamma\_surf-110l\_A2\_04w & \ce{Al_56 O_92 H_16} & R & -20.0 \\
 & $-1 \times$ & water & \ce{O H_2} & J & 15.2 \\
 & $1 / 2 \times$ & gamma\_surf-110l\_A2\_05w & \ce{Al_56 O_94 H_20} & B & -25.5 \\
\hline
\textbf{GSH} & $-1 / 12 \times$ & gamma\_surf-110l\_A2\_00w & \ce{Al_56 O_84} & R & -42.0 \\
 & $-1 \times$ & water & \ce{O H_2} & J & -17.4 \\
 & $1 / 12 \times$ & gamma\_surf-110l\_A2\_06w & \ce{Al_56 O_96 H_24} & B & -47.1 \\
\hline
\textbf{GSH} & $-1 / 10 \times$ & gamma\_surf-110l\_A2\_01w & \ce{Al_56 O_86 H_4} & R & -33.9 \\
 & $-1 \times$ & water & \ce{O H_2} & J & -12.8 \\
 & $1 / 10 \times$ & gamma\_surf-110l\_A2\_06w & \ce{Al_56 O_96 H_24} & B & -38.5 \\
\hline
\textbf{GSH} & $-1 / 8 \times$ & gamma\_surf-110l\_A2\_02w & \ce{Al_56 O_88 H_8} & R & -29.8 \\
 & $-1 \times$ & water & \ce{O H_2} & J & -5.1 \\
 & $1 / 8 \times$ & gamma\_surf-110l\_A2\_06w & \ce{Al_56 O_96 H_24} & B & -32.8 \\
\hline
\textbf{GSH} & $-1 / 6 \times$ & gamma\_surf-110l\_A2\_03w & \ce{Al_56 O_90 H_12} & R & -23.7 \\
 & $-1 \times$ & water & \ce{O H_2} & J & 0.6 \\
 & $1 / 6 \times$ & gamma\_surf-110l\_A2\_06w & \ce{Al_56 O_96 H_24} & B & -29.0 \\
\hline
\textbf{GSH} & $-1 / 4 \times$ & gamma\_surf-110l\_A2\_04w & \ce{Al_56 O_92 H_16} & R & -20.7 \\
 & $-1 \times$ & water & \ce{O H_2} & J & -31.3 \\
 & $1 / 4 \times$ & gamma\_surf-110l\_A2\_06w & \ce{Al_56 O_96 H_24} & B & -28.8 \\
\hline
\textbf{GSH} & $-1 / 2 \times$ & gamma\_surf-110l\_A2\_05w & \ce{Al_56 O_94 H_20} & R & -21.4 \\
 & $-1 \times$ & water & \ce{O H_2} & J & -77.7 \\
 & $1 / 2 \times$ & gamma\_surf-110l\_A2\_06w & \ce{Al_56 O_96 H_24} & B & -32.1 \\
\hline
\textbf{GSH} & $-1 / 2 \times$ & gamma\_surf-110l\_L2\_00w & \ce{Al_56 O_84} & R & -60.8 \\
 & $-1 \times$ & water & \ce{O H_2} & J & -15.8 \\
 & $1 / 2 \times$ & gamma\_surf-110l\_L2\_01w & \ce{Al_56 O_86 H_4} & B & -58.3 \\
\hline
\textbf{GSH} & $-1 / 4 \times$ & gamma\_surf-110l\_L2\_00w & \ce{Al_56 O_84} & R & -44.6 \\
 & $-1 \times$ & water & \ce{O H_2} & J & 10.5 \\
 & $1 / 4 \times$ & gamma\_surf-110l\_L2\_02w & \ce{Al_56 O_88 H_8} & B & -51.9 \\
\hline
\textbf{GSH} & $-1 / 2 \times$ & gamma\_surf-110l\_L2\_01w & \ce{Al_56 O_86 H_4} & R & -28.4 \\
 & $-1 \times$ & water & \ce{O H_2} & J & 36.8 \\
 & $1 / 2 \times$ & gamma\_surf-110l\_L2\_02w & \ce{Al_56 O_88 H_8} & B & -45.5 \\
\hline
\textbf{GSH} & $-1 / 6 \times$ & gamma\_surf-110l\_L2\_00w & \ce{Al_56 O_84} & R & -50.5 \\
 & $-1 \times$ & water & \ce{O H_2} & J & 25.8 \\
 & $1 / 6 \times$ & gamma\_surf-110l\_L2\_03w & \ce{Al_56 O_90 H_12} & B & -53.2 \\
\hline
\textbf{GSH} & $-1 / 4 \times$ & gamma\_surf-110l\_L2\_01w & \ce{Al_56 O_86 H_4} & R & -45.3 \\
 & $-1 \times$ & water & \ce{O H_2} & J & 46.6 \\
 & $1 / 4 \times$ & gamma\_surf-110l\_L2\_03w & \ce{Al_56 O_90 H_12} & B & -50.6 \\
\hline
\textbf{GSH} & $-1 / 2 \times$ & gamma\_surf-110l\_L2\_02w & \ce{Al_56 O_88 H_8} & R & -62.3 \\
 & $-1 \times$ & water & \ce{O H_2} & J & 56.5 \\
 & $1 / 2 \times$ & gamma\_surf-110l\_L2\_03w & \ce{Al_56 O_90 H_12} & B & -55.7 \\
\hline
\textbf{GSH} & $-1 / 8 \times$ & gamma\_surf-110l\_L2\_00w & \ce{Al_56 O_84} & R & -53.0 \\
 & $-1 \times$ & water & \ce{O H_2} & J & 24.9 \\
 & $1 / 8 \times$ & gamma\_surf-110l\_L2\_04w & \ce{Al_56 O_92 H_16} & B & -49.3 \\
\hline
\textbf{GSH} & $-1 / 6 \times$ & gamma\_surf-110l\_L2\_01w & \ce{Al_56 O_86 H_4} & R & -50.4 \\
 & $-1 \times$ & water & \ce{O H_2} & J & 38.5 \\
 & $1 / 6 \times$ & gamma\_surf-110l\_L2\_04w & \ce{Al_56 O_92 H_16} & B & -46.2 \\
\hline
\textbf{GSH} & $-1 / 4 \times$ & gamma\_surf-110l\_L2\_02w & \ce{Al_56 O_88 H_8} & R & -61.5 \\
 & $-1 \times$ & water & \ce{O H_2} & J & 39.3 \\
 & $1 / 4 \times$ & gamma\_surf-110l\_L2\_04w & \ce{Al_56 O_92 H_16} & B & -46.6 \\
\hline
\textbf{GSH} & $-1 / 2 \times$ & gamma\_surf-110l\_L2\_03w & \ce{Al_56 O_90 H_12} & R & -60.7 \\
 & $-1 \times$ & water & \ce{O H_2} & J & 22.1 \\
 & $1 / 2 \times$ & gamma\_surf-110l\_L2\_04w & \ce{Al_56 O_92 H_16} & B & -37.5 \\
\hline
\textbf{GSH} & $-1 / 2 \times$ & gamma\_surf-111\_D1\_03w & \ce{Al_32 O_54 H_12} & R & -60.7 \\
 & $-1 \times$ & water & \ce{O H_2} & J & -5.4 \\
 & $1 / 2 \times$ & gamma\_surf-111\_D1\_04w & \ce{Al_32 O_56 H_16} & B & -49.0 \\
\hline
\textbf{GSH} & $-1 / 4 \times$ & gamma\_surf-111\_D1\_03w & \ce{Al_32 O_54 H_12} & R & -49.5 \\
 & $-1 \times$ & water & \ce{O H_2} & J & 1.5 \\
 & $1 / 4 \times$ & gamma\_surf-111\_D1\_05w & \ce{Al_32 O_58 H_20} & B & -38.6 \\
\hline
\textbf{GSH} & $-1 / 2 \times$ & gamma\_surf-111\_D1\_04w & \ce{Al_32 O_56 H_16} & R & -38.4 \\
 & $-1 \times$ & water & \ce{O H_2} & J & 8.5 \\
 & $1 / 2 \times$ & gamma\_surf-111\_D1\_05w & \ce{Al_32 O_58 H_20} & B & -28.3 \\
\hline
\textbf{GSH} & $-1 / 6 \times$ & gamma\_surf-111\_D1\_03w & \ce{Al_32 O_54 H_12} & R & -42.9 \\
 & $-1 \times$ & water & \ce{O H_2} & J & 15.4 \\
 & $1 / 6 \times$ & gamma\_surf-111\_D1\_06w & \ce{Al_32 O_60 H_24} & B & -34.8 \\
\hline
\textbf{GSH} & $-1 / 4 \times$ & gamma\_surf-111\_D1\_04w & \ce{Al_32 O_56 H_16} & R & -34.0 \\
 & $-1 \times$ & water & \ce{O H_2} & J & 25.8 \\
 & $1 / 4 \times$ & gamma\_surf-111\_D1\_06w & \ce{Al_32 O_60 H_24} & B & -27.7 \\
\hline
\textbf{GSH} & $-1 / 2 \times$ & gamma\_surf-111\_D1\_05w & \ce{Al_32 O_58 H_20} & R & -29.6 \\
 & $-1 \times$ & water & \ce{O H_2} & J & 43.2 \\
 & $1 / 2 \times$ & gamma\_surf-111\_D1\_06w & \ce{Al_32 O_60 H_24} & B & -27.1 \\
\hline
\textbf{GSH} & $-1 / 2 \times$ & gamma\_surf-111\_D2\_03w & \ce{Al_32 O_54 H_12} & R & -50.3 \\
 & $-1 \times$ & water & \ce{O H_2} & J & 46.1 \\
 & $1 / 2 \times$ & gamma\_surf-111\_D2\_04w & \ce{Al_32 O_56 H_16} & B & -27.4 \\
\hline
\textbf{GSH} & $-1 / 4 \times$ & gamma\_surf-111\_D2\_03w & \ce{Al_32 O_54 H_12} & R & -44.4 \\
 & $-1 \times$ & water & \ce{O H_2} & J & 39.7 \\
 & $1 / 4 \times$ & gamma\_surf-111\_D2\_05w & \ce{Al_32 O_58 H_20} & B & -34.8 \\
\hline
\textbf{GSH} & $-1 / 2 \times$ & gamma\_surf-111\_D2\_04w & \ce{Al_32 O_56 H_16} & R & -38.5 \\
 & $-1 \times$ & water & \ce{O H_2} & J & 33.3 \\
 & $1 / 2 \times$ & gamma\_surf-111\_D2\_05w & \ce{Al_32 O_58 H_20} & B & -42.1 \\
\hline
\textbf{GSH} & $-1 / 6 \times$ & gamma\_surf-111\_D2\_03w & \ce{Al_32 O_54 H_12} & R & -39.4 \\
 & $-1 \times$ & water & \ce{O H_2} & J & 24.6 \\
 & $1 / 6 \times$ & gamma\_surf-111\_D2\_06w & \ce{Al_32 O_60 H_24} & B & -34.0 \\
\hline
\textbf{GSH} & $-1 / 4 \times$ & gamma\_surf-111\_D2\_04w & \ce{Al_32 O_56 H_16} & R & -33.9 \\
 & $-1 \times$ & water & \ce{O H_2} & J & 13.8 \\
 & $1 / 4 \times$ & gamma\_surf-111\_D2\_06w & \ce{Al_32 O_60 H_24} & B & -37.3 \\
\hline
\textbf{GSH} & $-1 / 2 \times$ & gamma\_surf-111\_D2\_05w & \ce{Al_32 O_58 H_20} & R & -29.4 \\
 & $-1 \times$ & water & \ce{O H_2} & J & -5.6 \\
 & $1 / 2 \times$ & gamma\_surf-111\_D2\_06w & \ce{Al_32 O_60 H_24} & B & -32.4 \\
\hline
\textbf{GSH} & $-1 / 2 \times$ & gamma\_surf-111\_P1\_1\_05w & \ce{Al_40 O_70 H_20} & R & -21.2 \\
 & $-1 \times$ & water & \ce{O H_2} & J & 1.7 \\
 & $1 / 2 \times$ & gamma\_surf-111\_P1\_1\_06w & \ce{Al_40 O_72 H_24} & B & -22.9 \\
\hline
\textbf{GSH} & $-1 / 2 \times$ & gamma\_surf-111\_P1\_2\_03w & \ce{Al_40 O_66 H_12} & R & -51.2 \\
 & $-1 \times$ & water & \ce{O H_2} & J & 47.5 \\
 & $1 / 2 \times$ & gamma\_surf-111\_P1\_2\_04w & \ce{Al_40 O_68 H_16} & B & -36.4 \\
\hline
\textbf{GSH} & $-1 / 4 \times$ & gamma\_surf-111\_P1\_2\_03w & \ce{Al_40 O_66 H_12} & R & -41.3 \\
 & $-1 \times$ & water & \ce{O H_2} & J & 34.9 \\
 & $1 / 4 \times$ & gamma\_surf-111\_P1\_2\_05w & \ce{Al_40 O_70 H_20} & B & -35.9 \\
\hline
\textbf{GSH} & $-1 / 2 \times$ & gamma\_surf-111\_P1\_2\_04w & \ce{Al_40 O_68 H_16} & R & -31.4 \\
 & $-1 \times$ & water & \ce{O H_2} & J & 22.3 \\
 & $1 / 2 \times$ & gamma\_surf-111\_P1\_2\_05w & \ce{Al_40 O_70 H_20} & B & -35.4 \\
\hline
\textbf{GSH} & $-1 / 6 \times$ & gamma\_surf-111\_P1\_2\_03w & \ce{Al_40 O_66 H_12} & R & -35.0 \\
 & $-1 \times$ & water & \ce{O H_2} & J & 24.3 \\
 & $1 / 6 \times$ & gamma\_surf-111\_P1\_2\_06w & \ce{Al_40 O_72 H_24} & B & -33.6 \\
\hline
\textbf{GSH} & $-1 / 4 \times$ & gamma\_surf-111\_P1\_2\_04w & \ce{Al_40 O_68 H_16} & R & -26.9 \\
 & $-1 \times$ & water & \ce{O H_2} & J & 12.7 \\
 & $1 / 4 \times$ & gamma\_surf-111\_P1\_2\_06w & \ce{Al_40 O_72 H_24} & B & -32.2 \\
\hline
\textbf{GSH} & $-1 / 2 \times$ & gamma\_surf-111\_P1\_2\_05w & \ce{Al_40 O_70 H_20} & R & -22.3 \\
 & $-1 \times$ & water & \ce{O H_2} & J & 3.1 \\
 & $1 / 2 \times$ & gamma\_surf-111\_P1\_2\_06w & \ce{Al_40 O_72 H_24} & B & -29.1 \\
\hline
\textbf{GSH} & $-1 / 2 \times$ & gamma\_surf-111\_P2\_1\_05w & \ce{Al_40 O_70 H_20} & R & -23.0 \\
 & $-1 \times$ & water & \ce{O H_2} & J & 60.2 \\
 & $1 / 2 \times$ & gamma\_surf-111\_P2\_1\_06w & \ce{Al_40 O_72 H_24} & B & -15.2 \\
\hline
\textbf{GSH} & $-1 / 2 \times$ & gamma\_surf-111\_P2\_2\_04w & \ce{Al_40 O_68 H_16} & R & -46.0 \\
 & $-1 \times$ & water & \ce{O H_2} & J & 29.6 \\
 & $1 / 2 \times$ & gamma\_surf-111\_P2\_2\_05w & \ce{Al_40 O_70 H_20} & B & -20.8 \\
\hline
\textbf{GSH} & $-1 / 4 \times$ & gamma\_surf-111\_P2\_2\_04w & \ce{Al_40 O_68 H_16} & R & -17.4 \\
 & $-1 \times$ & water & \ce{O H_2} & J & 3.1 \\
 & $1 / 4 \times$ & gamma\_surf-111\_P2\_2\_06w & \ce{Al_40 O_72 H_24} & B & -21.7 \\
\hline
\textbf{GSH} & $-1 / 2 \times$ & gamma\_surf-111\_P2\_2\_05w & \ce{Al_40 O_70 H_20} & R & 11.2 \\
 & $-1 \times$ & water & \ce{O H_2} & J & -23.5 \\
 & $1 / 2 \times$ & gamma\_surf-111\_P2\_2\_06w & \ce{Al_40 O_72 H_24} & B & -22.6 \\
\hline
\textbf{SUR} & $-1 / 32 \times$ & gamma\_bulk & \ce{Al_32 O_48} & R & 5.1 \\
 & $1 / 64 \times$ & gamma\_surf-001\_00w & \ce{Al_64 O_96} & J & 2.5 \\
 & & & & B & 4.9 \\
\hline
\textbf{SUR} & $-1 / 32 \times$ & gamma\_bulk & \ce{Al_32 O_48} & R & 5.7 \\
 & $1 / 80 \times$ & gamma\_surf-110b\_from\_boehm\_00w & \ce{Al_80 O_120} & J & 2.0 \\
 & & & & B & 3.6 \\
\hline
\textbf{SUR} & $-1 / 32 \times$ & gamma\_bulk & \ce{Al_32 O_48} & R & 7.7 \\
 & $1 / 64 \times$ & gamma\_surf-110b\_from\_bulk\_00w & \ce{Al_64 O_96} & J & 5.7 \\
 & & & & B & 7.3 \\
\hline
\textbf{SUR} & $-1 / 32 \times$ & gamma\_bulk & \ce{Al_32 O_48} & R & 9.3 \\
 & $1 / 48 \times$ & gamma\_surf-110l\_A1\_00w & \ce{Al_48 O_72} & J & 1.4 \\
 & & & & B & 7.9 \\
\hline
\textbf{SUR} & $-1 / 32 \times$ & gamma\_bulk & \ce{Al_32 O_48} & R & 7.4 \\
 & $1 / 56 \times$ & gamma\_surf-110l\_A2\_00w & \ce{Al_56 O_84} & J & 4.3 \\
 & & & & B & 6.6 \\
\hline
\textbf{SUR} & $-1 / 32 \times$ & gamma\_bulk & \ce{Al_32 O_48} & R & 7.2 \\
 & $1 / 56 \times$ & gamma\_surf-110l\_L2\_00w & \ce{Al_56 O_84} & J & 1.1 \\
 & & & & B & 5.2 \\
\hline

%% file: rmses.tex
\ce{Al-O}  & \si{\angstrom} & 0.08       & 0.35       & 0.07       & 1482 \\
\ce{Al-Al} & \si{\angstrom} & 0.37       & 0.33       & 0.07       & 1624 \\
\ce{O-H}   & \si{\angstrom} & 0.03       & 0.26       & 0.04       & 638 \\
\ce{O$\cdots$H} & \si{\angstrom} & 0.15       & 0.42       & 0.12       & 468 \\
\ce{Al-O-Al} & deg        & 22.0       & 9.3        & 3.3        & 1646 \\
\ce{Al-O-H} & deg        & 7.6        & 14.2       & 5.1        & 1077 \\
\ce{O-Al-O} & deg        & 32.2       & 11.0       & 3.2        & 3522 \\
\ce{Al-O$\cdots$H} & deg        & 8.6        & 12.8       & 4.9        & 803 \\
\ce{H-O$\cdots$H} & deg        & 15.8       & 18.2       & 8.1        & 484 \\
\ce{H$\cdots$O$\cdots$H} & deg        & 21.5       & 15.1       & 7.1        & 72 \\
\textbf{BSH} & \si{\kilo\cal \per \mol} & 11.8       & 38.1       & 4.7        & 5 \\
\textbf{GEH} & \si{\kilo\cal \per \mol} & 14.6       & 84.5       & 1.9        & 21 \\
\textbf{FOR} & \si{\kilo\cal \per \mol} & 10.3       & 22.4       & 2.4        & 11

%% file: 2-non-equilibrium-energies.tex
\section{Comparison of Non-Equilibrium ReaxFF and DFT energies}

A hydrated alumina slab (structure \texttt{gamma\_surf-110l\_A1\_06w} from the validation set) was used as a starting point for a constant-temperature molecular dynamics (MD) run with VASP using the same level of theory as the training set.
A Nosé-Hoover thermostat with a temperature of \SI{1000}{\kelvin} and a relaxation time of \SI{40}{\femto\second} was used to stimulate the desorption of water from the surface.
The simulation ran for 500 steps of \SI{1}{\femto\second}, and snapshots were taken every 10 steps for further analysis.
Two desorption events occurs during the first \SI{200}{\femto\second}, after which no additional water molecules desorb.

\begin{figure}
    \begin{center}
        \includegraphics{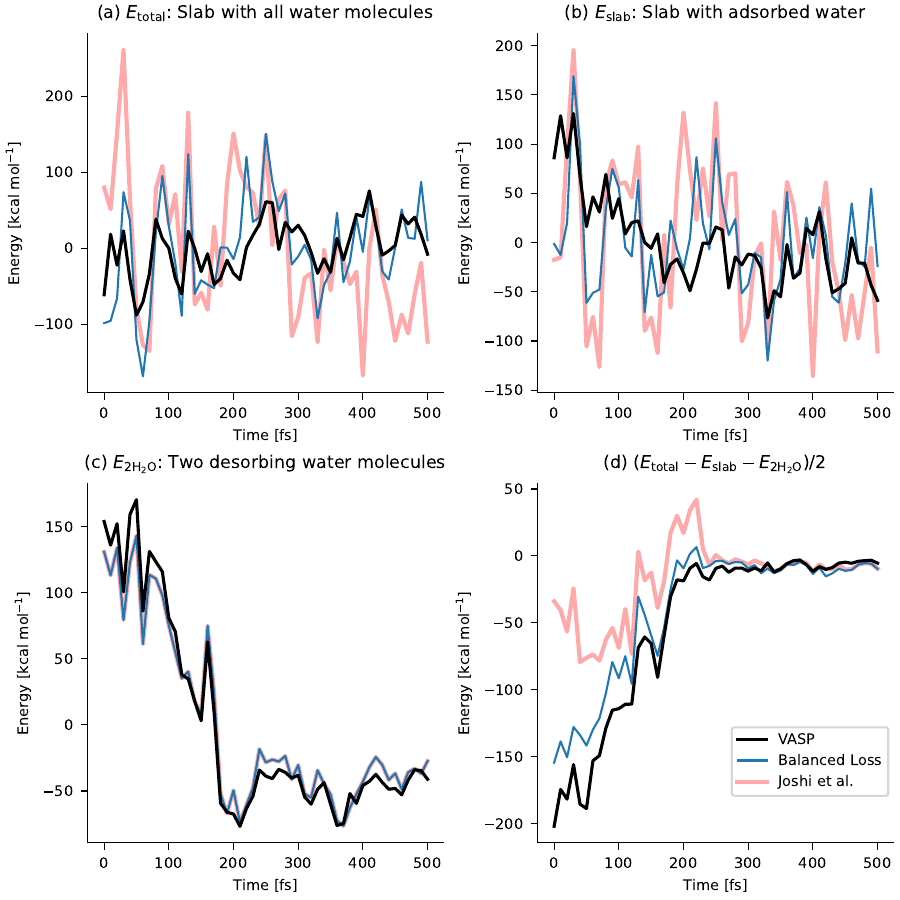}
    \end{center}
    \caption{
        Comparison of DFT and ReaxFF energies for snapshots taken from a \SI{1000}{\kelvin}
        DFT molecular dynamics simulation. (black: DFT, blue: ReaxFF with Balanced Loss parameters, red: ReaxFF with Joshi \textit{et al.} parameters\cite{joshi_reactive_2014}).
        Panel (a) contains the total energy of the system, panel (b) the energy of the slab and the water molecules that remain adsorbed, and panel (c) the energy of only the desorbing water molecules.
        Panel (d) depicts the instantaneous hydration energy as computed with Eq.~(3) in the main text.
    }
    \label{fig:nee}
\end{figure}

The total single-point energies, $E_\text{total}$, of the selected snapshots are compared in Fig.~\ref{fig:nee}(a).
The DFT energies are shown in black and ReaxFF results with the new parameters in blue.
ReaxFF energies obtained with the parameters of Joshi \textit{et al.}\cite{joshi_reactive_2014} are depicted in red.
The average is subtracted from both time series because these energies are only comparable up to a constant.
While the fluctuations in DFT and ReaxFF energy are correlated, they also show significant deviations.
This is expected, since the majority of the atoms are in the slab, not the water, and our training set emphasizes hydration reactions, not the vibrations in the alumina slab.

To show that our ReaxFF parameters can reproduce the relevant contribution to the DFT reference energy, additional single-point energies were calculated on the same snapshots from which some atoms were removed:
\begin{itemize}
    \item
    $E_\text{slab}$ is the energy of the slab and the water molecules that remain adsorbed, but without the atoms of the two water molecules that desorb during the simulation.
    \item
    $E_\text{2H$_2$O}$ is the energy computed for only the atoms of the two desorbing water molecules.
\end{itemize}
The energies of these two complementary subsystems are shown in Fig.~\ref{fig:nee}(b) and Fig.~\ref{fig:nee}(c), respectively.
(Since these are also absolute energies, the average is again subtracted in both plots.)
Finally, using these data, also the energy difference $(E_\text{total} - E_\text{slab} - E_\text{2H$_2$O}) / 2$ was computed and is shown in Fig.~\ref{fig:nee}(d).
In this case, no average was subtracted because the difference in energy has a chemically meaningful reference.

Fig.~\ref{fig:nee}(b) shows deviations between ReaxFF and DFT energies that are very similar to those in Fig.~\ref{fig:nee}(a), confirming that these deviations are due to the internal energy of the alumina slab.
The results in Fig.~\ref{fig:nee}(c) and Fig.~\ref{fig:nee}(d) show a fair agreement between the DFT and ReaxFF energies obtained with our new parameters.
Mainly Fig.~\ref{fig:nee}(d) is of interest, because it is closely related to the hydration energies in the training set: This energy difference is calculated similarly to energy training data (Eq.~(3) in the main text) but is now evaluated using non-equilibrium snapshots instead of optimized geometries.
Because of this similarity, it is reasonable to expect a correspondence of the energies in Fig.~\ref{fig:nee}(d).
The average of the relative error over the first \SI{200}{\femto\second} is \SI{25}{\percent} for the Balanced Loss parameters, which is comparable to relative errors on adsorption energies in the training set.
For Fig.~\ref{fig:nee}(a) and Fig.~\ref{fig:nee}(b), however, a similar agreement would have been coincidental, since no related data were used for training.
Also note that the new ReaxFF parameters show a clear improvement in Fig.~\ref{fig:nee}(d) with respect to the parameters of Joshi \textit{et al.},\cite{joshi_reactive_2014} for which the relative error is \SI{79}{\percent}.
In Fig.~\ref{fig:nee}(c) both ReaxFF parameterizations yield the same results because the parameters for water were not refitted.